\documentclass[11pt,a4paper]{article}
\pdfoutput=1
\usepackage{jcappub}

\usepackage[english]{babel}
\usepackage{xspace} 
\usepackage[tight]{subfigure} 

\usepackage{cosmodefs}

\newcommand*{\argstyle}[1]{\left( { \textstyle #1 } \right) }
\newcommand*{\kdom}{\mathcal{V}_\mathcal{T}}
\newcommand*{\modeQ}{\mathcal{Q}}
\newcommand*{\orthoR}{R}
\newcommand*{\sshape}{S}
\newcommand*{\Sshape}[2][\sshape]{{#1}_{#2}}
\newcommand*{\Sshaperm}[2][\sshape]{{#1}_\mathrm{#2}}
\newcommand*{\ashape}{\mathcal{A}}
\newcommand*{\Ashape}[1]{\ashape_{#1}}
\newcommand*{\Ashaperm}[1]{\ashape_\mathrm{#1}}
\newcommand*{\ishape}{\mathcal{I}}
\newcommand*{\Ishape}[1]{\ishape_{#1}}

\newcommand*{\normEtc}[1]{N\left[#1\right]}
\newcommand*{\Lop}[1]{L\left[#1\right]}
\newcommand*{\scaling}{\left(\frac{K_{111}}{2 K^3}\right)^{n_s - 1}}
\newcommand*{\cosAlpha}{\cos\frac{\boldalpha \pi}{2}}
\newcommand*{\bc}{{\bf c}}
\newcommand*{\es}{\epsilon_s}
\newcommand*{\ep}{\epsilon}
\newcommand*{\ct}{c_s^2}
\newcommand*{\ctInv}{c_s^{-2}}
\newcommand{\boldalpha}{\mbox{\boldmath$\alpha$}}

\title{Anatomy of bispectra in general single-field inflation -- modal expansions}
\author{Thorsten Battefeld and}
\author{Jan Grieb}

\affiliation{Institute for Astrophysics, University of Goettingen,\\
Friedrich-Hund-Platz 1, D-37077 Goettingen, Germany}

\emailAdd{jgrieb@astro.physik.uni-goettingen.de}
\emailAdd{tbattefe@astro.physik.uni-goettingen.de}

\abstract{
We discuss bispectra of single-field inflationary models described by general Lorentz invariant Lagrangians that are at most first order in field derivatives, including the fast-roll models investigated by Noller and Magueijo. Based on a factor analysis, we identify the least correlated basic contributions to the general shape and show quantitatively which templates provide a good approximation. We compute how relative contributions of basic shapes to the total bispectrum scale as slow roll is relaxed. To enable future comparison with CMB observations, we provide a modal expansion of these non-separable bispectra in Fourier space, employing the formalism by Shellard et~al. Convergence is rapid, usually better than ninety-five percent with less than thirty modes, due to the smoothness of these primordial shapes. 

Truncated polynomial modal expansions have restrictions, which we highlight using an example with slow convergence. The particular shape originates from particle production during inflation (common in trapped inflation) and entails  both localized and oscillatory features. We show that this shape can be recovered  efficiently using a Fourier basis and outline the prospect of future model parameter extraction and N-body simulations based on modal techniques.
}


\begin{document}
\maketitle

\section{Introduction}
\label{sec:intro}
Observations of primordial curvature perturbations by measurements of the cosmic microwave background radiation (CMB) and large-scale structure (LSS) are in good agreement with predictions of inflationary scenarios. However, the number of different models that predict almost scale-invariant, Gaussian fluctuations is vast. 
Non-Gaussianities (NG) allow for further discrimination if they are observed in current or upcoming CMB \cite{Planck_BlueBook,Ade:2011ah} and LSS \cite{SDSS,Giannantonio:2011ya} measurements (\ie\ via the scale dependent bias \cite{Dalal:2007cu,Matarrese:2008nc,Slosar:2008hx,Xia:2011hj}).

Up until recently, the main focus has been on predicting and measuring the amplitude of three separable shape templates of the bispectrum (see \cite{Huterer:2010en} for a collection of reviews): the \emph{local} one \cite{Gangui:1993tt,Verde:1999ij,Komatsu:2001rj}, easily recovered by means of the (non-linear) $\delta N$-formalism \cite{Lyth:2004gb,Lyth:2005fi}, the \emph{equilateral} one, often used to approximate the bispectrum of DBI inflation \cite{Alishahiha:2004eh} or scenarios with more general non-canonical kinetic terms \cite{Chen:2006nt}, and an orthogonal shape which is hard to generate directly (see however \cite{RenauxPetel:2011dv} for a possible source).  

Any detection of primordial NG would rule out simple single-field models, since the bispectrum is slow-roll suppressed \cite{Maldacena:2002vr}. The magnitude of the three templates in the data can be expressed by a single, scalar amplitude, the \emph{non-linearity parameter} $\fNL$. A recent CMB analysis based on the WMAP7 data set alone \cite{Komatsu:2010fb} yields at the $95\%$ confidence level
\begin{align}
 -10 \le \FNL{local} &\le 74, & -254 \le \FNL{equil} &\le 306, & -410 \le \FNL{orthog} \le 6,
 \label{eq:fNL_WMAP7}
\end{align}
consistent with a Gaussian spectrum. The data analysis, using estimators \cite{Creminelli:2005hu} that are optimal even in the presence of foreground contamination, depends crucially on separability of the bispectrum to reduce the dimensionality of integrals. 
Such an analysis is not feasible for non-separable shapes, since computing time of an optimal estimator scales with $\Ord{l_\mathrm{max}^5}$; here $l$ is the multipole number ($l_\mathrm{max}\sim 2000$ for PLANCK \cite{Planck_BlueBook,Ade:2011ah}).

Albeit convenient, relying on templates is problematic. 
First, a quantitative understanding of how well the three templates approximate real bispectra is needed.
In this paper we carry out this analysis for general single-field models (Lorentz invariant Lagrangians at most first order in field derivatives). These  models were first discussed in \cite{Seery:2005wm} under the assumptions of slow roll and small deviations from unity of the speed of sound. In \cite{Chen:2006nt,Chen:2010xka} an arbitrary speed of sound was allowed and in \cite{Noller:2011hd} slow roll was relaxed to yield the general bispectrum investigated further in this paper.\footnote{Deviations from slow roll can still be consistent with scale invariance of the power spectrum due to possible cancellations; in \cite{Noller:2011hd} bispectra of maximally slow-roll violating models (up to $\epsilon \sim 0.3$) were constructed.}
After dissecting the resulting general shape into a handful of different contributions/subclasses, we perform a factor analysis to identify the least correlated basic shape constituents; we compute the overlap with  templates and show that a restriction to common ones is usually sufficient for these models (see \cite{Ribeiro:2011ax} for related work on single-field Galileon models that appeared while completing this article; our conclusions are in line with \cite{Ribeiro:2011ax}).

Nevertheless, examples of non-separable shapes that show little overlap with the templates are known, \eg\ \cite{Barnaby:2010ke,Barnaby:2010sq}. Furthermore, by relying on a single number such as $f_{NL}$ to characterize non-Gaussianities, we waste the opportunity to tell individual sources apart (primordial/foregrounds/artefacts) since their respective shape function can act as a unique fingerprint.

To ameliorate these shortcomings, Shellard \etAl\ proposed an
expansion of the bispectrum into a complete set of separable modes (polynomials) \cite{Fergusson:2009nv}. The expansion coefficients of the late-time bispectrum in the CMB data can be calculated using a generalized version of the WMAP algorithm for separable shapes, which scales as $\Ord{l_\mathrm{max}^3}$.
Several concrete models proposed in the literature have been analyzed \cite{Fergusson:2008ra, Fergusson:2009nv, Fergusson:2010dm, Liguori:2010hx} and tested for convergence, with positive results. Among these are, in addition to the ones mentioned above, warm inflation \cite{Gupta:2002kn},
oscillatory features in the potential leading to resonances \cite{Chen:2008wn, Chen:2010bka} and ghost inflation \cite{ArkaniHamed:2003uz}.  A related mode expansion for the trispectrum was proposed in \cite{Regan:2010cn,Fergusson:2011sa} and an expansion into  Fourier modes has been proposed in \cite{Meerburg:2010ks} with the aim to enhance convergence for oscillatory shapes.

The formalism appears promising to perform a blind search for late-time NG signals, \eg\ caused by cosmic strings \cite{Fergusson:2009nv}, and can be used to tell apart foregrounds/artefacts from primordial signals in the PLANCK data analysis. Current developments entail the estimation of NG signals in the large-scale structure \cite{Fergusson:2010ia} and the generation of non-Gaussian initial conditions for N-body simulations \cite{Regan:2011zq,Steffen}.

In this paper we apply the (polynomial) modal expansion to the general single-field models described above.
We find that naively chosen basic shapes fall into three categories that cannot be separated further observationally at present, due to their strong correlation with templates. Using a factor analysis, we identify least correlated basic shape constituents that offer better observational separability. Focusing on concrete models,  we show that the three dominant contributions correspond to the three common templates; the remaining distinct subclasses are subdominant and hard to detect due to a reduced but still significant degree of correlation. In addition, we identify systematics that complicate reconstruction of subdominant constituents.

We further show how the relative contributions of the subclasses to the total bispectrum scale as slow roll is relaxed (the relaxation of slow roll requires fine-tuning to retain a nearly scale-invariant power spectrum). As a consequence, one could in principle pinpoint the concrete model if non-Gaussianities were observed. Particularly, if expansion coefficients were extracted from observations, one could perform a Markov chain Monte Carlo (MCMC) analysis to scan the parameter space and identify the best fit. To this end, the identification of well chosen basic shape constituents is crucial to perform this analysis efficiently: if basic shapes are chosen properly, the expansion coefficients of a general shape can be obtained from a weighted sum of basic shape coefficients (only weights depend on model parameters), which need to be computed only once.

As an aside, we comment on work in progress on applying modal techniques to N-body simulations.

However, truncated modal expansions have restrictions: we investigate a shape with both localized and oscillatory features in the bispectrum \cite{Barnaby:2010ke,Barnaby:2010sq} that has slow convergence for a polynomial expansion.
This shape originates from backscattering of particles produced during inflation from the inflaton condensate \cite{Barnaby:2009mc,Barnaby:2009dd}, a common phenomenon in trapped inflation \cite{Green:2009ds,Kofman:2004yc,Langlois:2009jp,Battefeld:2010sw,Battefeld:2011yj} or monodromy inflation \cite{Silverstein:2008sg}. However, we find faster convergence for this shape using the Fourier expansion of Meerburg \cite{Meerburg:2010ks}. Furthermore, Fourier expansion coefficients are better suited to identify key frequencies in the bispectrum, leading to improved model identification properties.

The paper is organized as follows:  we start with a pedagogical introduction to modal decompositions in \sectRef{sec:sepExp}. We focus on  the expansion of the bispectrum in separable polynomial modes as introduced by Shellard \etAl, and explain the considerably involved notation in detail. \SectRef{sec:genSingleFieldModels} contains a review of the generalized single-field models by Seery and Lidsey \cite{Seery:2005wm} (\sectRef{sec:conventionalSlowRoll}), Chen \etAl \cite{Chen:2006nt,Chen:2010xka} (\sectRef{sec:slowVariation}), as well as Noller and Magueijo \cite{Noller:2011hd} (\sectRef{sec:fastRoll}). Readers familiar with the modal decomposition may want to skip \sectRef{sec:sepExp}, while readers familiar with general single-field models may want to skip \SectRef{sec:genSingleFieldModels}. We separate general bispectra of singe-field models into subclasses, identify the least correlated basic shape constituents and perform a modal decomposition in \sectRef{sec:results}. Our results are summarized in \sectRef{sec:summary} and related to recent work in \sectRef{sec:recent_work}. We follow with an outlook to future work, especially the application of modal techniques to N-body simulations, in \sectRef{sec:outlook}. Fourier modal expansions are discussed in \sectRef{sec:oscFeat_altModes} and applied to non-Gaussianities from particle production during inflation. Some lengthy expressions are provided in the appendix.

\subsection{Non-Gaussianities, the bispectrum and the shape function}
\label{sec:NGintro}

Non-Gaussianities entail the study of higher order correlation functions beyond the two-point function, or power spectrum. In this paper, we focus on the three-point correlation function of the comoving curvature perturbation $\comovR$ using the modal expansion of \cite{Fergusson:2009nv,Fergusson:2010dm}. The Fourier transform of the thre-point function, the \emph{bispectrum of primordial fluctuations $\bspect{\comovR}(k_1,k_2,k_3)$}, is the object of interest; it is defined as
\begin{align}
 \average{ \comovR(\V k_1) \comovR(\V k_2) \comovR(\V k_3) } = (2 \pi)^3 \; \bspect{\comovR}(k_1,k_2,k_3) \; \delta( \V k_1 + \V k_2 + \V k_3 ),
\end{align}
where $(2 \pi)^3$ is a normalization factor.\footnote{Note that $\bspect{\comovR}$ is sometimes called $F$, for example in \cite{Senatore:2009gt}.} The Dirac delta function imposes a triangle condition onto the wavevectors $\V k_1$, $\V k_2$, and $\V k_3$, ensuring momentum conservation as required by translational invariance (homogeneity and isotropy of the universe). As a consequence, the bispectrum depends only on the magnitude of the wavevectors, the wavenumbers $k_1$, $k_2$, and $k_3$.
To simplify the analysis, we introduce a normalized, dimensionless shape function $\shape (k_1, k_2, k_3)$, where an overall scaling with $k^{-6}$ is removed,
\begin{align}
 \shape(k_1, k_2, k_3) \defeq \frac 1 N \; (k_1 k_2 k_3)^2 \; \bspect{\comovR}(k_1,k_2,k_3).
 \label{eq:shapeFunction}
\end{align}
Here, $N$ is chosen such that $\sshape(k,k,k) = 1$; this can be achieved for the most common scale-invariant templates with $N = \const$.

\begin{figure*}[t]
 \centering
 \subfigure[~tetrapyd domain $\kdom$]{
  \label{fig:tetrapyd}
  \includegraphics[width=.25\textwidth]{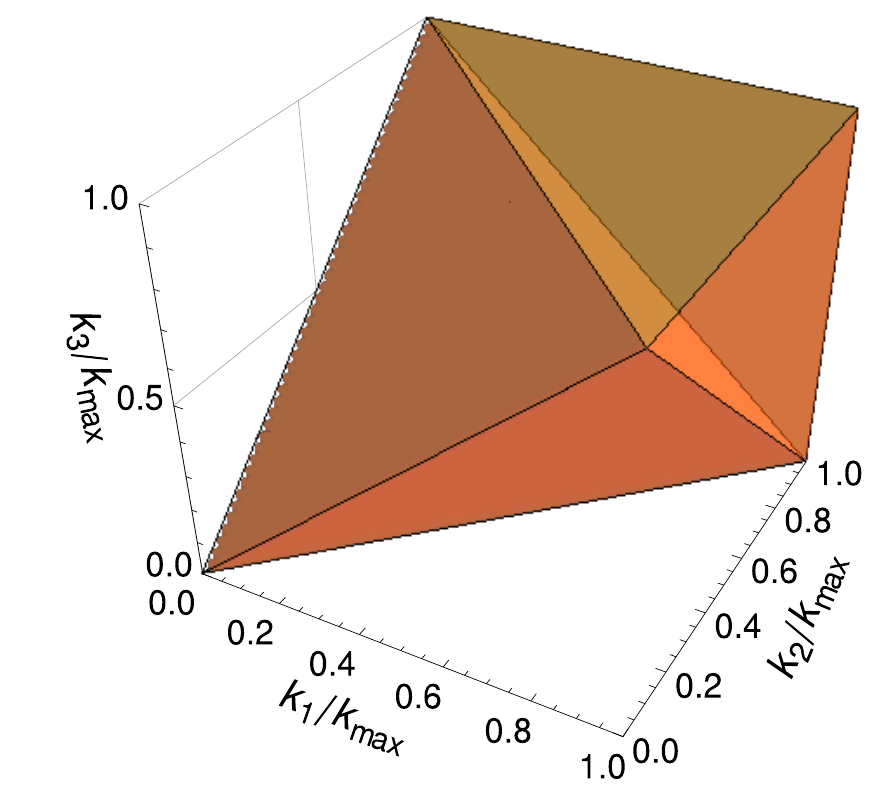}
  \hspace{1cm}
 }
 \subfigure[~NG shapes]{
  \label{fig:ng_shapes}
  \includegraphics[width=.4\textwidth]{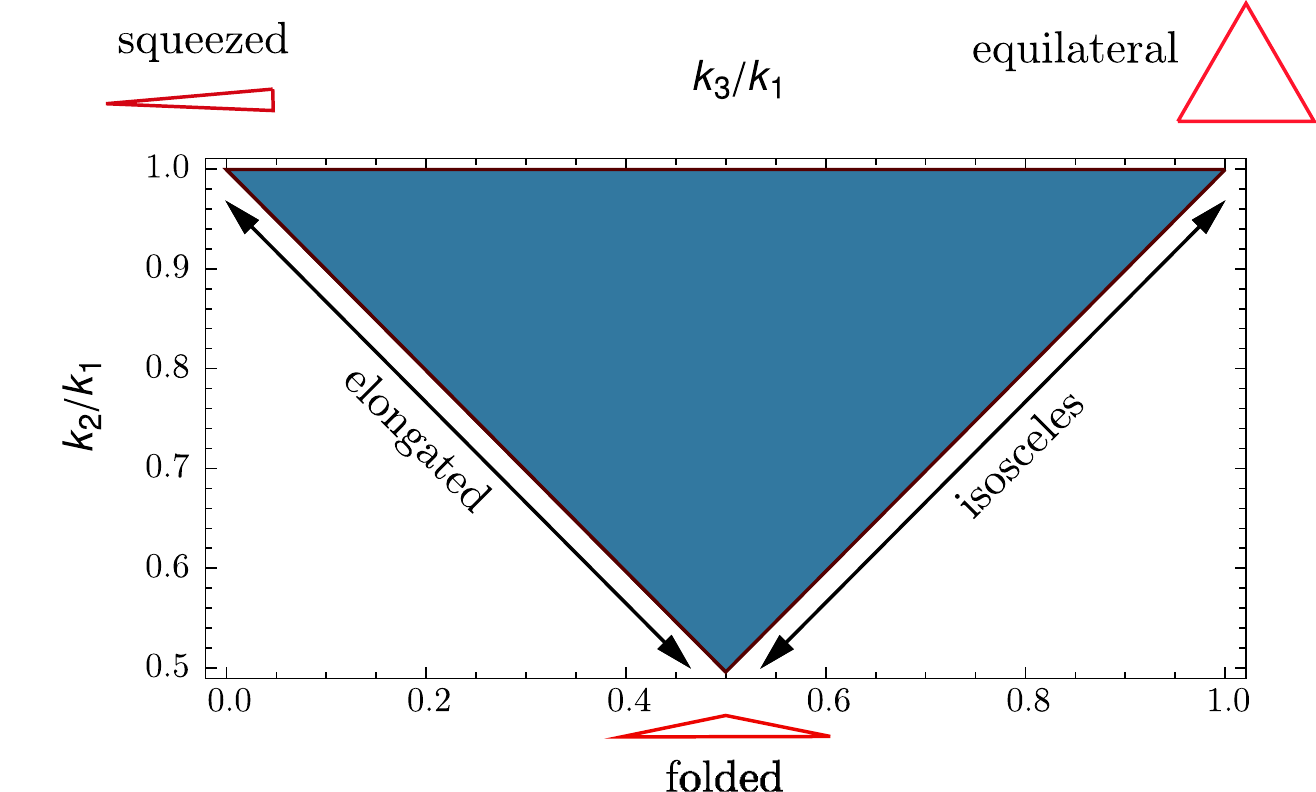}
 }
 \caption[The tetrapyd domain and common separable shape templates of non-Gaussianities]{\textbf{(a)} The \emph{tetrapyd} domain $\kdom$ as given by \eqRef{eq:tetrapyd}. \textbf{(b)} Shape templates, taken from \cite{Baumann:2009ds}, \fig{30}.}
 \label{fig:tetrapy_and_shapes}
\end{figure*}

The momentum space domain of $\shape$ is tetrahedral, but given the existence of a minimally observable length scale, the domain is capped by some  $k_\mathrm{max}$ resulting in a ``tetrapyd'',
\begin{equation}
 \begin{aligned}
  \kdom \colon \quad k_1 \le k_2 + k_3 \text{ for } k_1 &\ge k_2,k_3, \\
  k_2 \le k_1 + k_3 \text{ for } k_2 &\ge k_1,k_3, \\
  k_3 \le k_1 + k_2 \text{ for } k_3 &\ge k_1,k_2, \\
  k_1,k_2,k_3 &\le k_\mathrm{max},
 \end{aligned}
 \label{eq:tetrapyd}
\end{equation}
see \figRef{fig:tetrapyd}.

\subsection{Shape classes}
\label{sec:shape_classes}

\begin{figure*}[t]
 \centering
 \subfigure[~$\Sshaperm{equi}$]{
  \label{fig:equi_shape}
  \begin{minipage}[c]{4cm}
   \includegraphics[width=1\textwidth]{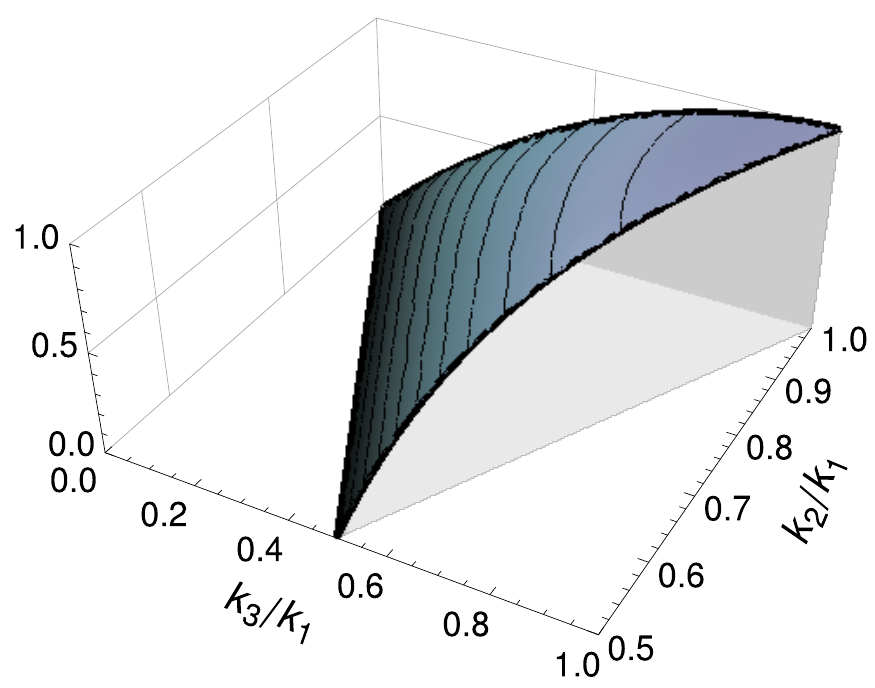}
  \end{minipage}
  \begin{minipage}[c]{6cm}
   \includegraphics[width=1\textwidth]{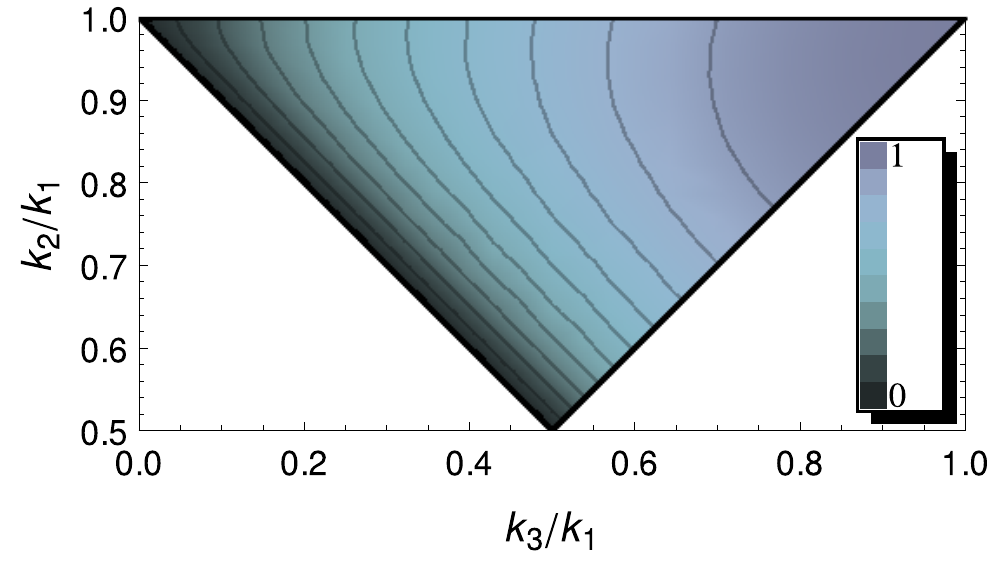}
  \end{minipage}
 }
 \subfigure[~$\Sshaperm{local}$]{
  \label{fig:local_shape}
  \begin{minipage}[c]{4cm}
   \includegraphics[width=1\textwidth]{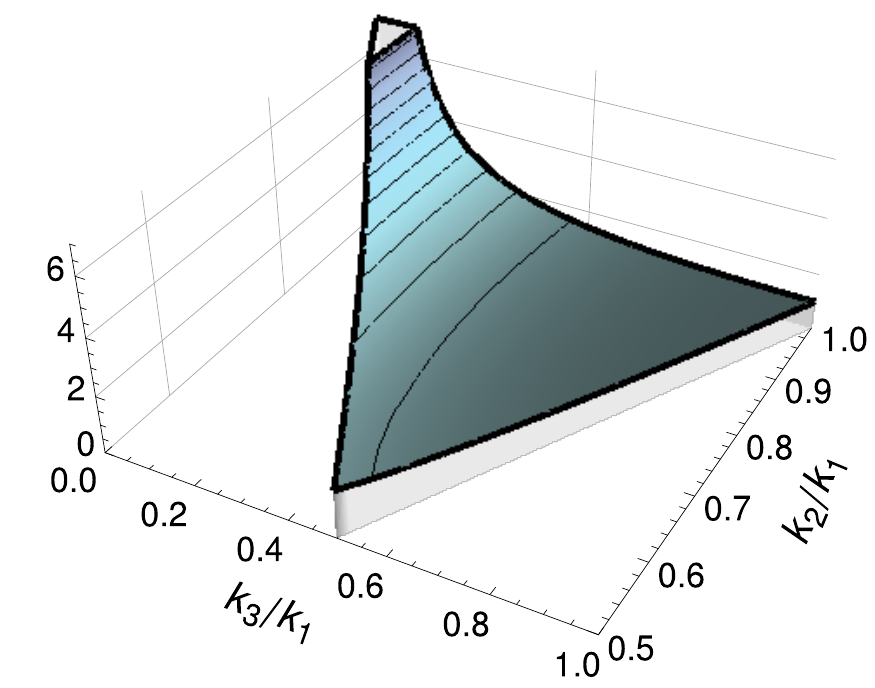}
  \end{minipage}
  \begin{minipage}[c]{6cm}
   \includegraphics[width=1\textwidth]{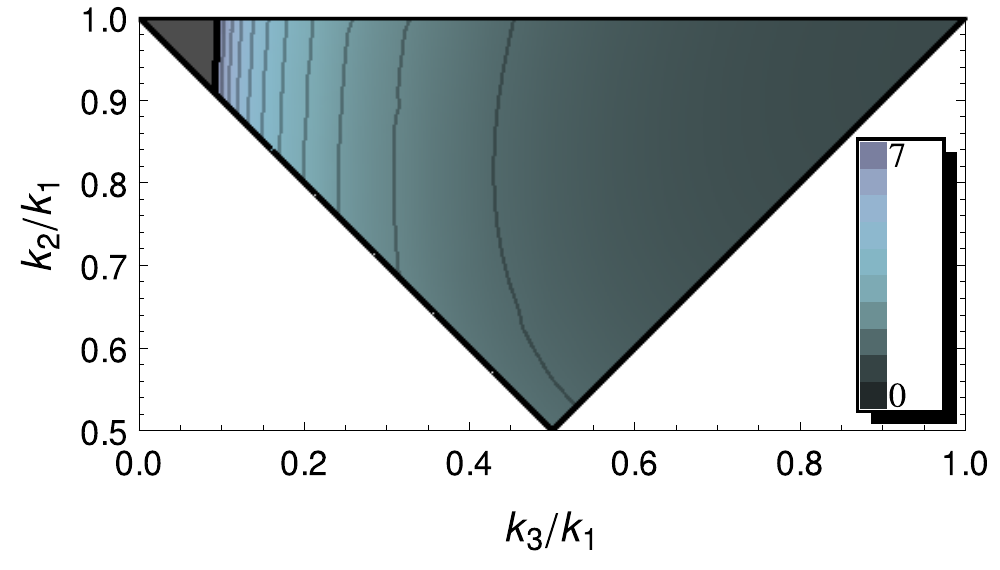}
  \end{minipage}
 }
 \subfigure[~$\Sshaperm{single}$]{
  \label{fig:single_shape}
  \begin{minipage}[c]{4cm}
   \includegraphics[width=1\textwidth]{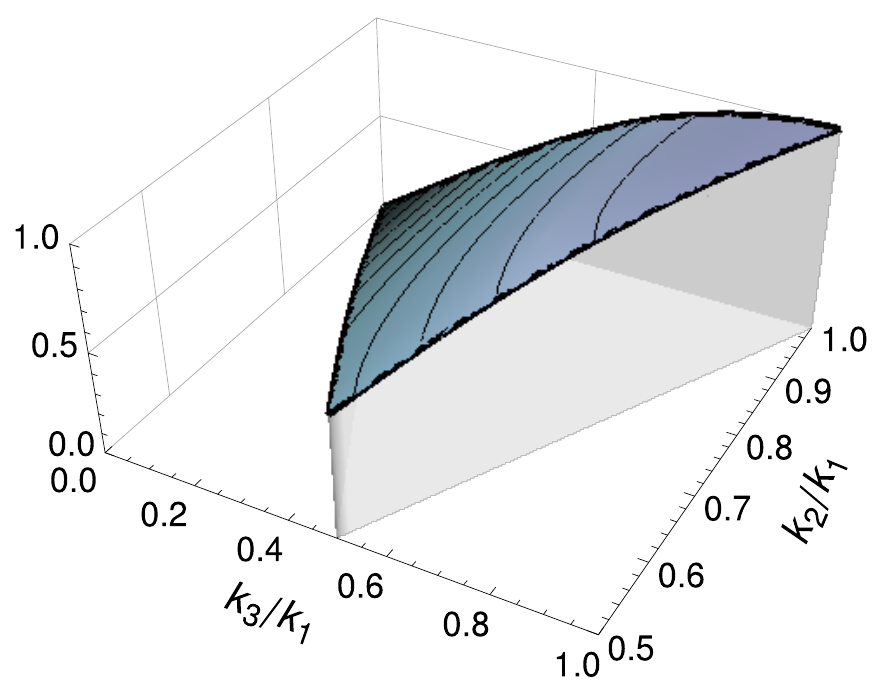}
  \end{minipage}
  \begin{minipage}[c]{6cm}
   \includegraphics[width=1\textwidth]{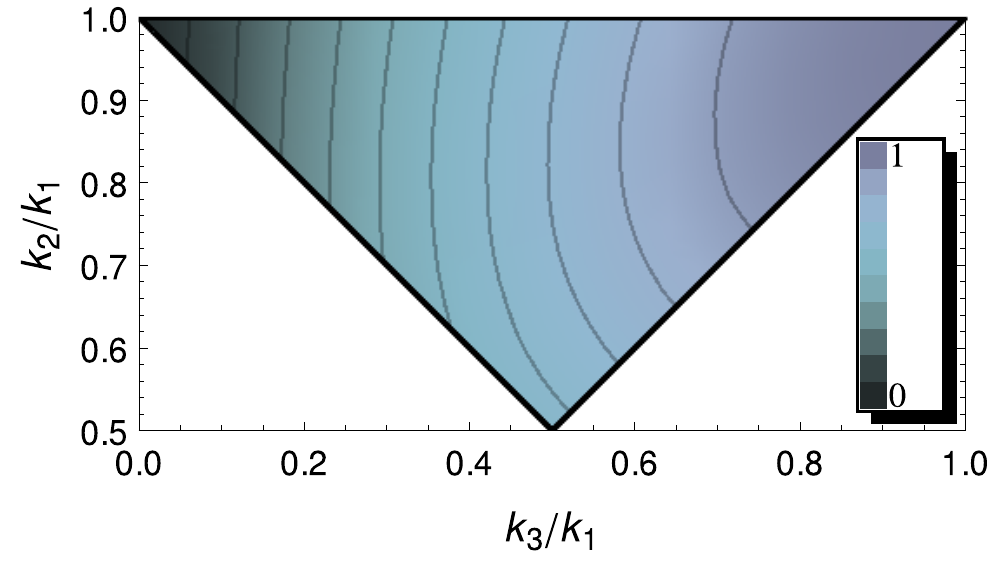}
  \end{minipage}
 }
 \caption{Plots of some illustrative shapes as introduced in \cite{Baumann:2009ds}; see \figRef{fig:ng_shapes} for interpretation. \textbf{(a)}~The equilateral shape. \textbf{(b)}~The local shape that peaks on squeezed triangles. \textbf{(c)}~The so-called ``single'' shape $\Sshaperm{single} = K_{111} / K_3$ that is non-zero for enfolded triangles. }
 \label{fig:common_shapes}
\end{figure*}

Do certain \emph{classes} of shapes exist? Since we will encounter many shapes in this paper, it is prudent to give a brief overview of known shapes,  
following \cite{Liguori:2010hx}. In \figRef{fig:common_shapes} we plot three illustrative shape functions $\sshape(1, k_2 / k_1, k_3 / k_1)$. Because of the symmetries of $\sshape$ on the tetrapyd domain, it is sufficient to plot momenta satisfying $k_3 / k_1 < k_2 / k_1 < 1$, see \eqRef{eq:tetrapyd}; the triangle inequality requires $k_2 / k_1 + k_3 / k_1 > 1$. All models discussed here are scale-invariant.

First, there are center-weighted models that peak on equilateral triangles. The most prominent shape function in this class is the equilateral template $\Sshaperm{equi}$, see \figRef{fig:equi_shape}, which is an approximation to the dominant bispectrum contribution in DBI inflation \cite{Silverstein:2003hf,Alishahiha:2004eh} (reviewed in \cite{McAllister:2007bg}).
We give an explicit formula for the shape in \eqRef{eq:equi_shape}. The orthogonal shape proposed by Smith \etAl\ \cite{Senatore:2009gt} is roughly given by the equilateral shape with a constant removed, rendering it similar to the so called enfolded shape proposed in \cite{Meerburg:2009ys} (see also \cite{Creminelli:2010qf,Burrage:2011hd} for closely related shapes.).

A second class encompasses corner-weighted models, which peak on squeezed triangles ($k_1 \ll k_2 \simeq k_3$ and permutations). They feature a divergence in the $k_i \to 0$ limit, as seen in \figRef{fig:local_shape}. This local shape is featured in many multi-field models \cite{Byrnes:2010em} but also more exotic proposals such as the ekpyrotic scenario \cite{Lehners:2010fy} among others; it can be recovered easily via the non-linear $\delta N$-formalism \cite{Lyth:2004gb,Lyth:2005fi}.

We complete our brief overview with edge-weighted models that peak on flattened triangles. There are only few examples known to have a dominant contribution of this type, for example models with non-Bunch Davies initial conditions \cite{Holman:2007na}. They also constitute one of the non-suppressed but unobservable small contributions in standard slow-roll single-field models. Fergusson and Shellard call this contribution the $\Sshaperm{single}$ template, plotted in \figRef{fig:single_shape}. We address this shape in \sectRef{sec:systKaysNM}.

Various other shapes exist in the literature, some of which include explicit scale dependence or localized  features, highlighting the need for a unified approach to accommodate general shapes.

\section{Separable, polynomial expansion of the bispectrum}
\label{sec:sepExp}

To deal with general primordial shapes,\footnote{We do not perform a decomposition in spherical harmonic space, which requires an evolution of the shape to later times; the latter is needed to make direct contact with CMB observations \cite{Fergusson:2010dm}. We give an outlook to the late-time CMB bispectrum in \sectRef{sec:late-time}.} we briefly review the separable mode expansion of the bispectrum using tetrahedral polynomials, as introduced in \cite{Fergusson:2009nv,Fergusson:2010dm}. Our goal is to analyze a general shape function $\shape(k_1, k_2, k_3)$ by a decomposition into orthonormal or separable mode functions. The definition of either basis is explained in \sectRef{sec:orthoBasis}. The conversion matrix between respective expansion coefficients is provided in \eqRef{eq:alphaConvMatrix}. 

Since we need to compute the correlation between shapes, we start be defining an integration over the tetrapyd domain $\kdom$
\begin{align}
 \mathcal{T}[f] \defeq \int_{\kdom} f(k_1, k_2, k_3) \, w(k_1,k_2,k_3) \dint \kdom,
\end{align}
employing a weight function $w(k_1,k_2,k_3)$ that should be chosen to reflect the scaling of the CMB bispectrum estimator.
To remain compatible with work done by Fergusson, Shellard and Meerburg, 
we employ $w = 1$ for the analysis on the primordial wavenumber domain.\footnote{For the CMB bispectrum, the weight in \cite{Fergusson:2009nv}, \eq{48}, should be employed. Note that there is a mistake in one of the equations in \cite{Fergusson:2009nv} that we correct in appendix \ref{sec:lSpaceExp}. In order to prepare the analysis of primoridial bispectra for the multipole expansion, one should use the weight $w = 1/(k_1 + k_2 + k_3)$ as mentioned by Meerburg in footnote 4 of \cite{Meerburg:2010ks}.}
An inner product is imposed by letting 
\begin{align}
 \iprod{f}{g} \defeq \mathcal{T}[f \conj{g}],
\end{align}
 where a star denotes complex conjugation (not needed for polynomials, but required for the Fourier modes of \cite{Meerburg:2010ks}, see \sectRef{sec:MeerburgModes}). 
The correlation between two shapes $\Sshape{1}$ and $\Sshape{2}$ can then be measured by the normalized inner product on the tetrapyd
\begin{align}
 C_{12} \defeq \frac{\iprod{\Sshape{1}}{\Sshape{2}}}{\sqrt{\iprod{\Sshape{1}}{\Sshape{1}}} \sqrt{\iprod{\Sshape{2}}{\Sshape{2}}}}.
 \label{eq:correlation}
\end{align}
In \cite{Babich:2004gb} and related works, this correlation coefficient is called the ``cosine'' between shapes.

\subsection{Construction of an orthonormal basis}
\label{sec:orthoBasis}

\begin{figure*}[t]
 \centering
 \includegraphics[width=0.6\textwidth]{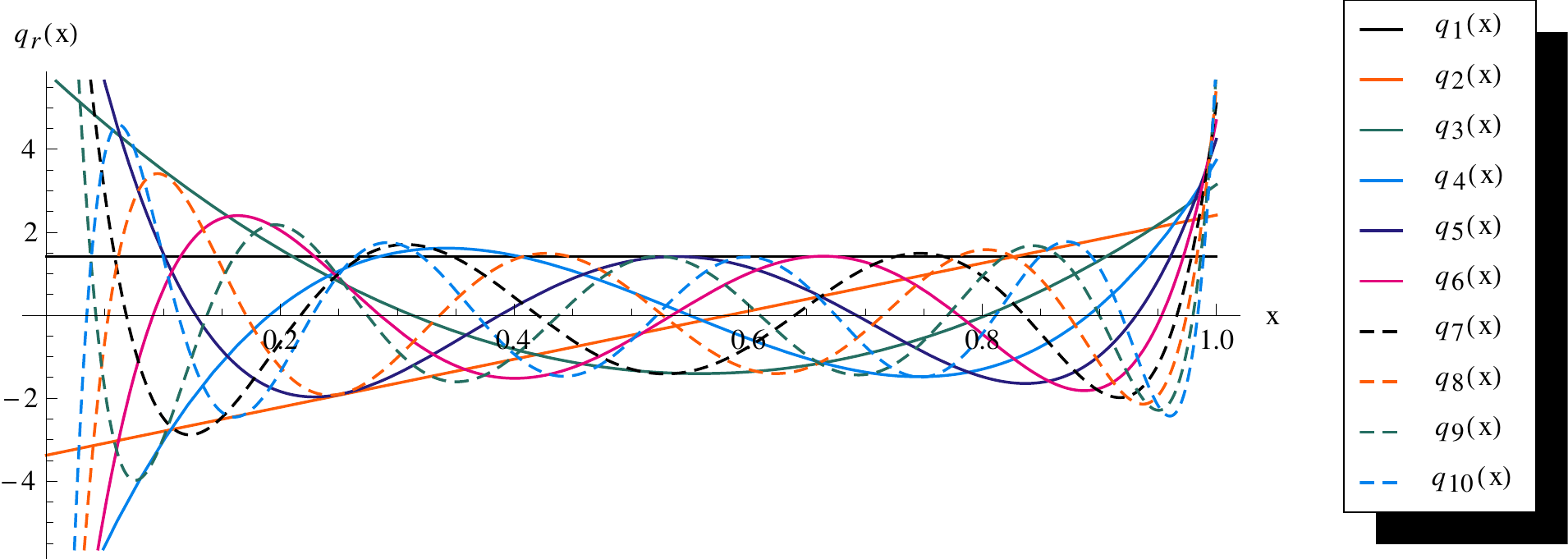}
 \caption{Plot of the first ten one-dimensional basis functions, $q_r(x)$ (compare to \cite{Fergusson:2009nv}, \fig{5}).}
 \label{fig:poly_qn}
\end{figure*}

Let us rescale wavenumbers by\footnote{Unless scale-dependent bispectra are discussed, the maximal wavenumber is irrelevant.}
\begin{align}
 x_i \defeq k_i / k_\mathrm{max}
 \label{eq:rescaled}
\end{align}
and generate orthonormal polynomials in a one-dimensional subspace as a starting point. To this end, we define the reduced weight function
\begin{align}
 \tilde w(x) \defeq \int_{\mathcal{V}_{\mathcal{T},x_2,x_3}} w(x,x_2,x_3) \dint \mathcal{V}_{\mathcal{T},x_2,x_3},
 \label{eq:wtilde}
\end{align}
where we integrated out $x_2$ and $x_3$, as well as a reduced, one-dimensional integral
\begin{align}
 \mathcal{T}_x[f] \defeq \int_0^1 f(x) \; \tilde w(x) \dint x.
\end{align}
Using $w(x,y,z) = 1$, we get $\tilde w(x) = \frac 1 2 x(4-3x)$. As the basis for the polynomials we choose monomials, 
\begin{align}
 e_r \defeq x^r, \quad r \in \N_0.
 \label{eq:1dMonomialBasis}
\end{align}
Defining
\begin{align}
 w_r \defeq \mathcal{T}_x[e_r] = \frac{r+6}{2(r+3)(r+2)},
\end{align}
we arrive at orthonormal polynomials (\wrt\ the reduced weight function) via
\begin{align}
 q_r(x) \defeq \frac{1}{\mathcal{N}} \left| \begin{array}{cccc}
  w_0 & w_1 & \ldots & w_r \\
  w_1 & w_2 & \ldots & w_{r+1} \\
  \ldots & \ldots & \ldots & \ldots \\
  w_{r-1} & w_r & \ldots & w_{2r-1} \\
  e_0 & e_1 & \ldots & e_r
 \end{array} \right|.
\end{align}
We choose $\mathcal{N}$ such that $\mathcal{T}_x[q^2_r]=1$ for all $r$, so that the polynomials $q_r(x)$ fulfill
\begin{align}
 \iprod{q_r}{q_s}_x \defeq \mathcal{T}_x[q_r q_s^\ast] = \delta_{rs}.
\end{align}
Analytic expressions for the first five functions can be found in \cite{Fergusson:2009nv}, \eq{54}. See \figRef{fig:poly_qn} for a plot of the first ten orthonormal polynomials.

Due to the permutation symmetry on the tetrypyd domain, we define a three\nbd{}dimensional complete set of functions as a symmetrized superposition
\begin{equation}
 \begin{aligned}
  \modeQ_n(x,y,z) \defeq q_{\{p} q_{r} q_{s\}} &= \frac{1}{6 \mathcal{N}} \big[q_p(x) q_r(y) q_s(z) + q_r(x) q_s(y) q_p(z) + q_s(x) q_p(y) q_r(z) \\
  &\quad + q_p(x) q_s(y) q_r(z) + q_s(x) q_r(y) q_p(z) + q_r(x) q_p(y) q_s(z) \big].
 \end{aligned}
 \label{eq:Qsymmetrization}
\end{equation}
The definition of the super-index, the mode number $n$, is arbitrary;\footnote{One may want to investigate a different ordering to improve rapid convergence for common shapes. The chosen order has proven sufficient, but is not the result of any optimization.} we map $n$ to the triplet $\{p,r,s\}$ using the ``slicing'' ordering, where the $\{p,r,s\}$ are sorted in ascending order of the quantity $p^2 + q^2 + r^2$. To clarify the ordering, we print the $n \leftrightarrow (p,r,s)$ mapping up to $n=53$ in appendix \ref{sec:slicingOrdering}. We choose $\mathcal{N}$ such that $\mathcal{T}[\modeQ^2_r]=1$ for all $n$.

\begin{figure*}[t]
 \centering
 \subfigure[]{
  \label{fig:poly_gamma}
  \includegraphics[width=0.4\textwidth]{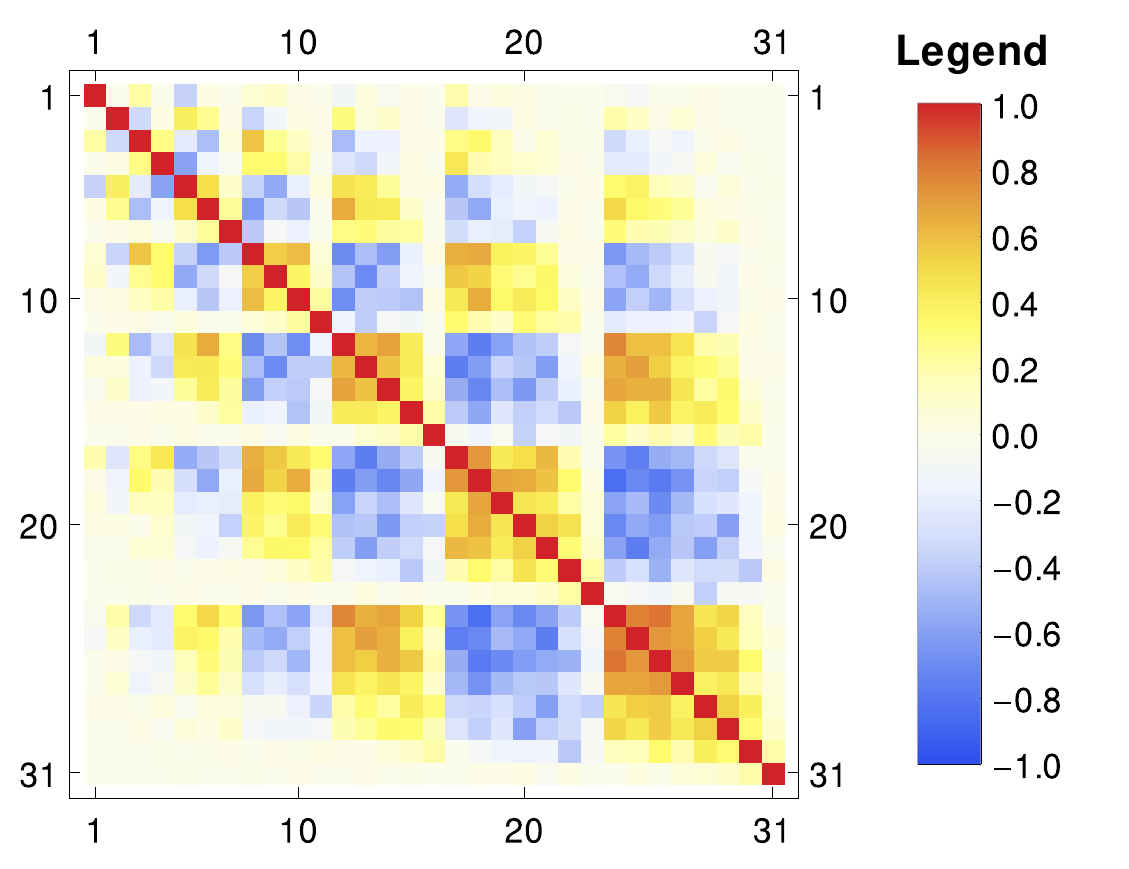}
 }
 \subfigure[]{
  \label{fig:poly_lambdaPrime}
  \includegraphics[width=0.4\textwidth]{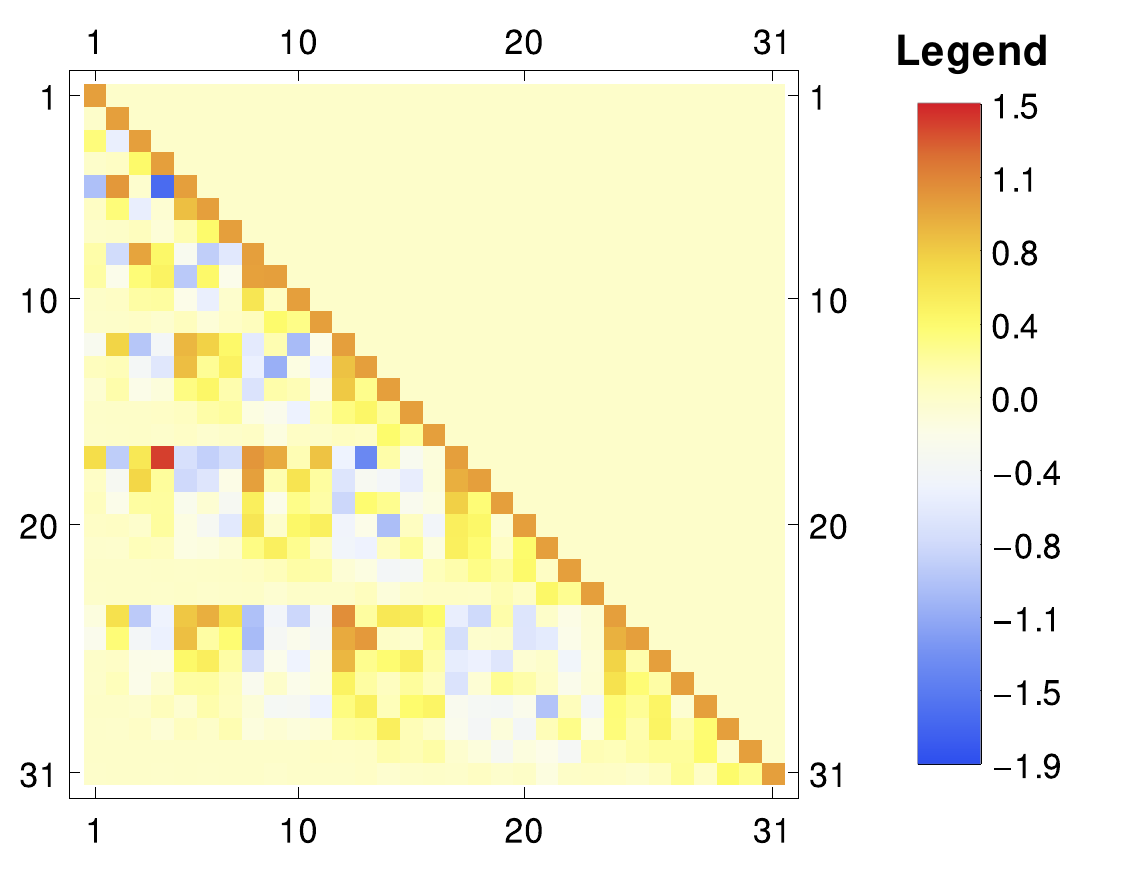}
 }
 \caption{\textbf{(a)}~Plot of the inner product matrix $\Gamma$ defined in \eqRef{eq:inderProductMatrix} of the first 31 modes (compare \cite{Fergusson:2009nv}, \fig{7}). \textbf{(b)}~Plot of the conversion matrix $\Lambda'$ defined in \eqRef{eq:conversionMatrix} for the first 31 modes (compare \cite{Fergusson:2009nv}, \fig{7}).}
 \label{fig:poly_gamma_lambdaPrime}
\end{figure*}

The number $d_N$ of independent symmetrized products $\modeQ_n(x,y,z)$ of polynomial order $N$ is
\begin{align}
 \{ d_N \} = \{ 1, 1, 2, 3, 4, 5, \ldots \}, \quad d_N = 1 + d_{N-2} + d_{N-3} - d_{N-5}.
 \label{eq:numIndepPolys}
\end{align}
Thus, we have $53$ mode functions if we include all polynomials up to order $9$. These polynomials are separable, but not orthonormal; nevertheless, their inner products
\begin{align}
 \Gamma = (\gamma_{nm}), \quad \gamma_{nm} = \iprod{\modeQ_n}{\modeQ_m}
 \label{eq:inderProductMatrix}
\end{align}
show a predominantly diagonal structure, inherited from the orthonormal one-dimensional functions. In \figRef{fig:poly_gamma} we plot the color-coded matrix components for the first 31 mode functions (up to polynomial order $7$). The ``checker-board'' pattern stems from sets of polynomials of the same order with lengths in \eqRef{eq:numIndepPolys}. It is worth noting that the $\modeQ_n$ become more correlated the higher the mode number $n$ is.

In a third step we obtain an orthonormal basis $\{ \orthoR_n \}$,
\begin{align}
 \iprod{\orthoR_n}{\orthoR_m} = \delta_{nm},
\end{align}
using \emph{Gram-Schmidt orthonormalization}. Let
\begin{align}
 \Lambda' = (\lambda'_{nm}), \quad \lambda'_{nm} := \begin{cases} -\iprod{\modeQ_n}{\orthoR_m'}, & n > m \\ 1, & n = m \\ 0, & \text{otherwise} \end{cases},
 \label{eq:conversionMatrix}
\end{align}
where
\begin{align}
 \orthoR'_n = \sum_{m=1}^n \lambda'_{nm} \modeQ_m.
\end{align}
We obtain the final orthonormal functions by letting
\begin{align}
 \orthoR_n \defeq \frac{ \orthoR'_n }{ \sqrt{ \iprod{ \orthoR'_n }{ \orthoR'_n} } }.
\end{align}
The color-coded matrix entries of $\Lambda'$ for the first 31 mode functions can be found in \figRef{fig:poly_lambdaPrime}.

\subsection{Mode expansion of bispectra}
\label{sec:modeExp}

Equipped with the symmetrized, orthonormal mode functions on the tetrapyd, we can express the bispectrum $\bspect{\comovR}$ of any theoretical model by a series of mode coefficients $\alpha_n$ of the shape function $\shape(k_1, k_2, k_3)$ in \eqRef{eq:shapeFunction}. The full shape function is recovered as the formal sum
\begin{align}
 \shape(k_1,k_2,k_3) = \sum_{n=1}^\infty \alpha^\orthoR_n \; \orthoR_n(x_1,x_2,x_3),
\end{align}
where the momenta are rescaled as in \eqRef{eq:rescaled} and the coefficients $\alpha^\orthoR_n$ are given by
\begin{align}
 \alpha^\orthoR_n = \iprod{\orthoR_n}{\sshape}.
 \label{eq:orthoExpand}
\end{align}
The matrix
\begin{align}
 \Lambda = (\Lambda_{nm}), \quad \lambda_{nm} = \frac{ \lambda_{nm}' }{ \sqrt{ \iprod{ \orthoR'_n }{ \orthoR'_n} } },
 \label{eq:alphaConvMatrix}
\end{align}
can be used to perform the conversion between the expansion coefficients $\alpha_n^\orthoR$ and the coefficients of the expansion in the separable $\modeQ_n$ basis, denoted $\alpha_n^\modeQ$. Since the conversion matrix needs to be computed only once, we retain the full advantage of separability in computations, given that the expansion series is sufficiently convergent (see \sectRef{sec:convProp}); in this manner, shape functions can be expressed by a few numbers. In the following, we obtain mode coefficients $\alpha_n^\Idx{model}$ numerically up to $n_\mathrm{max} = 53$ for several shapes.\footnote{We only use the $\alpha^\orthoR$ coefficients in the text and figures. Hence, we drop the superscript $\mathcal R$ on $\alpha$ in favor of a superscript to indicate the shape function used.}

\begin{figure*}[t]
 \centering
 \includegraphics[width=0.9\textwidth]{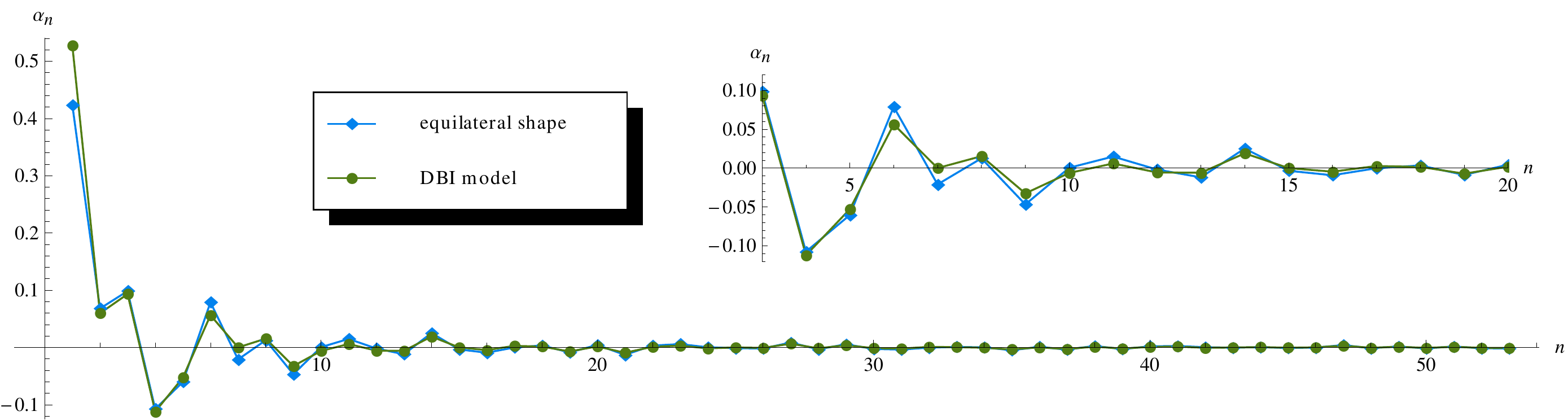}
 \caption{Plot of the expansion coefficients $\alpha_n^\Idx{equi}$ and $\alpha_n^\Idx{DBI}$ for the equilateral shape (blue), as defined in \eqRef{eq:equi_shape}, and the DBI model (green), as defined in \eqRef{eq:dbi_shape}; compare \cite{Fergusson:2009nv}, \fig{10}.}
 \label{fig:poly_alphaEquiVsDBI}
\end{figure*}

To illustrate the modal decomposition, let us investigate the shape originating in DBI inflation \cite{Silverstein:2003hf,Alishahiha:2004eh}. 
 This stringy inflationary model incorporates a speed limit in fields space via a modified kinetic term in the Lagrangian and is a special case of the more general Lagrangians that we discuss in \sectRef{sec:results}.
The resulting bispectrum led to the definition of the equilateral template. Thus, the expansion coefficients of these two shapes, namely\footnote{Equivalent expressions used in the literature are
\begin{align*}
 \Sshaperm{equi}(k_1,k_2,k_3) \defeq \frac{\prod_i (K - 2 k_i)}{K_{111}}, \quad \text{and} \quad \Sshaperm{DBI}(k_1, k_2, k_3) = \frac{1}{K_{111} K^2} \left( K_5 + 2 K_{14} - 3 K_{23} + K_{113} - 4 K_{122} \right).
\end{align*}
}
\begin{align}
 \Sshaperm{equi}(k_1,k_2,k_3) &\defeq \frac{1}{K_{111}} \left( K_{12} - K_3 - 2 K_{111} \right), \quad \text{and}
 \label{eq:equi_shape} \\
 \Sshaperm{DBI}(k_1, k_2, k_3) &\defeq  \frac{1}{K_{111}} \left(- \frac{K_{22}}{K} + \frac{K_{23}}{2 K^2} + \frac{K_3}{8} \right),
 \label{eq:dbi_shape}
\end{align}
should have a strong correlation. The shorthand notation $K_{\dots}$ indicates simple polynomials in $\{k_i\}$ and is explained in appendix  \ref{sec:shapeShorthand}.  We find the mode coefficients shown in \figRef{fig:poly_alphaEquiVsDBI}, recovering the analysis by Shellard et al. How good, quantitatively speaking, is the convergence and correlation between these shapes?

\subsubsection{Convergence properties}
\label{sec:convProp}

We would like to use the partial sum
\begin{align}
 \Sshape{n_\mathrm{max}} \defeq \sum_{n=1}^{n_\mathrm{max}} \alpha_n \; \orthoR_n,
\end{align}
to approximate a given bispectrum. In order to quantify the level of convergence of that sum, we use Parseval's theorem,
\begin{align}
 \iprod{S}{S} = \sum_{n=1}^\infty \alpha_n \sconj{\alpha_n},
 \label{eq:parsevalTheorem}
\end{align}
and investigate how fast the limit
\begin{align}
 \frac{1}{\iprod{S}{S}} \sum_{n=1}^{n_\mathrm{max}} \alpha_n \sconj{\alpha_n} \limarrow{n_\mathrm{max} \to \infty} 1
 \label{eq:alpha_convergence}
\end{align}
is approached.

\begin{figure*}[t!]
 \centering
 \includegraphics[width=0.9\textwidth]{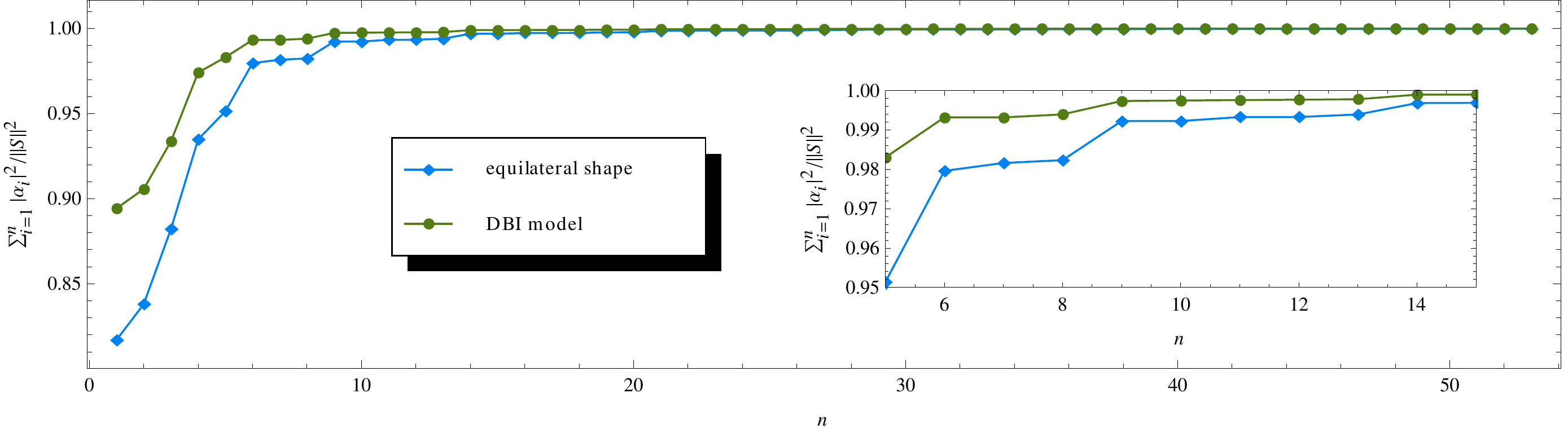}
 \caption{Plot of the sum of squared coefficients in \eqRef{eq:alpha_convergence} for the equilateral shape and the DBI model (see \cite{Fergusson:2009nv}, \fig{9}). The inset shows that $\Ord{15}$ modes are sufficient to achieve better than $99.5\%$ convergence.}
 \label{fig:poly_alpha_convergence}
\end{figure*}

We consider a series to be sufficiently convergent if the series truncated at order $n_\mathrm{max}$ reaches $1 - \varepsilon$, where $\varepsilon$ is a small parameter. Our results for the mode expansions of the equilateral shape and the DBI model are shown in \figRef{fig:poly_alpha_convergence}. Using $n_\mathrm{max} = 53$ modes, we get a convergence as good as $\varepsilon \approx 0.0002$. Fifteen modes are sufficient to achieve $\varepsilon \approx 0.0031$. In the literature, a minimal convergence of $95\%$ to $98\%$ is required \cite{Fergusson:2010dm}, but a higher convergence may be needed for comparison with PLANCK data.

\subsubsection{Correlation between shapes}
\label{sec:shapeCorr}

\begin{figure*}[t!]
 \centering
 \includegraphics[width=0.8\textwidth]{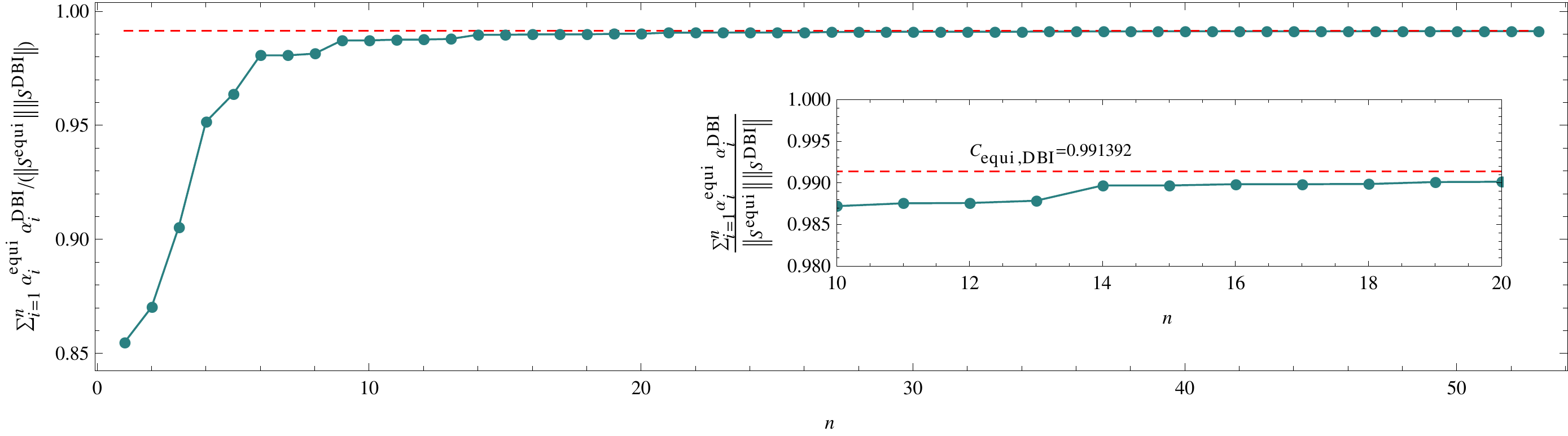}
 \caption{Plot of partial correlation sum for the equilateral shape and the DBI model.}
 \label{fig:poly_alpha_correlation}
\end{figure*}

The correlator between two shapes is defined in \eqRef{eq:correlation}. For the equilateral shape and the DBI model we get $C_{\mathrm{equi},\mathrm{DBI}} \approx 0.991391$, an expectedly high correlation. In \figRef{fig:poly_alpha_correlation}, we show the partial correlation sum
\begin{align}
 \sum_{i=1}^{n} \alpha^\mathrm{(equi)}_i \conj{\alpha^\mathrm{(DBI)}_i}, \quad n \le n_\mathrm{max},
\end{align}
which converges to the final result $C_{\mathrm{equi},\mathrm{DBI}}$ rather quickly, because the sum over the squared mode coefficients in \eqRef{eq:alpha_convergence} converges satisfactorily after only 15 modes for both shapes.

We observe that certain modes improve correlation more than others, which is expected since no physical input went into the ordering of the basis functions. If we had a particular shape to match, we could reorder modes to achieve faster convergence. Such a procedure should be performed if observational data (CMB or LSS) is analyzed, in contrast to the approach in \cite{Fergusson:2010dm} which used the same polynomials as above. 

\subsubsection{Classes of shapes by correlation}

Given the polynomial decomposition, which other shapes have good convergence properties? Fergusson and Shellard \cite{Fergusson:2008ra} give a selective overview of common bispectrum classes. We find good convergence for the center-weighted models, such as the DBI shape, that peak on equilateral triangles; this is not surprising, since all models in this class are highly correlated with the equilateral shape $\Sshaperm{equi}$.

The expansion series of corner-weighted models that peak on squeezed triangles do not converge due to the divergence of these models in the $k_i \to 0$ limit. Fergusson and Shellard \cite{Fergusson:2008ra} suggest a clipping of such shapes based on a minimal physically relevant wavenumber $k_\mathrm{min} \approx (2 / l_\mathrm{max}) k_\mathrm{max}$. Commonly discussed shapes in this group are correlated among each other by varying degree. 
It should be noted that the local shape is also correlated with the equilateral shape and not orthogonal to it. Thus members of this class are often distinguishable, but not completely independent. 

The correlation of shapes among members of these classes is summarized in table \ref{tab:classes}, whose entries stem from table I \& II in \cite{Liguori:2010hx} and were reproduced using our framework.

\begin{table}[hbtp]
 \begin{center}
  \begin{tabular}{l|c|c|c}
   \bfseries{Model} & equilateral type & local type   & feature type \\
   \hline
   equilateral type & at least $86\%$  & up to $46\%$ & up to $36\%$ \\
   local type       &                  & up to $62\%$ & up to $44\%$ \\
  \end{tabular}
 \end{center}
 \caption{Shape correlation according to \eqRef{eq:correlation} among members of three shape classes: the equilateral template, the DBI model, and the ghost model make up the class of equilateral-type NG. The class of the local-type NG consists of the local template, the warm model, and the flat model. The feature model is defined in \eqRef{eq:Sfeat}. Values are take from \cite{Liguori:2010hx}, tables I \& II, which we reproduced.}
 \label{tab:classes}
\end{table}

\subsection{Divergent shapes and the clipping mechanism}
\label{sec:divShapes}

The shape derived from the local ansatz,\footnote{The local ansatz is defined in \eqRef{eq:localAnsatz}.}
\begin{align}
 \Sshaperm{local} = \frac 1 3 \frac{K_3}{K_{111}},
 \label{eq:localShape}
\end{align}
does not have a convergent expansion series due to the divergence seen in \figRef{fig:local_shape}. To ameliorate similar problems with divergent shapes, one can try to remove any local contribution from the discussed shapes or, if a divergence remains, clip the shape.

\begin{figure*}[t]
 \centering
 \includegraphics[width=8cm]{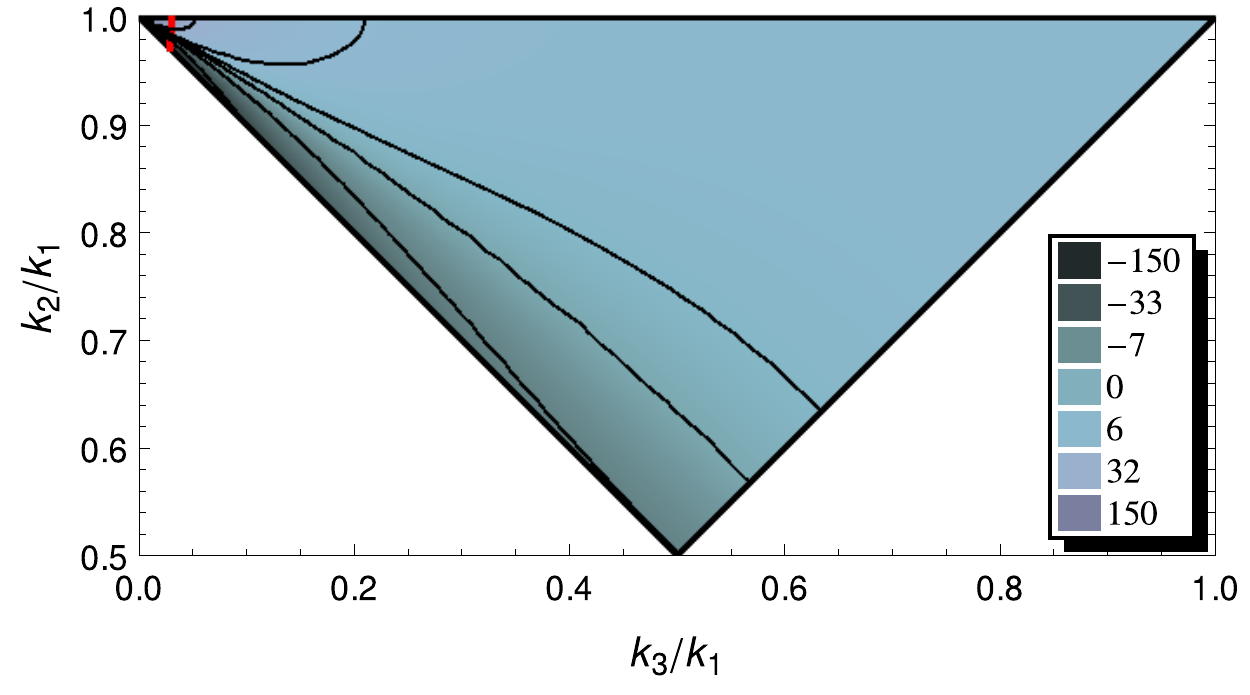}
 \caption{Plot of the warm shape, where the cut-off (top left corner) in \eqRef{eq:cutoffLimit} is marked by a red line.}
 \label{fig:cut_warm_shape}
\end{figure*}

Several corner-weighted shapes that peak for squeezed triangles are discussed in \cite{Fergusson:2008ra}, \sect{III.E}. For example, the warm shape \cite{Moss:2007cv} produced during warm inflation (inflation in the presence of a radiation bath, see \cite{Berera:2008ar} for a review),\footnote{Next to the warm shape, non-Gaussianity of the local/constant type is also produced if damping terms are allowed to depend on the temperature \cite{Moss:2011qc}.}
\begin{align}
 \Sshaperm{warm} \propto \frac{K_{45} - K_{27} + 2 K_{225}}{K_{333}},
\end{align}
is one of the shapes that require a cut-off in order to be treatable by the mode expansion algorithm. Based on the flat sky approximation
\begin{align}
 k \approx \frac{l}{\tau_0 - \tau_R},
\end{align}
where $\tau_0$ is the age of the universe and $\tau_R$ the time from recombination till today,
Fergusson and Shellard \cite{Fergusson:2008ra} argue that a natural cut-off exist at 
\begin{align}
 k_\mathrm{min} \approx \frac{2}{l_\mathrm{max}} k_\mathrm{max}.
 \label{eq:kminFergShellard}
\end{align}
Accordingly, the divergent part can be removed by requiring \cite{Fergusson:2008ra} \ie
\begin{align}
 \frac{k_1}{k_2 + k_3} < 0.015,
 \label{eq:cutoffLimit}
\end{align}
where the exact value of the cut-off is chosen by hand. The contour of this cut is marked red in \figRef{fig:cut_warm_shape}. We provide a discussion of the clipping mechanism's validity in appendix \ref{sec:clipping_mechanism}.

However, the warm shape is still hard to treat with the modal expansion. First, the effects of the cut-off need to be kept small. This can be achieved, as proposed in \cite{Fergusson:2008ra}, by smoothing the shape after truncation, \ie by applying a Gaussian window function. Second, the divergence of the warm model in the limit $k_3 \to 0$ has a direction-depending sign, contrary to the local shape in \eqRef{eq:localShape}, to which it is only $33\%$ correlated.

In the following, we will encounter shapes that are likewise divergent, but most of them are less problematic: subtraction of the local contribution, 
\begin{align}
 \sshape^\Idx{conv} \defeq \sshape^\Idx{div} - 3 \FNL[\hat f]{local} \Sshaperm{local},
\end{align}
renders these shapes convergent. Here, we defined
\begin{align}
 \FNL[\hat f]{local} \defeq \lim_{k_3 \to 0} \atFrac{ (k_1 k_2 k_3) \; \sshape^\Idx{div}(k_1, k_2, k_3) }{ k_1^3 + k_2^3 + k_3^3 }{k_2=k_1},
\end{align}
which is an analog of the non-linearity parameter $\FNL{local}$ for shape functions $S$.\footnote{Note that $\FNL[\hat f]{local}$ and $\FNL{local}$ differ by a shape-dependent multiplicative factor due to the normalization $S(k,k,k) = 1$.}

\subsection{Scale-dependent features}
\label{sec:scaleDepFeat}

\begin{figure*}[t]
 \centering
 \subfigure[~shape]{
  \label{fig:shape_feature}
  \includegraphics[width=0.27\textwidth]{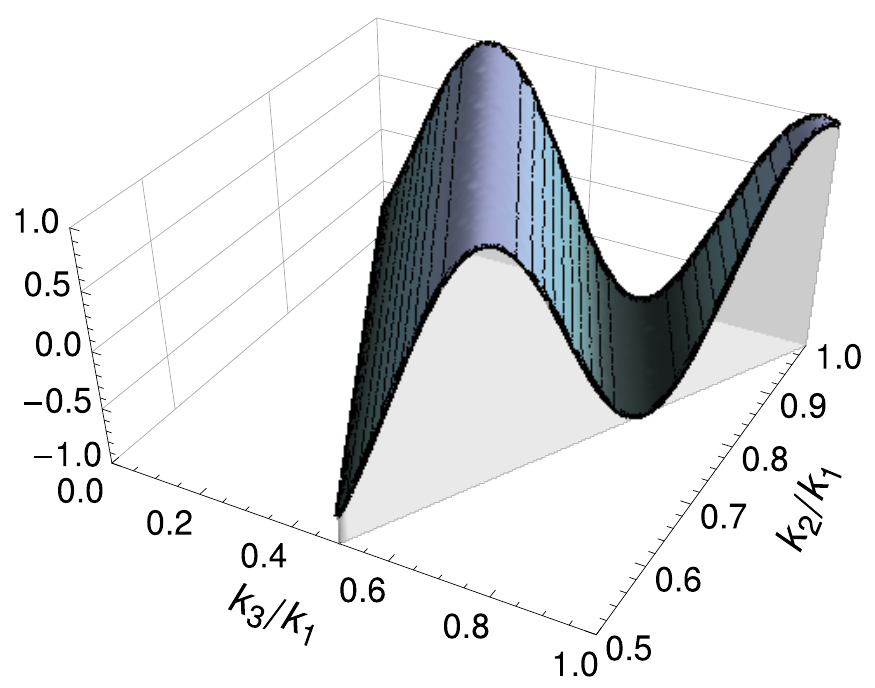}
 }
 \subfigure[~convergence]{
  \label{fig:poly_alphaFeat_nonconvergence}
  \includegraphics[width=0.68\textwidth]{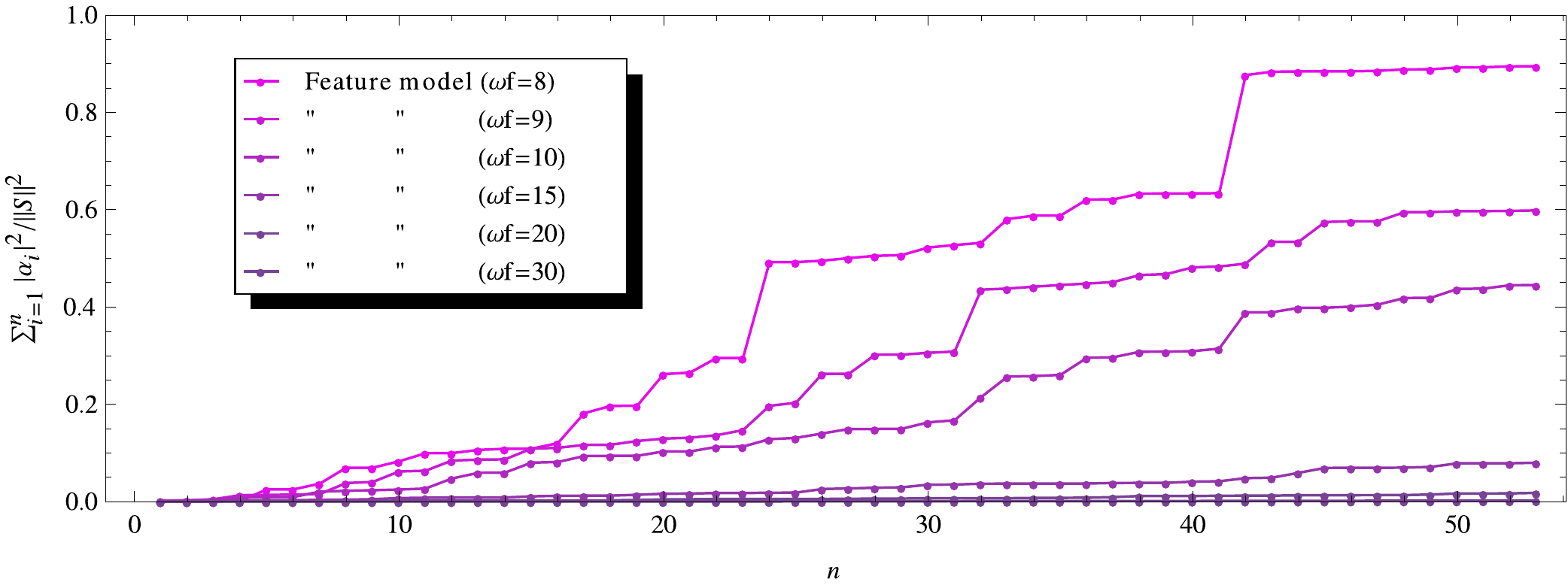}
 }
 \caption{\textbf{(a)} 3D plot of the feature model's shape function for $k_\ast = \FRAC{k_\mathrm{max}}{\omega_f}$, where $\omega_f = 9$, and $\delta=0$. \textbf{(b)} Plot of the sum of squared coefficients for the feature model using a polynomial expansion (see \cite{Meerburg:2010ks}, \fig{3}); convergence is slow (compare to \figRef{fig:fourier_alpha_convergence}, where faster convergence is achieved by using Fourier modes).}
 \label{fig:poly_feature}
\end{figure*}

Bispectra with scale-dependent features occur occasionally, for example the \emph{sharp feature model},
\begin{align}
 \Shape{feat}(k_1, k_2, k_3) = \sin \left( \frac{k_1 + k_2 + k_3}{k_\ast} + \delta \right),
 \label{eq:Sfeat}
\end{align}
found by Chen \cite{Chen:2008wn,Chen:2010bka}, see also \cite{Hannestad:2009yx}. Here, $k_\ast$ defines the scale of horizon exit when the inflaton field encounters a feature in the potential, and $\delta$ is a phase.

The convergence of such oscillatory bispectra \wrt\ the polynomial expansion is rather slow, see \figRef{fig:poly_alphaFeat_nonconvergence}. Meerburg \cite{Meerburg:2010ks} proposed an alternative mode expansion using Fourier modes for such oscillatory models, to which we come back in \sectRef{sec:oscFeat_altModes}. He found faster convergence of the expansion series for certain models such as the feature model, see \figRef{fig:fourier_alpha_convergence}, resonant type non-Gaussianities \cite{Chen:2008wn, Flauger:2009ab, Flauger:2010ja} and models with non-standard vacua \cite{Meerburg:2009ys,Meerburg:2009fi}. 

These shapes highlight a common problem of truncated modal expansions: the choice of basis functions renders a truncated expansion more sensitive to certain shapes, while becoming blind to others. We investigate an example of a problematic shape in \sectRef{sec:partProd} that cannot be recovered rapidly with a polynomial expansion, while Fourier modes work well.

\section[General single-field inflation models]{Review of general single-field inflationary models with non-standard kinetic terms}
\label{sec:genSingleFieldModels}

In this section we review the analysis of general single-field models based on the most general Lagrangian that is Lorentz invariant and whose kinetic term contains first derivatives in the field only. The reader might want to skip sections \ref{sec:conventionalSlowRoll} and \ref{sec:slowVariation} if he/she is familiar with general slow-roll single-field models. In addition, readers familiar with the slow-roll violating models by Noller and Magueijo \cite{Noller:2011hd} might want to skip \sectRef{sec:fastRoll}.
                                                                                                                                                                                                                                                                                                                                                                                                                                                                                                                                                                                                                                                                                                                                                     
The single-field inflation model currently favoured by data, \ie\ the simplest model that is not excluded by observations, has a standard kinetic term, is minimally coupled to gravity and follows a slow-roll trajectory. Such a model only produces a negligible amount of non-Gaussianity \cite{Maldacena:2002vr}. Since future observations \cite{Planck_BlueBook} may detect a NG signal (note the nearly $2\sigma$ signal in \eqRef{eq:fNL_WMAP7} for $\FNL{local}$) a plethora of more complicated inflationary models allowing for larger non-Gaussianities have gained popularity. For example, the DBI model, as introduced above, invokes a non-standard kinetic term and naturally leads to observable non-Gaussianities. A general Lagrangian encompassing DBI inflation among other models was proposed by Garriaga and Mukhanov in the context of \emph{k-Inflation} \cite{ArmendarizPicon:1999rj, Garriga:1999vw}. Here,  a speed of sound different from unity is possible so that slow-roll conditions of the field can be violated without spoiling the flatness of the power spectrum. These are the models we wish to discuss in \sectRef{sec:fastRoll}. 

Even more general models, dropping the requirement of first order field derivatives in the Lagrangian in favour of at most second order equations of motion \cite{Horndeski:1974}, go under the name of Horndeski models. Of particular interest are Galileon models \cite{Burrage:2010cu, Ribeiro:2011ax} where the self-interactions are protected by a covariant generalization of the Galilean shift symmetry (corresponding to a specific choice of coefficients in the Horndeski action); some of these were investigated by Seery and Ribeiro in \cite{Ribeiro:2011ax}, who found a ``new'' Galileon shape which is however strongly correlated with a known shape appearing in radiatively stable Galileon models with higher order derivative operators (discussed in \cite{Creminelli:2010qf}), and to a lesser degree to the one found in \cite{Burrage:2011hd} brought forth by second order slow-roll corrections. 

We focus on the narrower class explained above, with the aim of providing a complete anatomy of the resulting bispectrum. To this end, we slowly increase the degree of complexity, introduce key concepts and set our notation in three steps: first, we summarize the framework of
Seery and Lidsey \cite{Seery:2005wm} in \sectRef{sec:conventionalSlowRoll}, who calculated the bispectrum for a general single-field scenario at first order in slow roll under the assumption that the speed of sound is close to one. They found that the bispectrum can be expressed in terms of the standard slow-roll parameters and three additional parameters that measure the non-trivial kinetic structure, the speed of sound and the rate of change of the speed of sound.

We follow with a summary of Chen \etAl\ \cite{Chen:2006nt,Chen:2010xka} in \sectRef{sec:slowVariation}, who treat a more general case, sometimes referred to as \emph{slow\  variation} in contrast to the conventional slow-roll case. They calculate the bispectrum to leading order in slow roll, but allow for $c_s \ll 1$, and find a bispectrum depending on five free parameters. The interesting case $c_s \ll 1$ can generate non-negligible amounts of non-Gaussianities, leading to a bispectrum dominated by two possible qualitative shapes; the corresponding amplitudes depend on two parameters.

Lastly, we focus on the work by Noller and Magueijo \cite{Noller:2011hd} in \sectRef{sec:fastRoll}, which is motivated by the insight that deviations from the usual slow-roll conditions do not always break scale-invariance of the power spectrum \cite{ArmendarizPicon:2003ht, Khoury:2008wj}. Their maximally slow-roll violating models allow up to $\epsilon \sim 0.3$ consistent with observations.

The violation of slow roll has also been investigated in the case of the Galileon models (\emph{G-inflation}, \cite{Kobayashi:2011pc}).

\subsection{Non-standard kinetic terms in single-field inflation assuming slow roll}
\label{sec:conventionalSlowRoll}

Following \cite{Seery:2005wm} we treat a single-field inflationary model that allows for non-standard kinetic terms by letting the Lagrangian $\Lagr$ be a general function of the form
\begin{align}
 \Lagr = P(X, \phi), \quad \text{with} \quad X = -g^{ab} \covdiff_a \phi \covdiff_b \phi,
\end{align}
so that it contains at most first derivatives in the scalar field and is Lorentz invariant. The field $\phi$ is minimally coupled to gravity leading to the action
\begin{align}
 S &= \Intdn{4}{x} \sqrt{-g} \left[ \frac R 2 + \Lagr \right].
\end{align}
 The scalar field has a speed of sound
\begin{align}
 \ct = \frac{P_\p{X}}{P_\p{X} + 2 X P_\p{XX}},
\end{align}
which can differ from one. Here, $(\dots)_{,X}$ denotes a partial derivative w.r.t.~$X$.

\subsubsection{The slow-roll approximation}

Slow roll can be imposed by requiring\footnote{Often an alternative definition $\eta_H = 2 H''/H$ is used, which corresponds to $\eta_V = V'' / V$ for single-field models with a canonical kinetic term, leading to $\eta = -2 \eta_V + 4 \epsilon_V$ at first order in slow-roll.}
\begin{align}
 \epsilon &\defeq - \frac{\dot H}{H^2} = \frac{X P_\p{X}}{H^2} \ll 1, & \eta &\defeq \frac{\dot \epsilon}{\epsilon H} \ll 1.
\end{align}
These \emph{flow parameters} describe the evolution in parameter space of inflationary models. We decompose the first flow parameter as
\begin{align}
 \epsilon = - \frac{\dot \phi}{H^2} \Partdiff{H}{\phi} - \frac{\dot X}{H^2} \Partdiff{H}{X} \eqdef \epsilon_\phi + \epsilon_X.
\end{align}
Smallness of $\epsilon_\phi$ induces $\dot \phi^2 \ll H^2$. With these definitions, the field equation leads to $\dot X = - 6 H \ct X ( 1 - \epsilon_\phi / \epsilon)$, enabling the replacement of derivatives \wrt\ $t$ with derivatives \wrt\ the field's kinetic term $X$.

The speed of sound is best described in terms of an additional parameter $u$, 
\begin{align}
 u &\defeq 1 - \ctInv = -2 X \frac{P_\p{XX}}{P_\p{X}}, & s &\defeq \frac 1 H \frac{\dot c_s}{c_s}.
\end{align}
Here, we introduce a new ``slow-roll'' parameter $s$ to express the rate of change of $c_s$, so that $\dot u = 2 H s (1 - u)$. We require the speed of sound to be slowly changing in order to retain an almost scale-invariant spectrum.  
In \cite{Seery:2005wm} $\epsilon$ and $\eta$ are assumed to be of the same order ($\epsilon \sim \eta$) and constant, consistent with $\dot \epsilon, \dot \eta \sim \Ord{\epsilon^2}$. Demanding $u = \Ord{\epsilon}$ implies a small deviation from $\ct = 1$ leading to $s = \Ord{\epsilon^2}$; it follows that $\epsilon_X = \Ord{\epsilon^2}$ \cite{Seery:2005wm} so that $\epsilon_X$ is subdominant to $\epsilon_\phi$. To simplify the description of the bispectra in this framework, we define
\begin{align}
 \Sigma &\defeq X P_\p{X} + 2 X^2 P_\p{XX},
  \label{eq:bispect_param_Sigma} 
  \\
 \lambda &\defeq X^2 P_\p{XX} + \frac 2 3 X^3 P_\p{XXX},
  \label{eq:bispect_param_lambda}
\end{align}
which can be rewritten as
\begin{align}
 \Sigma &= \frac{H^2 \epsilon}{\ct} = H^2 \epsilon( 1 - u), & \lambda &= \frac \Sigma 6 \left[ \frac 2 3 \frac{\epsilon}{\epsilon_X} (1 - u) s - u \right].
 \label{eq:bispect_params_slowroll}
\end{align}
These identities hold in general and are not contingent on the slow-roll approximation.

\subsubsection{The power spectrum}

Seery and Lidsey \cite{Seery:2005wm} analyze the action $S[\comovR]$ for the comoving curvature perturbation $\comovR$ 
\emph{to leading order in slow roll} using the ADM formalism. 
Keeping only terms up to quadratic order in $\comovR$ in the action, the Fourier transform of the two-point correlation function, the power spectrum, reads
\begin{align}
 \average{ \comovR(\V k_1) \comovR(\V k_2) } &= (2 \pi)^3 \delta(\V k_1 + \V k_2) P(k_1),
  \label{eq:twoPointLidseySeery}
  \\
 P(k) &= \frac{H^2}{4 \epsilon} \frac{1}{k^3}.
  \label{eq:powSpectLidseySeery}
\end{align}
The primordial power spectrum becomes
\begin{align}
 \pspect{\comovR}(k) &:= \frac{k^3}{2 \pi^2} P(k) = \frac{H^2}{2 \epsilon (2 \pi)^2},
\end{align}
and the scalar spectral index can be computed to
\begin{align}
 n_s - 1 \defeq \Diff{\log \pspect{\comovR}}{\log k}  \simeq -2 \epsilon - \eta = -2 \epsilon_\phi - 2 \epsilon_X - \eta.
 \label{eq:spectIndexLidseySeery}
\end{align}

\subsubsection{The bispectrum}

In \cite{Seery:2005wm} the bispectrum is calculated by expanding the action up to third order in $\comovR$, restricted to leading order in slow roll. 
Seery and Lidsey apply a path integral formalism for the three-point function,
\begin{align}
 \Braket{\Omega}{\comovR(t)^3}{\Omega} = \frac{ \int [\dabl \comovR] \; \comovR(t)^3 \ee{\ii \int_\mathcal{C} L} }{ \int [\dabl \comovR] \ee{\ii \int_{\mathcal C} L} },
\end{align}
and analyze the tree level contributions from all possible interaction vertices ($\dot \comovR^3$, $\dot \comovR \partial^{-2} \dot \comovR$, $\dot \comovR \partial^2 \comovR \partial^{-2} \dot \comovR$). Combining these terms, 
the Fourier transform of the three-point correlation function is (assuming $k_1 \sim k_2 \sim k_3$)
\begin{align}
 \average{ \comovR(\V k_1) \comovR(\V k_2) \comovR(\V k_3) } = (2 \pi)^7 \; \delta^{(3)} \argstyle{ \sum_i \V k_i } \pspect{\comovR}^2 \frac{\ashape}{4 \prod_i k_i^3},
 \label{eq:bispectrumSeeryLidsey}
\end{align}
where (\cite{Seery:2005wm}, \eq{82f})
\begin{align}
 \ashape = \frac 4 K (u + \epsilon) K_{22} - \frac{4}{K^3} \left( u + \frac{\epsilon}{\epsilon_X} \frac s 3 \right) K_{222} - \frac{2 u}{K^2} K_{23} + \frac 1 2 (\eta - u - \epsilon) K_3 + \frac \epsilon 2 K_{12}.
 \label{eq:seeryLidsey}
\end{align}
We employ the notation introduced in appendix \ref{sec:shapeShorthand}, which simplifies the identification of shape templates in the bispectrum amplitude. We can construct a shape function as defined in \sectRef{sec:sepExp} by letting $\sshape = \ashape / ( K_{111} \, N)$, where $N$ should be set such that $\sshape(k,k,k) = 1$.

As an aside, we would like to comment on the validity of the consistency relation found by Creminelli and Zaldarriaga \cite{Creminelli:2004yq}. In the squeezed limit $k_3 \to 0$, which 
corresponds to a renormalization of the background,
\begin{align}
 \average{ \comovR(\V k_1) \comovR(\V k_2) \comovR(\V k_3) } = (2 \pi)^3 \; \delta^{(3)} \argstyle{ \sum_i \V k_i } \; (1 - n_s) P(k_1) P(k_3)
 \label{eq:CreminelliZaldarriaga}
\end{align}
is valid for general Lagrangians \cite{Cheung:2007sv}, and initial conditions \cite{Kundu:2011sg}. This \emph{consistency relation} shows the existence of local-type NG for any inflationary model with $n_s\neq 1$. The main reason for the existence of this relation is the presence of a single dynamical degree of freedom that acts as a physical \emph{clock} for exiting the quasi-de Sitter phase.

The amplitude $\ashape$ in the $\epsilon = \eta = 0$ case, namely \eq{88} in \cite{Seery:2005wm},
\begin{align}
 \at{\ashape}{\epsilon = \eta = 0} = 4 u \frac{K_{22}}{K} - 4 u \frac{K_{222}}{K^3} - 2 u \frac{K_{23}}{K^2} - \frac u 2 K_3,
 \label{eq:pureFieldNG}
\end{align}
highlights the NG contributions that arise purely from the inflaton field, \ie\ without the coupling to gravity. In particular, $\at{\ashape}{\epsilon = \eta = 0} \limarrow{k_3 \to 0} 0$ follows in line with the consistency relation (independent of the speed of sound).

\subsubsection{Bispectrum amplitudes and \altPdfText{$\fNL$}{fNL}}

To search for the presence of non-Gaussianity, it is useful to define a single quantity which provides a measure of the overall amplitude. Naturally, any shape and k-dependence is lost by focusing on just one number (see \cite{Byrnes:2010ft} for a proposal to go beyond the amplitude by considering the running of the bispectrum). The most common approach consists of picking a separable shape template, such as the local or equilateral one, whose magnitude can be easily extracted from the data. Theoretical bispectra are then projected onto the template to enable a comparison with the measured value, which can be problematic if the correlation is weak. A better definition of an amplitude that does not rely on templates was recently employed in \cite{Fergusson:2010dm} and uses the modal decomposition in spherical harmonic space to compare theory with observations. 
However, since the latter approach is still being optimized (choice/ordering of basis functions) and the forward evolution of shapes by means of the radiation transfer functions is non-trivial, we shall use the standard definitions of the \emph{non-linearity parameters}, as defined in \eqs\ \eqref{eq:fNLlocalSL} and \eqref{eq:fNLequiSL}, at this point. 

Consider first the so-called local non-Gaussianity, corresponding to $k_1 \sim k_2 \gg k_3$, whose amplitude is expressed by the non-linearity parameter $\FNL{local}$. It was introduced in a (spatially) local ansatz for non-Gaussianities \cite{Gangui:1993tt,Verde:1999ij,Komatsu:2001rj} caused by the square of the Gaussian scalar potential\footnote{Note that the sign convention of Seery and Lidsey \cite{Seery:2005wm} is opposite to the more common one by Komatsu \etAl\ \cite{Komatsu:2010fb}. We employ the latter in order to keep comparability with the WMAP results.}
\begin{align}
 \comovR = \comovR_G + \frac 3 5 \fNL \left( \comovR_G^2 - \average{\comovR_G^2} \right).
 \label{eq:localAnsatz}
\end{align}
The corresponding three-point correlation function is
\begin{align}
  \average{ \comovR(\textbf{k}_1) \comovR(\textbf{k}_2) \comovR(\textbf{k}_3) } = (2\pi)^7 \; \delta^3 \argstyle{ \sum_i \V k_i } \frac{3}{10} \fNL \; \pspect{\comovR}^2(k) \frac{\sum_i k_i^3}{\prod_i k_i^3}.
 \label{eq:fNL_bispectrum}
\end{align}
Given the general single-field bispectrum to leading order in slow roll in \eqRef{eq:seeryLidsey}, the prediction 
\begin{align}
 \FNL{local} &\defeq \frac 5 6 \at{ \frac{\ashape}{\sum_i k_i^3} }{k_1 = k_2, k_3 \to 0} = - \frac{5}{12} (n_s - 1)
 \label{eq:fNLlocalSL}
\end{align}
is a direct consequence of \eqRef{eq:CreminelliZaldarriaga}.

The other commonly used limit is the equilateral one  $k_1 \sim k_2 \sim k_3$, leading to the prediction \cite{Seery:2005wm}
\begin{align}
 \FNL{equi} &\defeq \frac 5 6 \at{ \frac{\ashape}{\sum_i k_i^3} }{k_1 = k_2 = k_3} = \frac{275}{972} u - \frac{10}{729} \frac{\epsilon}{\epsilon_X} s + \frac{55}{36} \epsilon + \frac{5}{12} \eta.
 \label{eq:fNLequiSL}
\end{align}
This reduces to  Maldacenas \cite{Maldacena:2002vr} result $\FNL{equi} = - \frac{5}{12} (n_s - 1) - \frac{25}{72} (n_t - 1)$ for $\ct = 1$.

Thus, the overall amplitude of non-Gaussianities for these templates is of the order of the slow-roll parameters and not observable, because foreground effects are expected to yield an $\abs{\fNL{}}$ of $\Ord{1}$ to $\Ord{10}$ (see \cite{Bartolo:2004if,Bartolo:2010qu}). In reaching this conclusion, we assumed that at least one (or both) of the templates have a good correlation with the primordial shape, which is indeed the case as we shall see later on.

\subsection{Towards a violation of slow roll: a small speed of sound}
\label{sec:slowVariation}

The discussion in the previous section on general single-field, slow-roll inflationary models with $\ct \approx 1$ showed that non-Gaussian signals are small, because $\FNL{}$ is slow-roll suppressed. But we have also seen in \eqRef{eq:pureFieldNG} that a pure field contribution arises for $\ct < 1$. Further, the DBI model in \sectRef{sec:modeExp} gave a first indication that a low speed of sound can indeed lead to observably large non-Gaussianities. Thus, let us turn our attention to models with a small speed of sound, following \cite{Chen:2006nt,Chen:2010xka}. 

\subsubsection{The power spectrum} 

In the more general class of models in \cite{Chen:2006nt}, the power spectrum has an additional  scaling $\propto c_s^{-1}$ in comparison to the result of Lidsey and Seery in \eqRef{eq:powSpectLidseySeery},\footnote{In order to keep our equations comparable to the original work, we express the spectra in terms of the curvature perturbation on uniform density hypersurfaces $\zeta$ in this section; the latter is identical to the comoving curvature perturbation $\comovR$ on superhorizon scales.}
\begin{align}
 \pspect{\zeta}(k) = \frac{1}{2 c_s \epsilon} \powerfrac{H}{2 \pi}{2}.
 \label{eq:powSpectChenEtAl}
\end{align}
Here, the (sound) horizon exit takes place at $c_s k = a H$, and the spectral index
\begin{align}
 n_s - 1 = - 2 \epsilon - \eta - s
 \label{eq:spectIndexChenEtAl}
\end{align}
has an additional dependence on the variation of $c_s$ via $s$. The tensor-to-scalar ratio is also altered to $r = \pspect{T} / \pspect{\zeta} = - 8 c_s n_T = 16 c_s \epsilon$, where $n_T$ is the tensor spectral index. Evidently, a deviation of the speed of sound from unity enhances the amplitude of the scalar curvature fluctuations relative to the amplitude of tensor fluctuations.

\subsubsection{The bispectrum}

In order to keep the power spectrum almost scale-invariant, Chen \etAl\ assume that all of the slow-variation parameters $\epsilon, \eta, s$ and $l$ in \eqRef{eq:slowVariationParams} are of the same order $\Ord{\epsilon}$ and small $\epsilon \ll 1$.  Using Maldacena's approach as in \cite{Seery:2005wm}, they derive the three-point correlation function from the effective action of the curvature perturbation $\zeta$ expanded up to $\Ord{\zeta^3}$. After a field redefinition to absorb a term proportional to the variation of the quadratic Lagrangian, they find that for an arbitrary speed of sound some leading order terms are \emph{not} slow-roll suppressed and, thus, potentially observable. Since these terms vanish in the limit $c_s \to 1$,  subleading corrections need to be kept. By these means, they obtain the complete bispectrum up to $\Ord{\epsilon}$
\begin{align}
 \langle \zeta(\textbf{k}_1)\zeta(\textbf{k}_2)\zeta(\textbf{k}_3)\rangle = \; &(2\pi)^7 \delta^3(\textbf{k}_1+\textbf{k}_2+\textbf{k}_3) P^2_\zeta(K) \frac{1}{\prod_i k_i^3} \notag \\
  &\times \left( \Ashape{\lambda} + \Ashape{c} + \Ashape{o} + \Ashape{\epsilon} + \Ashape{\eta} + \Ashape{s} \right),
  \label{eq:chenEtAlBispect}
\end{align}
where the bispectrum amplitude has been decomposed into 
{\allowdisplaybreaks\begin{align}
 \Ashape{\lambda} &= \left( \frac{1}{\ct}-1 - \frac{\lambda}{\Sigma}[ 2 - (3-2\bc_1)l ] \right)_K \frac{3 K_{222}}{2 K^3},
  \label{eq:Alambda}
  \\
 \Ashape{c} &= \left( \frac{1}{\ct} - 1 \right)_K \left( - \frac{K_{22}}{K} + \frac{K_{23}}{2 K^2} + \frac{K_3}{8} \right),
  \label{eq:Ac}
  \\
 \Ashape{o} &= \left( \frac{1}{\ct} - 1 - \frac{2\lambda}{\Sigma} \right)_K
  \left( \epsilon F_{\lambda\epsilon} + \eta F_{\lambda\eta} + s F_{\lambda s} \right) + \left( \frac{1}{\ct} - 1 \right)_K
  \left(\epsilon F_{c\epsilon} + \eta F_{c\eta} + s F_{cs} \right),
  \label{eq:Ao}
  \\
 \Ashape{\epsilon} &= \epsilon \left( - \frac{K_3}{8} + \frac{K_{12}}{8} + \frac{K_{22}}{K} \right),
  \label{eq:Aepsilon}
  \\
 \Ashape{\eta} &= \eta \left( \frac{K_3}{8} \right),
  \label{eq:Aeta}
  \\
 \Ashape{s} &= s F_s.
  \label{eq:As}
\end{align}
Here, $\bc_1$ is the Euler constant. The shape functions $F_{\dots}$ in $\Ashape{o}$ and $\Ashape{s}$ are given in appendix B.1 of \cite{Chen:2006nt} as unevaluated integral expression; explicit formulae can be found in appendix D of \cite{Burrage:2010cu}. The definitions of the four slow-variation parameters are
\begin{align}
 \epsilon &\defeq -\frac{\dot H}{H^2}, & \eta &\defeq \frac{\dot \epsilon}{\epsilon H}, & s &\defeq \frac{\dot c_s}{c_s H}, & l &\defeq \frac{\dot \lambda}{\lambda H}.
 \label{eq:slowVariationParams}
\end{align}
The additional parameters $\lambda$ and $\Sigma$ are defined in \eqs{\eqref{eq:bispect_param_Sigma}} and \eqref{eq:bispect_param_lambda}. It is worth noting that the amplitudes of the bispectrum shapes are determined by 5 parameters only: $\epsilon, \eta, s, c_s$ and $\lambda / \Sigma$. The parameter $l$ is only contained in subleading contributions (see \cite{Chen:2006nt}, \sect{4.4}) leading to $\Ord{\epsilon^2}$ effects.

$P_\zeta(K)$ is the power spectrum in \eqRef{eq:powSpectChenEtAl}. Note that the definition of the bispectrum amplitude $\ashape$ by Chen \etAl\ differs by a factor of $4$ from the one by Seery and Lidsey in \cite{Seery:2005wm}.

\subsubsection{Bispectrum amplitudes and \altPdfText{$\fNL$}{fNL}}

Chen \etAl\ \cite{Chen:2006nt} use the same definition for  non-linearity parameters as  in \eqRef{eq:fNL_bispectrum}; focusing on equilateral triangles ($k_1 \sim k_2 \sim k_3$) gives\footnote{Note that there is a mistaken parenthesis in \cite{Chen:2006nt}, \eq{5.3}. Note that the sign must be inverted for comparability with the original work.}
\begin{equation}
 \begin{aligned}
  \FNL{\lambda} &= \frac{5}{81} \left( \frac{1}{c_s^2}-1-\frac{2\lambda}{\Sigma} + (3-2\bc_1) \frac{l \lambda}{\Sigma} \right), \\
  \FNL{c} &= -\frac{35}{108} \left(\frac{1}{c_s^2}-1 \right), \\
  \abs{\FNL{o}} &= \Ord{ \frac{\epsilon}{c_s^2}, \frac{\epsilon\lambda}{\Sigma} }, \\
  \abs{\FNL{\epsilon,\eta,s}} &= \Ord{\epsilon}.
 \end{aligned}
 \label{eq:fNLequiChenEtAl}
\end{equation}
We can learn from these amplitudes that observable non-Gaussianities can arise in these inflationary models if $c_s$ is much smaller than one, and/or $\FRAC \lambda \Sigma$ gives a non-negligible contribution to $\fNL$.

Can we already constrain $c_s$ based on current limits on non-Gaussianities? Senatore \etAl\ \cite{Senatore:2009gt}, who employ a slightly different formalism for their general single-field inflation model, use the equilateral and an \emph{orthogonal} shape orthogonal to the equilateral shape \wrt\ to the scalar product in \eqRef{eq:correlation}). Using these separable shapes, they were able to provide (weak) constraints on the speed of sound: $c_s > 0.011$ at 95\% CL or $c_s$ is so small that the higher-derivative kinetic term dominates at horizon crossing.

\subsubsection{Recovering slow-roll results}

In the case of conventional slow-roll inflation, \ie\ for $u \ll 1$, Chen \etAl\ recover the result $\Ashaperm{SL} / 4$ (the factor of 4 is due to different conventions) from \eqRef{eq:seeryLidsey}. To this end, they neglect $\Ashape{s}$ and $\Ashape{o}$, which are second order in the conventional slow-roll parameters.
\begin{equation}
 \begin{aligned}
  \ashape \equiv \frac 1 4 \Ashaperm{SL} &= -\left( \frac{\epsilon}{3\epsilon_X}s + u \right) \frac{K_{222}}{K^3} - u \left( -\frac{K_{22}}{K} + \frac{K_{23}}{2K^2} + \frac{K_3}{8} \right) \\
  &+ \epsilon \left( -\frac{K_3}{8} + \frac{K_{12}}{8} + \frac{K_{22}}{K} \right) + \eta \left( \frac{K_3}{8} \right).
 \end{aligned}
\end{equation}
Furthermore, for $u=0$ and $s=0$ (\ie\ $c_s = 1 = \Const$), they find agreement with Maldacena's result up to another prefactor of 8:\footnote{Due to yet another convention, there is an additional factor of $2$, so that in total, Maldacena's convention and the one used here differ by a factor $8$}
\begin{align}
 \at{\ashape}{u=s=0} \equiv \frac 1 8 \Ashaperm{Mald} &= \frac{ 3 n_T - 2 n_s - 1}{16} K_3 + \frac{1 - n_T}{16} \left[ K_{12} + \frac{8 K_{22}}{K} \right].
 \label{eq:Amald}
\end{align}

\subsection{Allowing as much slow roll violation as possible}
\label{sec:fastRoll}

Noller and Magueijo took inspiration from work by Khoury and Piazza \cite{Khoury:2008wj}, who derived the bispectra for exactly scale-invariant spectra without restricting the parameter space by slow roll.
Dropping the requirement of exact scale invariance, Noller and Magueijo \cite{Noller:2011hd} compute the general single-field bispetrum using the variables
\begin{align}
 \boldalpha_1 &\defeq  n_s - 1 = \frac{2 \epsilon + \es}{\es + \epsilon - 1}, & \boldalpha_2 &\defeq \frac{2 \epsilon - \es}{\es + \epsilon - 1},
 \label{eq:alphas}
\end{align}
where $\epsilon_s\equiv s=\dot{c}_s/(c_s H)$ and instead of $\epsilon_X$ they employ
\begin{align}
 f_X \defeq \frac{\ep \es}{3 \ep_X}.
\end{align}
 Note that the parameter $\lambda$ can be written as
\begin{align}
 \lambda = \frac{\Sigma}{6}\left( \frac{2 f_X - 1}{\ct} - 1 \right),
\end{align}
according to \eqRef{eq:bispect_params_slowroll}.
They assumed that $\epsilon$ and $\es$ dominate over $\eta$ and $\eta_s=\dot{\epsilon}_s/(\epsilon_sH)$ (as well as higher order parameters), which are set to zero in the following.

\subsubsection{The bispectrum}

Following the same line of thought as outlined above, an effective action of cubic order in $\zeta$ can be obtained, but in contrast to Chen \etAl, no slow-variation conditions are assumed. Computing the full bispectrum in this way for each term in the action, they obtain
\begin{equation}
 \begin{aligned}
  \Ashape{\dot\zeta^3} &= \frac{1}{2 \ct} \scaling
   \left[ (\epsilon + \epsilon_s - 1) (f_X - 1) \Ishape{\dot\zeta^3}(\boldalpha_2) + \ct (\epsilon + \epsilon_s - 1) \Ishape{\dot\zeta^3}(\boldalpha_1) \right], \\
  \Ashape{\zeta \dot\zeta^2} &= \frac{1}{4 \ct} \scaling
   \left[ (\epsilon - 3) \Ishape{\zeta \dot\zeta^2}(\boldalpha_2) + 3 \ct \Ishape{\zeta \dot\zeta^2}(\boldalpha_1) \right], \\
  \Ashape{\zeta(\partial \zeta)^2} &= \frac{1}{8 \ct} \scaling
   \left[ (\epsilon - 2 \es + 1) \Ishape{\zeta(\partial \zeta)^2}(\boldalpha_2) - \ct \Ishape{\zeta(\partial \zeta)^2}(\boldalpha_1) \right], \\
  \Ashape{\dot \zeta \partial \zeta \partial \chi} &= \ \frac{1}{4 \ct} \scaling
   \left[ -\epsilon \, \Ishape{\dot \zeta \partial \zeta \partial \chi}(\boldalpha_2) \right], \\
  \Ashape{\epsilon^2} &= \ \frac{1}{16 \ct} \scaling
   \left[ \epsilon^2 \, \Ishape{\epsilon^2 }(\boldalpha_2)\right],
 \end{aligned}
 \label{eq:NollerMagueijoShape}
\end{equation}
where
\begin{align}
 \Ishape{\dot\zeta^3}(\boldalpha) &= \cosAlpha \Gamma(3+\boldalpha) \frac{K_{222}}{K^3}, \notag \\
 \Ishape{\zeta \dot\zeta^2}(\boldalpha) &= \cosAlpha \Gamma(1+\boldalpha)
  \left[ (2+\boldalpha) \frac{K_{22}}{K} - (1+\boldalpha) \frac{K_{23}}{K^2} \right], \notag \\
 \Ishape{\zeta(\partial \zeta)^2}(\boldalpha) &= - \cosAlpha \Gamma(1+\boldalpha)
  K_2 \left[ \frac{K}{\boldalpha - 1} + \frac{K_{11}}{K} + K_{111} \frac{1 + \boldalpha}{K^2} \right] \notag \\
  &= \cosAlpha \Gamma(1+\boldalpha)
  \left[\frac{K_3}{1-\boldalpha} + 2 \frac{2+\boldalpha}{K} K_{22} - 2 \frac{1+\boldalpha}{K^2} K_{23} + \frac{\boldalpha}{1-\boldalpha} K_{12} - \boldalpha K_{111} \right], \notag \\
 \Ishape{\dot \zeta \partial \zeta \partial \chi}(\boldalpha) &= \cosAlpha \Gamma(1+\boldalpha)
  \left[K_3 + \frac{\boldalpha - 1}{2} K_{12} - 2 \frac{1+\boldalpha}{K^2} K_{23} - 2 \boldalpha K_{111} \right], \notag \\
 \Ishape{\epsilon^2} (\boldalpha) &= \cosAlpha \Gamma(1+\boldalpha)(2 + \boldalpha/2)
  \left[ K_3 - K_{12} + 2 K_{111} \right].
 \label{eq:NollerMagueijoShape2}
\end{align}
These amplitudes reduce to the slow-roll limit discussed in the previous sections.

\subsubsection{Bispectrum amplitudes and \altPdfText{$\fNL$}{fNL}}

The non-linearity parameter for the equilateral shape is given by
\begin{align}
 \fNL = 30 \at{ \frac{\ashape}{K^3} }{ k_1 = k_2 = k_3 },
\end{align}
equivalent to the definition in \eqRef{eq:fNLequiSL} (up to a factor of 4, because Noller and Mageuijo's amplitude $\ashape$ corresponds to Chen \etAl's convention). The full result can be expressed as \cite{Noller:2011hd}
\begin{align}
 \fNL = {\cal C}_1(n_s,\epsilon) \left( 1 - {\cal C}_2(n_s, \epsilon, f_X) c_s^{-2} \right),
 \label{eq:fNLequiNM}
\end{align}
where the ${\cal C}_i$ are functions of the slow-roll parameters and $f_X$ (given in the appendix of \cite{Noller:2011hd}). In \cite{Noller:2011hd}, the dependence of $\fNL$ on the parameters is discussed for some examplary models. For instance, the case $\fNL \gg 1$ leads to
\begin{align}
 \fNL \sim \Ord{c_s^{-2}} + \Ord{\frac \lambda \Sigma}.
\end{align}
The running of the non-linearity parameter
\begin{align}
 n_{\fNL} = \Diff{\log \abs{\fNL}}{\log k},
\end{align}
derived from the parametrization in \eqRef{eq:fNLequiNM}, is investigated in detail in \cite{Noller:2011hd}.

\section[Bispectrum correlation and mode decomposition in general single-field models]{Correlation properties and mode decomposition of bispectrum shapes in general single-field models}
\label{sec:results}

This section contains our first main results. We analyze the basic terms that arise in the bispectrum amplitudes of single-field models, followed by a discussion of some concrete fast-roll models.

\subsection{Parameter extraction and the CMB bispectrum}
\label{sec:paramExtract}

The motivation behind a series expansion of the bispectrum is the easier comparison to the data for general not necessarily separable shapes. To this end, the predicted CMB bispectrum $B_{l_1 l_2 l_3}$ needs to be expanded in terms of the multipole moments (not wavenumbers) for each theoretical model (after evolving it forward in time using radiation transfer functions) and compare the resulting coefficients to the observed ones. Based on WMAP5 data and the polynomial modal decomposition, Fergusson \etAl\ \cite{Fergusson:2010dm} were able to constrain the amplitudes for certain commonly discussed models and plan to incorporate the WMAP7 data release (and subsequently PLANCK data). It is desirable to optimize the choice of mode functions as well as their ordering, \ie\ via a principal component analysis, which has not been done yet. Therefore, we work with the primordial shapes $\sshape(k_1, k_2, k_3)$ and discuss if the contribution of different shapes could in principle be disentangled and whether a parameter extraction appears feasible.

As a first case study, we compare the general slow-variation single-field shape constituents in \eqs\ \eqref{eq:Alambda} to \eqref{eq:As}. If the series coefficients of an observed primordial bispectrum were available with a high signal-to-noise ratio and the expansion coefficients of the relevant shape functions were at hand, the model parameters $\epsilon$, $\eta$, $c_s$, $s$ and $\lambda / \Sigma$ can be extracted by a non-linear regression or using MCMC analysis.

\subsection{Correlation of reduced, non-divergent shape contributions}
\label{sec:exampleShapeCorrelation}

The simplest constituents of the shape functions $\ashape$ that we encountered so far are of the form
\begin{align}
 K_3; \quad \frac{K_{pq}}{K^{p+q-3}}, \; \text{where} \; p \ge 1, q \ge \max\{2,p\}; \quad \text{and} \quad \frac{K_{rst}}{K^{r+s+t-3}}, \; \text{where} \; t \ge s \ge r \ge 1.
 \label{eq:allowedKays}
\end{align}
All of these contributions to the Seery and Lidsey result in \eqRef{eq:seeryLidsey} are divergent on their own, except for $\Ashaperm{const} \propto K_{111}$ and $\Ashaperm{single} \propto K_{222} / K^3$. At this point, bear in mind that the conversion rule to the dimensionless shape function gives $\Sshaperm{single} \propto \Ashaperm{single} / K_{111} = K_{111} / K^3$. In the following, we switch between $\ashape$ and $\sshape$ depending on which one is more convenient to use.

Before analyzing these basic constituents, we would like to remind the reader that the general single-field result does not contain the separable equilateral shape $\Ashaperm{equi} = K_{12} - K_3 - 2 K_{111}$, see \figRef{fig:equi_shape}. It closely approximates, however, the divergent-free part of the $\Ashape{\epsilon}$ shape, namely
\begin{align}
 \Ashape{\epsilon} - 2 K_3 \propto K_{12} - 3 K_3 + 8 \frac{K_{22}}{K}.
\end{align}
The previously mentioned DBI model is
\begin{align}
 \Ashaperm{DBI} \propto \Ashape{c} \propto K_3 - 8 \frac{K_{22}}{K} + 4 \frac{K_{23}}{K},
\end{align}
which is closely approximated by the equilateral shape (see \figRef{fig:poly_alpha_correlation}). These three shapes are examples of linear combinations of the $K_{\ldots}$ terms where the divergent contributions cancel.

\subsubsection[Correlation matrix of basic Noller and Magueijo contributions]{Correlation matrix of basic contributions to the Noller and Magueijo model}
\label{sec:systKaysNM}

The $K_{\ldots}$ terms in \eqRef{eq:allowedKays} are not independent of each other, see appendix \ref{sec:KtermInterdep}, but we can identify five independent, basic, convergent constituents of the general (scale-invariant) bispectrum shapes by Noller and Magueijo in \eqRef{eq:NollerMagueijoShape2} in addition to the local ($\propto K_3$) and constant ($\propto K_{111}$) ones.  
We choose
\begin{equation}
 \begin{aligned}
  \Ashape{1} &= K_{12}, & \Ashape{2} &= \Ashaperm{const} = K_{111}, & \Ashape{3} &= \frac{K_{22}}{K}, \\
  \Ashape{4} &= \frac{K_{23}}{K^2}, & \Ashape{5} &= \frac{K_6}{K^3}, & \Ashape{6} &= \Ashaperm{single} = \frac{K_{222}}{K^3},
 \end{aligned}
 \label{eq:defSystematicKaysNM}
\end{equation}
and the local shape as a complete set of such basic terms (up to polynomial order 6), \ie we can express all other terms with the right scaling and symmetry properties as linear superpositions of the terms in \eqRef{eq:defSystematicKaysNM}.

Let $\normEtc{\cdot}$ be the operator that removes the divergent (local) part proportional to $K_3$, subtracts the constant mode from a certain shape and normalizes the result, that is
\begin{align}
 \FNL{local}\left( \normEtc{\Ashape{i}} \right) &= 0, & 
 \iprod{\Ashape{2}}{\normEtc{\Ashape{i}}} &= 0, & \iprod{\normEtc{\Ashape{i}}}{\normEtc{\Ashape{i}}} &= 1,
 \label{eq:Nop}
\end{align}
where the non-linearity parameter of local-type non-Gaussianity is given by
\begin{align}
 \FNL{local}(\ashape) = (-5/6) \lim_{k_3 \to 0} \left[ \ashape(k, k, k_3)/(2 k^3 + k_3^3) \right].
\end{align}
Since most of the correlation between $\Ashaperm{equi}$ and $\Ashaperm{DBI}$ is due to the constant mode, we achieve better shape discrimination after removal of $\Ashape{2}$ (which is then orthogonal to every $\normEtc{\cdot}$ shape).
The resulting six shapes are plotted in \figRef{fig:base_NM_kay_shapes}. The shapes $\Ashape{1}$, $\Ashape{3}$, $\Ashape{6}$ and the mirrored $\Ashape{4}$ have an overall similarity, while $\Ashape{5}$ is distinct. 

\begin{figure}[t]
 \centering
 \subfigure[~$\normEtc{K_{12}}$]{
  \label{fig:NMkay_shape_1}
  \includegraphics[width=0.24\textwidth]{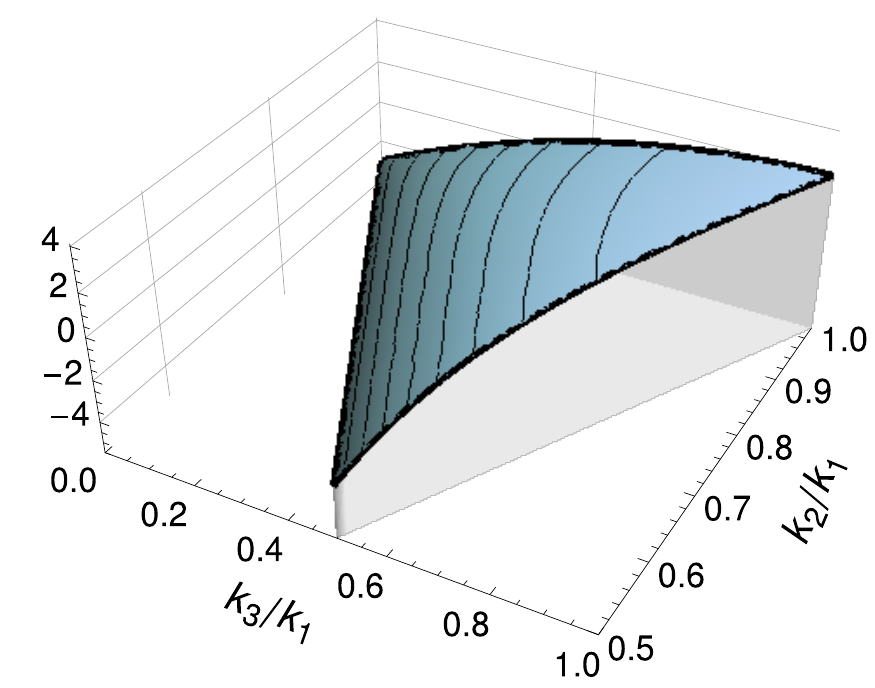}
 }
 \subfigure[~$\normEtc{K_{111}}$]{
  \label{fig:NMkay_shape_2}
  \includegraphics[width=0.24\textwidth]{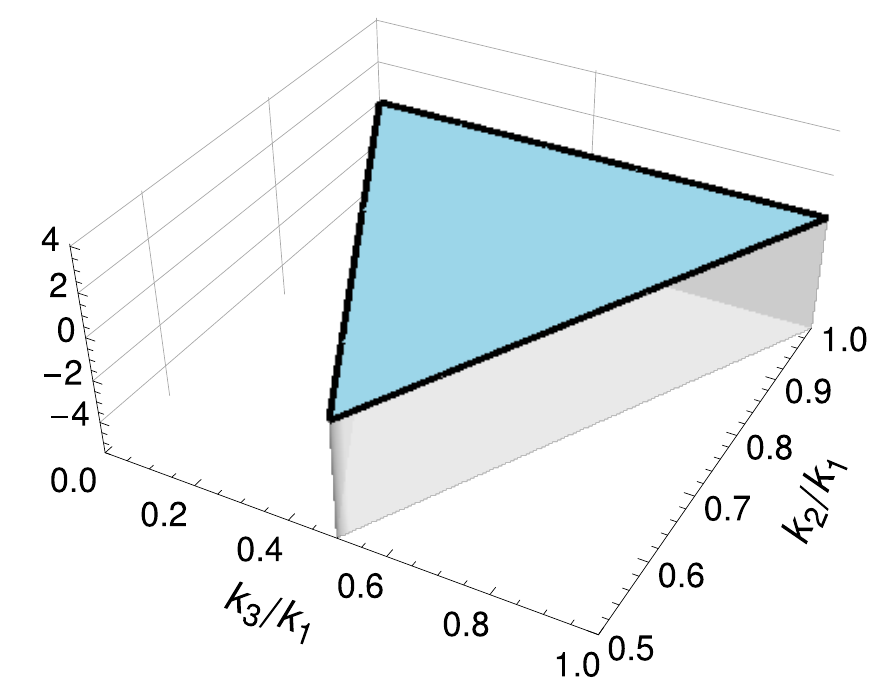}
 }
 \subfigure[~$\normEtc{K_{22} / K}$]{
  \label{fig:NMkay_shape_3}
  \includegraphics[width=0.24\textwidth]{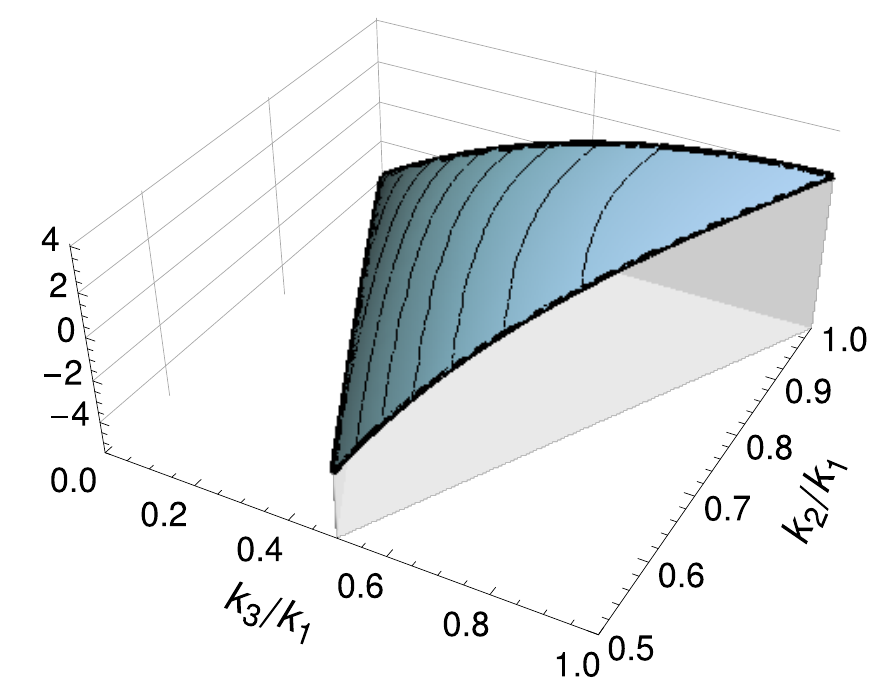}
 }
 \subfigure[~$\normEtc{K_{23} / K^2}$]{
  \label{fig:NMkay_shape_4}
  \includegraphics[width=0.24\textwidth]{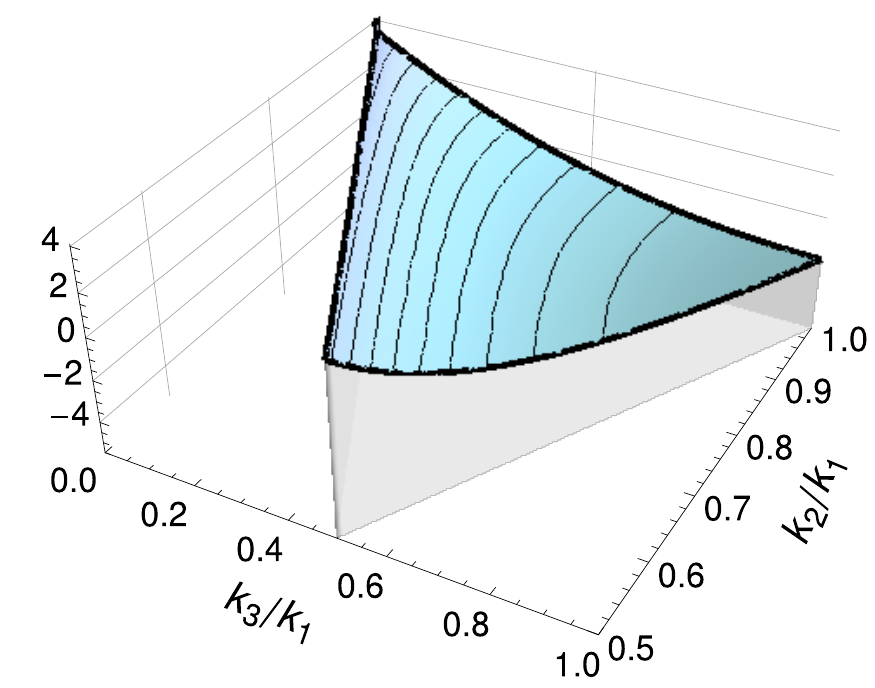}
 }
 \subfigure[~$\normEtc{K_6 / K^3}$]{
  \label{fig:NMkay_shape_5}
  \includegraphics[width=0.24\textwidth]{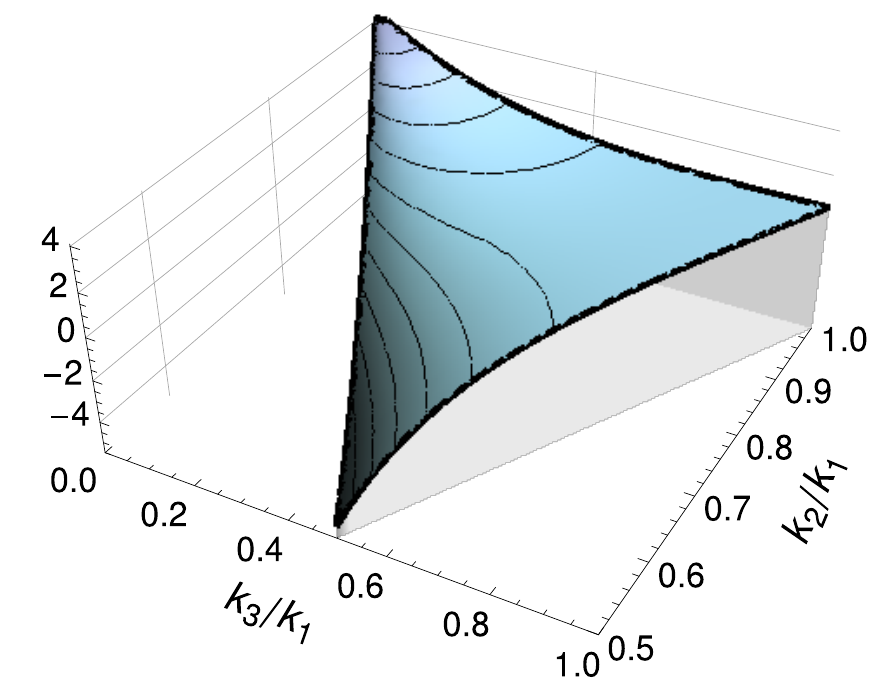}
 }
 \subfigure[~$\normEtc{K_{222} / K^3}$]{
  \label{fig:NMkay_shape_6}
  \includegraphics[width=0.24\textwidth]{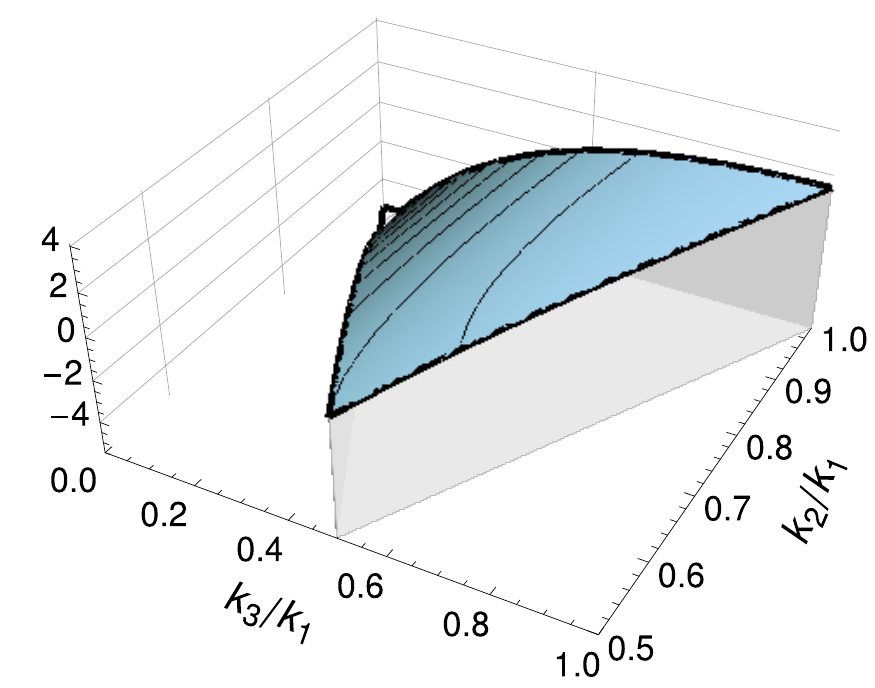}
 }
 \caption{Plots of the basic normalized and divergent-free contributions $\normEtc{\Ashape{i}}$, where $i = 1,\ldots,6$, which are the result of the $\normEtc{\cdot}$ operator, \cf\ \eqRef{eq:Nop}, on the terms in \eqRef{eq:defSystematicKaysNM}.}
 \label{fig:base_NM_kay_shapes}
\end{figure}

The mode expansion coefficients,
\begin{align}
 \alpha^\idx{i}_n = \iprod{\orthoR_n}{\Sshape[\tilde \sshape]{i}},
 \label{eq:alphaSystematicKayNM}
\end{align}
of the shape functions,
\begin{align}
 \Sshape[\tilde \sshape]{i} = \normEtc{\Ashape{i}} / K_{111},
\end{align}
corresponding to the normalized and divergent-free basic terms in \eqRef{eq:defSystematicKaysNM}, are plotted in \figRef{fig:base_NM_kays_conv}. The convergence of the expansion series is good (as expected), because we removed the divergent local contribution. As the plot shows, $\Ord{20}$ modes are sufficient to achieve better than $95\%$ convergence, just a few more as in the case of the equilateral or DBI shape  (\cf\ \figRef{fig:poly_alpha_convergence}). Thus, all shapes can be recovered rapidly by the polynomial modal expansion, an important requirement for future comparison with data.

\begin{figure}[t]
 \centering
 \includegraphics[width=0.9\textwidth]{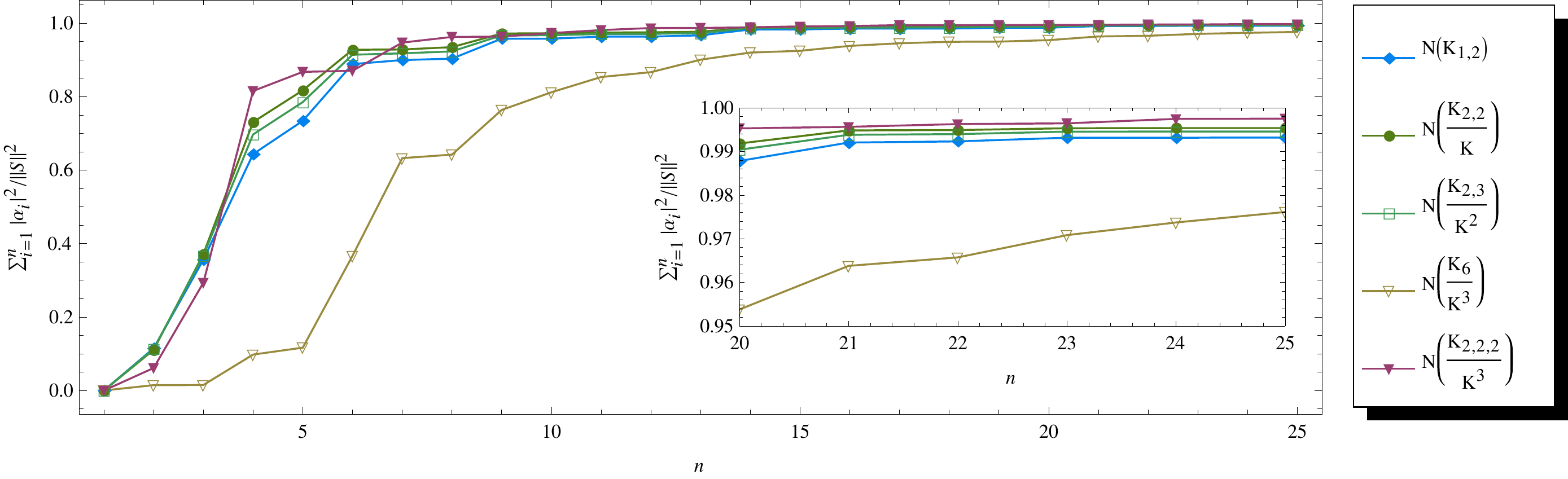}
 \caption{Convergence of the expansion series, \ie\ $\sum_{n=0}^{n_\mathrm{max}} \alpha^\idx{i}_n \conj{\alpha^\idx{i}_n} / \iprod{ \Sshape[\tilde \sshape]{i} }{ \Sshape[\tilde \sshape]{i} }$, after the $\normEtc{\cdot}$ operation (removal of divergent terms, substraction of a constant mode and normalization, \cf\ \eqRef{eq:Nop}) on the remaining five basic shapes in \eqRef{eq:defSystematicKaysNM}. The inset shows that $\Ord{20}$ shapes are sufficient to achieve better than $95\%$ convergence. 
 }
 \label{fig:base_NM_kays_conv}
\end{figure}

However, the strong correlation, visible to the naked eye, between four shapes in \figRef{fig:base_NM_kay_shapes} indicates that our ad-hoc choice is not well suited for subsequent discrimination. We can make this statement quantitative by computing the correlation coefficients between shapes defined via
\begin{align}
  C_{ij} \defeq \frac{ \iprod{ \Sshape[\tilde \sshape]{i} }{ \Sshape[\tilde \sshape]{i} } }{ \sqrt{ \iprod{ \Sshape[\tilde \sshape]{i} }{ \Sshape[\tilde \sshape]{i} }} \sqrt{ \iprod{ \Sshape[\tilde \sshape]{i} }{ \Sshape[\tilde \sshape]{i} } } },
 \label{eq:correlation_ij}
\end{align}
using the inner product on the tetrapyd in \sectRef{sec:shapeCorr}. The correlation coefficients $C_{ij}$ constitute the components of the correlation matrix in \figRef{fig:base_NM_kays_corr}.

\begin{figure}[t]
 \centering
 \subfigure[~unordered]{
  \label{fig:base_NM_kays_corr_unordered}
  \includegraphics[width=0.29\textwidth]{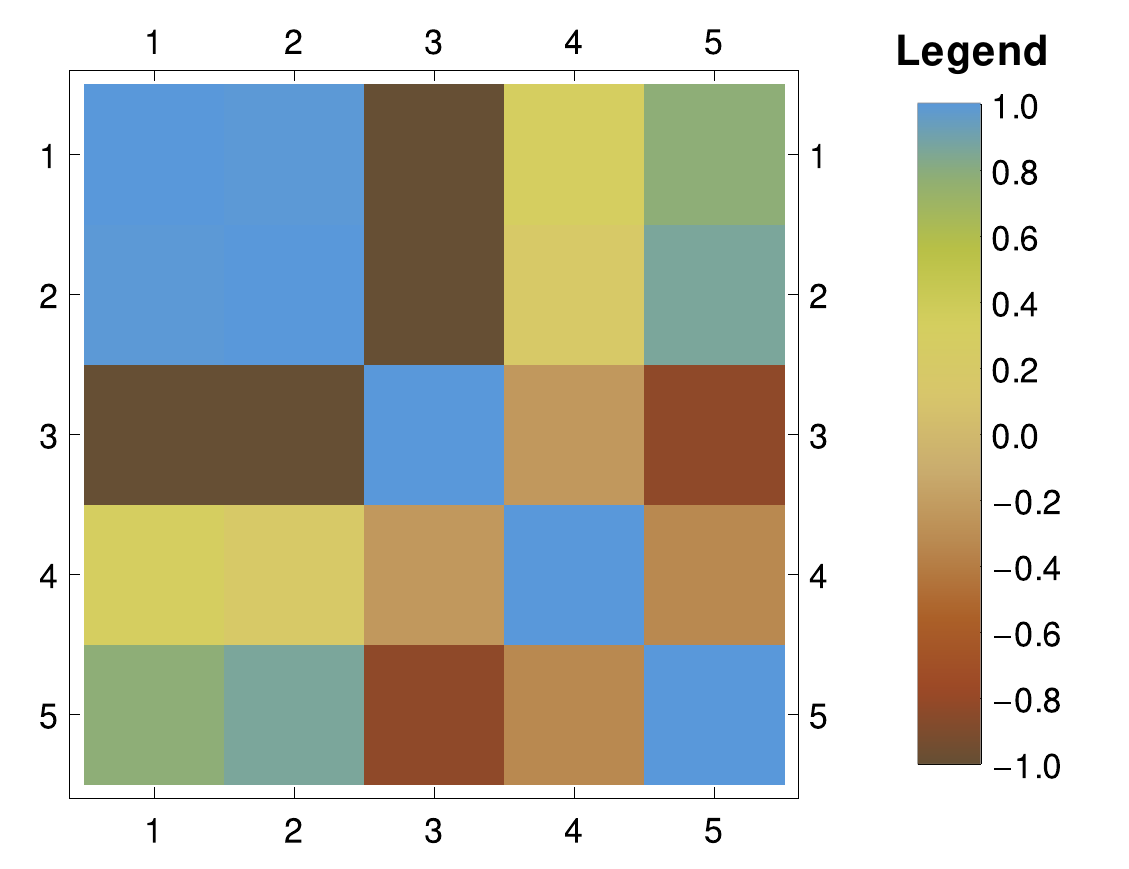}
 }
 \subfigure[~ordered]{
  \label{fig:base_NM_kays_corr_ordered}
  \includegraphics[width=0.29\textwidth]{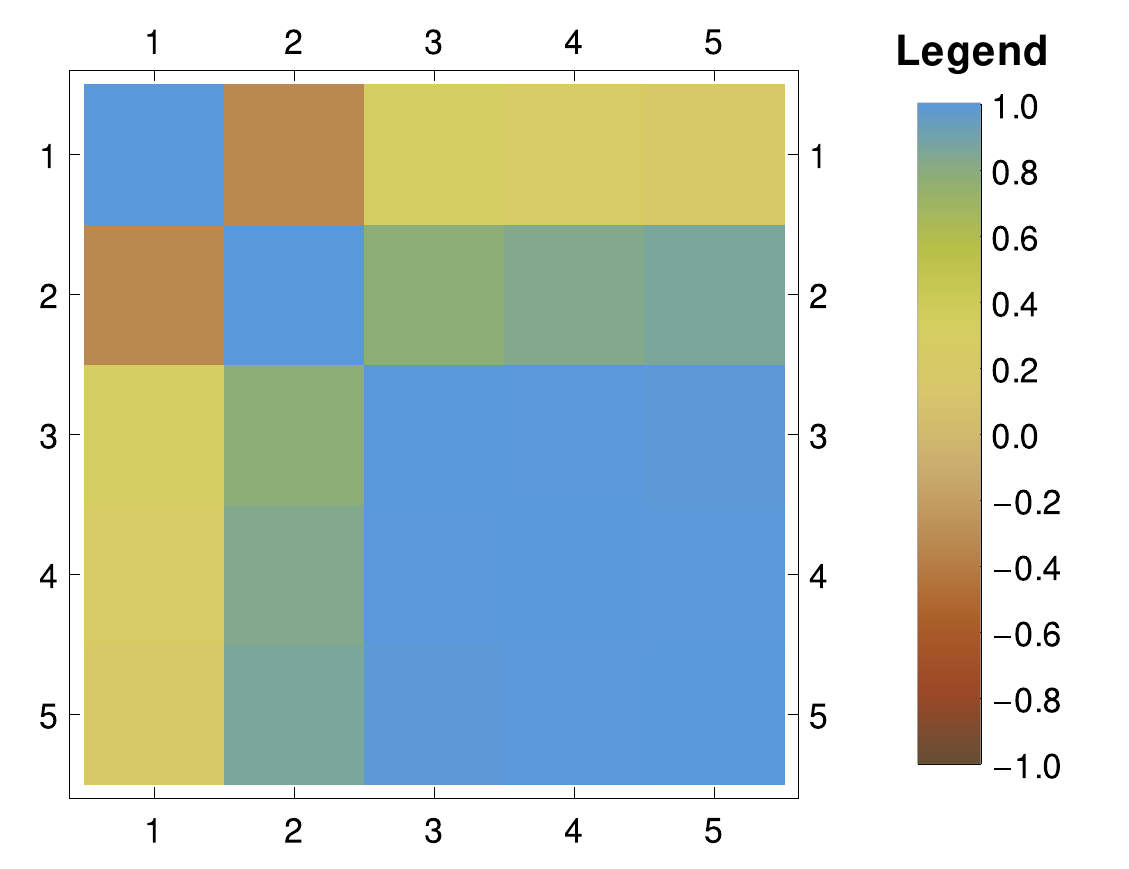}
 }
 \caption{Correlation matrix $\{ C_{ij} \}_{i=1,\ldots,5, \, j=1,\ldots,5}$ of the basic contributions after the $\normEtc{\cdot}$ operation (removal of divergent terms, subtraction of a constant mode and normalization, \cf\ \eqRef{eq:Nop}) on the remaining five basic shapes in \eqRef{eq:defSystematicKaysNM}. Ordering: \textbf{(a)} $\Ashape{1}\rightarrow 1$, $\Ashape{3}\rightarrow 2$, $\Ashape{4}\rightarrow 3$, $\dots$ and \textbf{(b)} as in \eqRef{eq:systKaysNMorder}, where $\Ashape{4}$ was mirrored, turning correlation into anti-correlation and vice versa for this shape. }
 \label{fig:base_NM_kays_corr}
\end{figure}

Ordering the shapes such that the structure in the correlation matrix becomes more visible, namely\footnote{One can for example start with an ordering according to the loads of the first eigenvalue and then fine-tune the ordering. If extended, this method would lead to a factor analysis or principal component analysis (PCA) of the shapes.}
\begin{align}
 1 \leftrightarrow \normEtc{\Ashape{5}}; \quad 2 \leftrightarrow \normEtc{\Ashape{6}}; \quad 3 \leftrightarrow \normEtc{\Ashape{1}}, \; 4 \leftrightarrow -\normEtc{\Ashape{4}}, \; \text{and} \; 5 \leftrightarrow \normEtc{\Ashape{3}},
 \label{eq:systKaysNMorder}
\end{align}
we can identify groups of correlated shapes, \figRef{fig:base_NM_kays_corr_ordered}, also indicated in the previous line by the spacing and the punctuation.

This grouping agrees with our intuitive eye-balling of the shapes in \figRef{fig:base_NM_kay_shapes}. The constant mode is, by construction, completely separated. Interestingly, we observe that $\Ashape{6}$ has significantly less correlation with the other shapes, while the correlation among the last four terms is at least $78.1\%$, and among the last three terms above $99.0\%$.

As a guide to compare with commonly known shapes, we would like to emphasize that the equilateral and orthogonal shapes,
\begin{align}
 \Ashaperm{equi} &\propto K_{12} - K_3 - 2 K_{111} & \Ashaperm{ortho} &\propto \Ashaperm{equi} - \frac 2 3 \Ashaperm{const} \propto K_{12} - K_3 - \frac 8 3 K_{111}
 \label{eq:equiAndOrthoInKays}
\end{align}
(see \cite{Liguori:2010hx}, \page{12}) are just linear combinations of the $\normEtc{\Ashape{1}}$ and $\Ashaperm{const}$ shapes.  Hence, a better choice could involve these two shapes as we already observed their orthogonality. Further, it turns out that most of the prefactors of our chosen basic shapes, see \eqs{\eqref{eq:seeryLidsey}} and \eqref{eq:Alambda} to \eqref{eq:As}, are slow-roll suppressed.

Hence, for exactly scale-invariant ($n_s=1$) single-field models for which the bispectrum is built from the basic terms in \eqRef{eq:defSystematicKaysNM}, the application of the commonly used shape templates provides a good approximation in almost all cases.\footnote{Such models are for example the scale-invariant models in \cite{Khoury:2008wj} and the dominant contributions in slow-roll type models as in \cite{Chen:2006nt}.}

However, being constrained by templates is unsatisfactory for more general/realistic models. To elucidate this point, consider the dependent $K$-terms in \eqRef{eq:kayInterdep} (without considering their potentially slow-roll suppressed prefactors): the correlation with our six basic terms is not always high, even though they are a linear combination of the basic contributions. This is not surprising, since the basic shapes we chose are not orthogonal. Building an orthonormal basis from the latter, as we have done for the separable mode functions $\modeQ_n(x,y,z)$ (see \sectRef{sec:modeExp}), is clearly an option \cite{Ribeiro:2011ax}, but the resulting basic shapes would be non-separable and increasingly intricate (\ie\ after application of Gram-Schmidt to the set in \eqRef{eq:systKaysNMorder}).

\subsubsection{Finding the least correlated, independent \altPdfText{$K$}{K}-terms}
\label{sec:leastCorrShapes}

\begin{figure}[t]
 \centering
 \subfigure[~$\normEtc{\FRAC{K_4}{K}}$]{
  \label{fig:kay_shape_3}
  \includegraphics[width=0.24\textwidth]{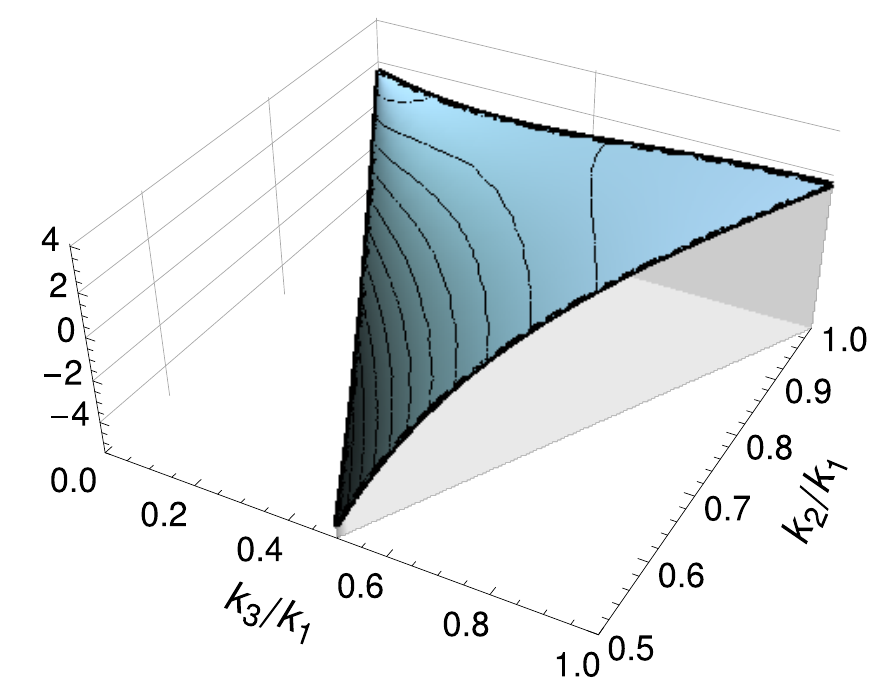}
 }
 \subfigure[~$\normEtc{\FRAC{K_{14}}{K^2}}$]{
  \label{fig:kay_shape_4}
  \includegraphics[width=0.24\textwidth]{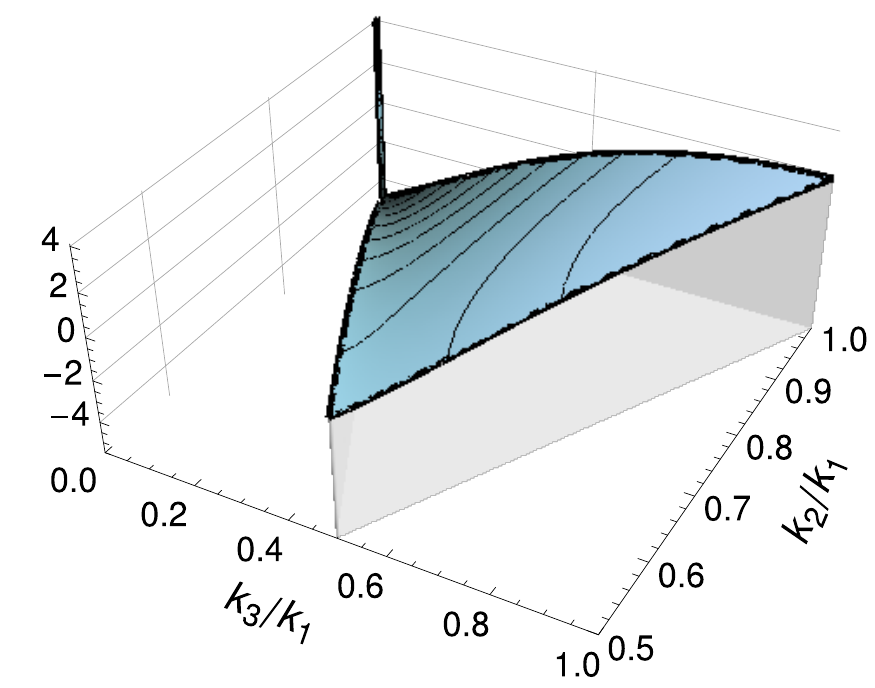}
 }
 \subfigure[~$\normEtc{\FRAC{K_{123}}{K^3}}$]{
  \label{fig:kay_shape_6}
  \includegraphics[width=0.24\textwidth]{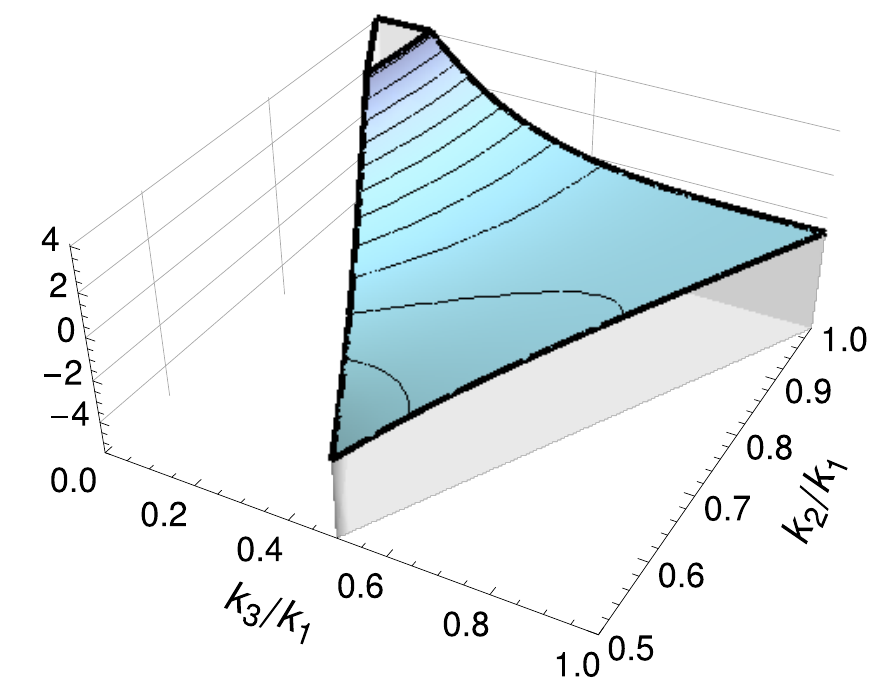}
 }
 \caption{Plots of the basic normalized and divergent-free contributions $\normEtc{\Ashape{i}^\Idx{lc}}$, where $i = 1,\ldots,6$, which are the result of the $\normEtc{\cdot}$ operator, \cf\ \eqRef{eq:Nop}, on the terms in \eqRef{eq:defBasisKay} and not yet plotted in \figRef{fig:base_NM_kay_shapes}. These shapes are chosen to minimize the correlation between the basic shapes.}
 \label{fig:base_kay_shapes}
\end{figure}

Without invoking an orthogonalization procedure, that is without combining the allowed shapes in \eqRef{eq:allowedKays}, we would like to identify the least correlated, independent $K$-terms in order to estimate their contribution to a given expansion series.  We can achieve this goal by applying a  factor analysis to the full set of possible $K$-terms in \eqRef{eq:allowedKays}; the resulting \emph{rotated basis}
\begin{equation}
 \begin{aligned}
  \Ashape{1}^\Idx{lc} &= K_{12}, & \Ashape{2}^\Idx{lc} &= \Ashaperm{const} = K_{111}, & \Ashape{3}^\Idx{lc} &= \frac{K_4}{K}, \\
  \Ashape{4}^\Idx{lc} &= \frac{K_{14}}{K^2}, & \Ashape{5}^\Idx{lc} &= \frac{K_6}{K^3}, & \Ashape{6}^\Idx{lc} &= \frac{K_{123}}{K^3},
  \label{eq:defBasisKay}
 \end{aligned}
\end{equation}
and $K_3$ (local shape) is a much better candidate to discriminate observationally between shapes without introducing increasingly cumbersome (but fully orthogonal) templates. The new terms are $\Ashape{3}^\Idx{lc}$, $\Ashape{4}^\Idx{lc}$, and $\Ashape{6}^\Idx{lc}$, which are plotted in \figRef{fig:base_kay_shapes}. Since the new shapes remain simple, we find the same rapid convergence for a polynomial mode expansion, see \figRef{fig:base_kays_conv} (only $\Ord{20}$ modes are needed for better than $95\%$ convergence).

\begin{figure}[t]
 \centering
 \includegraphics[width=0.9\textwidth]{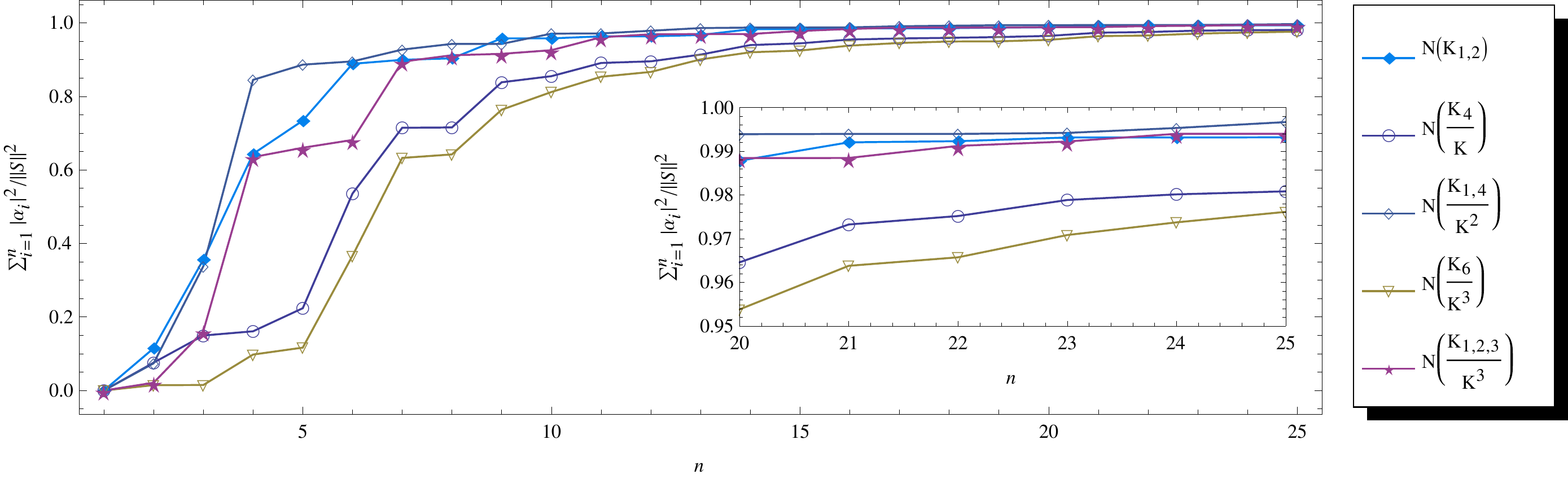}
 \caption{Convergence of the expansion series, \ie\ $\sum_{n=0}^{n_\mathrm{max}} \alpha^\idx{i}_n \conj{\alpha^\idx{i}_n} / \iprod{ \Sshape[\tilde \sshape]{i} }{ \Sshape[\tilde \sshape]{i} }$, after the $\normEtc{\cdot}$ operation (removal of divergent terms, subtraction of a constant mode and normalization, \cf\ \eqRef{eq:Nop}) on the remaining five basic shapes in \eqRef{eq:defBasisKay}, who are chosen to minimize the correlation between the basic shapes. The inset shows that $\Ord{20}$ shapes are sufficient to achieve better than $95\%$ convergence.}
 \label{fig:base_kays_conv}
\end{figure}

\begin{figure}[t]
 \centering
 \includegraphics[width=0.29\textwidth]{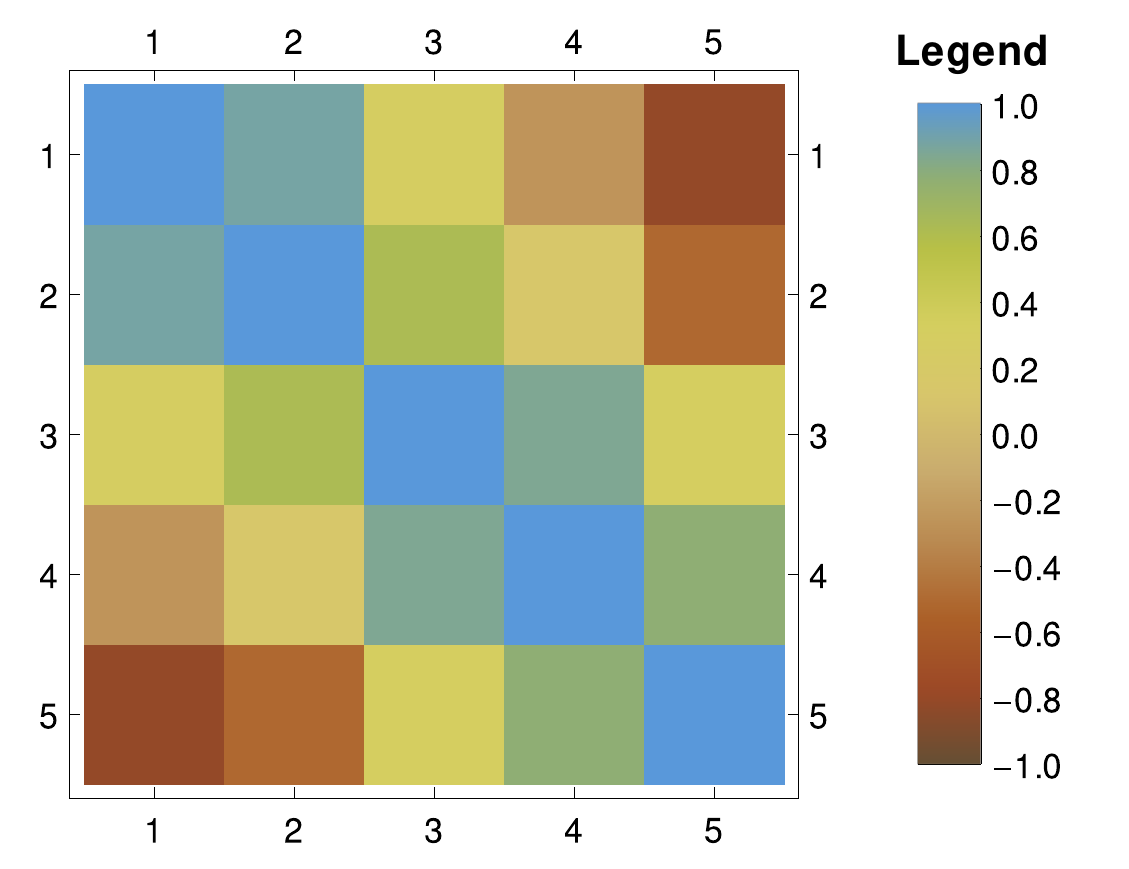}
 \caption{Correlation matrix $\{ C_{ij}^\Idx{lc} \}_{i=1,\ldots,5, \, j=1,\ldots,5}$ of the basic contributions after the $\normEtc{\cdot}$ operation (removal of divergent terms, subtraction of a constant mode and normalization, \cf\ \eqRef{eq:Nop}) on the remaining five basic shapes in \eqRef{eq:defBasisKay}, ordered as in \eqRef{eq:basisKayCorrOrder}. These shapes are chosen to minimize the correlation between the basic shapes.}
 \label{fig:base_kays_corr}
\end{figure}

The correlation matrix
\begin{align}
 C_{ij}^\Idx{lc} \defeq \frac{ \iprod{ \Sshape[\tilde \sshape]{i}^\Idx{lc} }{ \Sshape[\tilde \sshape]{i}^\Idx{lc} } }{ \sqrt{ \iprod{ \Sshape[\tilde \sshape]{i}^\Idx{lc} }{ \Sshape[\tilde \sshape]{i}^\Idx{lc} } } \sqrt{ \iprod{ \Sshape[\tilde \sshape]{i}^\Idx{lc} }{ \Sshape[\tilde \sshape]{i}^\Idx{lc} } } }, \quad \text{where} \quad \Sshape[\tilde \sshape]{i}^\Idx{lc} = \normEtc{\Ashape{i}^\Idx{lc}} / K_{111},
 \label{eq:correlation_ij_lc}
\end{align}
for the ordering
\begin{equation}
 \begin{aligned}
  1 \leftrightarrow -\normEtc{\Ashape{4}^\Idx{lc}}, \; 2 \leftrightarrow -\normEtc{\Ashape{6}^\Idx{lc}}, \; 3 \leftrightarrow -\normEtc{\Ashape{5}^\Idx{lc}}, \; 4 \leftrightarrow \normEtc{\Ashape{3}^\Idx{lc}}, \; \text{and} \; 5 \leftrightarrow -\normEtc{\Ashape{1}^\Idx{lc}},
 \end{aligned}
 \label{eq:basisKayCorrOrder}
\end{equation}
is plotted in \figRef{fig:base_kays_corr}: correlation between $62.4\%$ and $88.1\%$ on the off-diagonal and $81.0\%$ anti-correlation between $\normEtc{\frac{K_{14}}{K^2}}$ and $\normEtc{K_{12}}$ is evident, a significant improvement compared to \figRef{fig:base_NM_kays_corr}.

We would like to highlight the usefulness of such a least correlated, but simple choice of basic shapes. A general  (scale-invariant) bispectrum is a linear superposition of the basic shapes
\begin{align}
\ashape = \sum_i c_i \; \normEtc{\Ashape{i}^\Idx{lc}} + c_\mathrm{const} \; \Ashaperm{const} + c_\mathrm{local} \; \Ashaperm{local}.
\end{align}
By construction, model parameters, such as $c_s,\es, \dots$, enter only via the coefficients $c_i$, $c_\mathrm{const}$ and $c_\mathrm{local}$, not in $\Ashape{i}^\Idx{lc}$. Let's assume we would like to identify the most likely model parameters given a measurement of modal expansion parameters.  To scan the parameter space, \eg\ via MCMC simulations,  modal expansions for many $\Ashape{}$s are needed; fortunately, this requires a one-time modal expansion of the basic shapes only. At this step, any choice of simple basic shapes suffices, but if the chosen basic shapes are only weakly correlated, a resulting bispectrum is less likely the result of a near perfect cancellation of some highly correlated shapes (the latter would amplify errors). Furthermore, it is possible to pinpoint the basic shape that dominates the bispectrum, enabling physical insight into the origin of non-Gaussianities; the latter would be obfuscated if a cumbersome orthogonal set were chosen.  
 
Roughly speaking, the least correlated basic shapes provide the middle ground between the orthogonal mode functions on the tetrapyd, whose expansion coefficients are optimally constrained by the data without, however, offering direct physical insight, and the full primordial shape function, which is the direct imprint of a cosmological model, but unfortunately not directly accessible in observations.

\subsection{Expanding slow-roll violating models}
\label{sec:slowRollViolatingModeExpansion}

Equipped with a thorough understanding of general shapes in single-field slow-roll models, we come to the main focus of this paper, the slow-roll violating single-field models introduced in \sectRef{sec:fastRoll}. We are interested in models with a bispectrum that is a mixture of different commonly discussed shapes (\cf\ \cite{Noller:2011hd}, \fig{2}). To this end, the discussion of shapes in \sectRef{sec:leastCorrShapes} is directly applicable as we use the least correlated basic shapes for an optimal reconstruction of the shape coefficients using the modal decomposition. We constrain our investigation to models that are consistent with the newest WMAP data, that is we assume an almost scale-invariant power spectrum with a slight red tilt. The WMAP7 result for the spectral index is \cite{Komatsu:2010fb}
\begin{align}
 n_s \approx 0.968 \pm 0.012 \quad (68\% \mbox{\text{ CL}}) \quad \text{(WMAP7 only)}.
 \label{eq:WMAP7ns}
\end{align}

As concrete examples, we would like to focus on two models: \emph{the general DBI model} with $f_X^\Idx{dbi} = 1 - c_s^2$, and \emph{the negative $f_X$ model}, where we chose $f_X = -100$ as in \cite{Noller:2011hd}, \fig{4}. In both models, we chose a small speed of sound, namely $c_s = 1/20$, in order to have a non-negligible $\FNL{equi}$. The slow-roll parameter $\epsilon$ is chosen in the range $[0, 0.3]$, so that observations remain in line with CMBR observations (\cf\ \cite{Noller:2011hd}, \eq{2.9}).

\begin{table}[hbtp]
 \begin{center}
  \begin{tabular}{l|c|c|c|c}
   \bfseries{Model}     & $\epsilon$                                  & $c_s$  & $f_X$       & $n_s$  \\
   \hline
   General DBI model    & $\{0.3, 0.2, 0.1, 0.075, 0.05, 0.04, 0.03,$ & $0.05$ & $1 - c_s^2$ & $0.96$ \\
                        & $0.025, 0.02, 0.015, 0.01, 0.005, 0.001\}$  &        &             &        \\
   Negative $f_X$ model & $\{0.3, 0.15, 0.060, 0.03,$                 & $0.05$ & $-100$      & $0.96$ \\
                        & $0.015, 0.006, 0.003, 0.001\}$              &        &             &        \\
  \end{tabular}
 \end{center}
 \caption{Parameter table for two exemplary slow-roll violating models to which we apply the modal expansion.}
 \label{tab:model_parameters}
\end{table}

\begin{figure}[t]
 \centering
 \subfigure[~general DBI model]{
  \label{fig:NM_fNLequi_GenDBI}
  \includegraphics[width=0.4\textwidth]{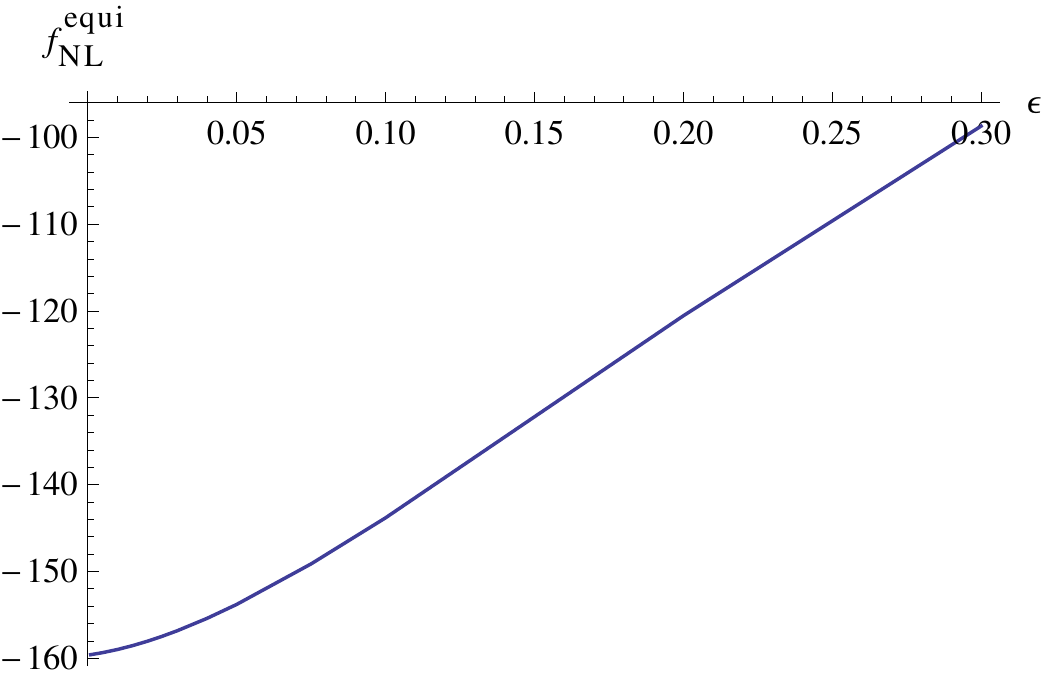}
 }
 \subfigure[~negative $f_X$ model]{
  \label{fig:NM_fNLequi_NegFX}
  \includegraphics[width=0.4\textwidth]{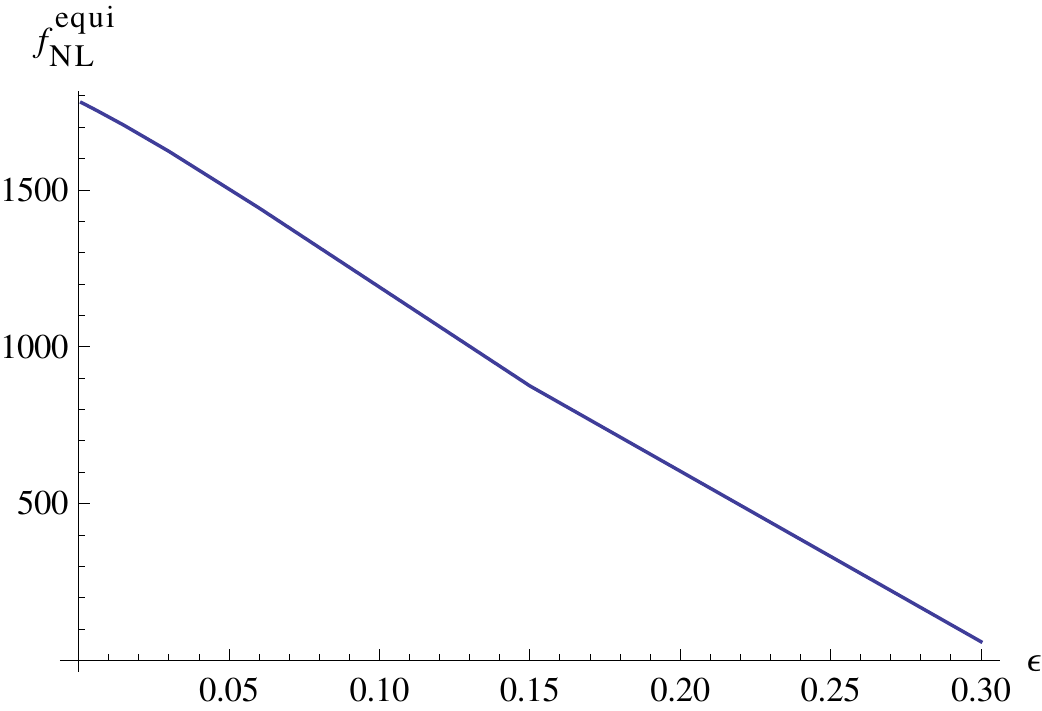}
 }
 \caption{$\FNL{equi}$ for the two concrete examples of the Noller and Magueijo shape. \textbf{(a)} The general DBI model with $n_s = 0.96$, $c_s = 1/20$ and $f_X^\Idx{dbi} = 1 - cs^2$. \textbf{(b)} The negative $f_X$ model with $n_s = 0.96$, $c_s = 1/20$ and $f_X = -100$.}
 \label{fig:NM_fNLequi}
\end{figure}

Our first model corresponds to the first concrete example chosen in \cite{Noller:2011hd}, \sect{4.1} (\emph{fast-roll DBI inflation}). For models in this class, the first term in the bispectrum, $\Ashape{\dot\zeta^3}$, is negligible compared to the other amplitudes in \eqRef{eq:NollerMagueijoShape}.

The negative $f_X$ model is an example from the class of models discussed in \cite{Noller:2011hd}, \sect{4.2} ($\abs{\lambda / \Sigma} \gg 1$). The equilateral shape is the most prominent shape in this class, dominating over the remaining ones such as the local contribution caused by the violation of scale invariance.

In both cases deviations from slow roll decrease the magnitude of $\FNL{equi}$, see \figRef{fig:NM_fNLequi}.\footnote{We employ, just as in \sectRef{sec:genSingleFieldModels} and in \cite{Noller:2011hd}, the $\fNL$ sign convention by Komatsu \etAl\ \cite{Komatsu:2010fb}.}

\subsubsection{Approximative scale invariance}
\label{sec:approx_si}

\begin{figure}[t]
 \centering
 \subfigure[~$\Lop{K_{12}}$]{
  \label{fig:log_kay_shape_1}
  \includegraphics[width=0.24\textwidth]{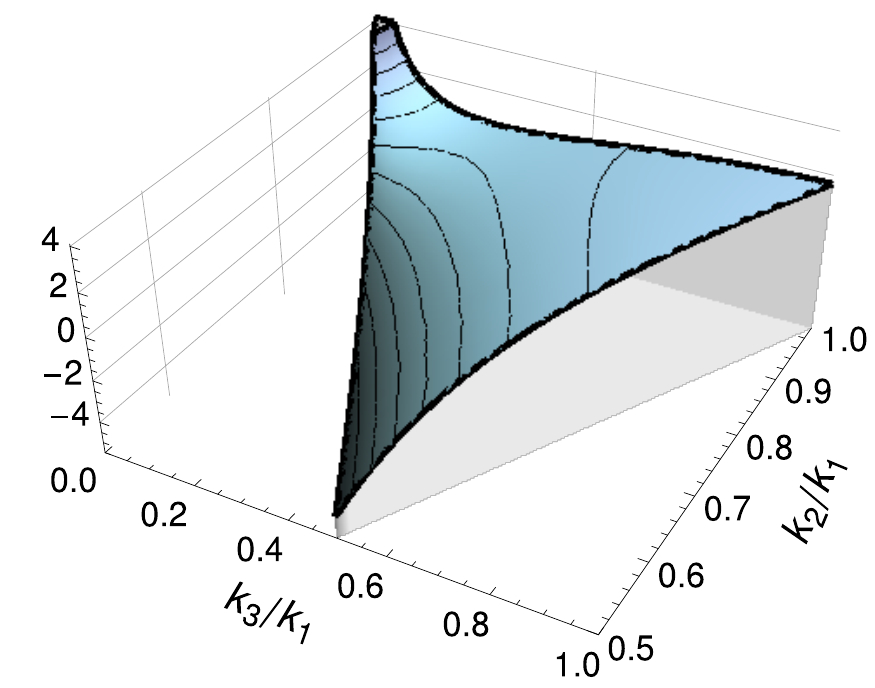}
 }
 \subfigure[~$\Lop{K_{111}}$]{
  \label{fig:log_kay_shape_2}
  \includegraphics[width=0.24\textwidth]{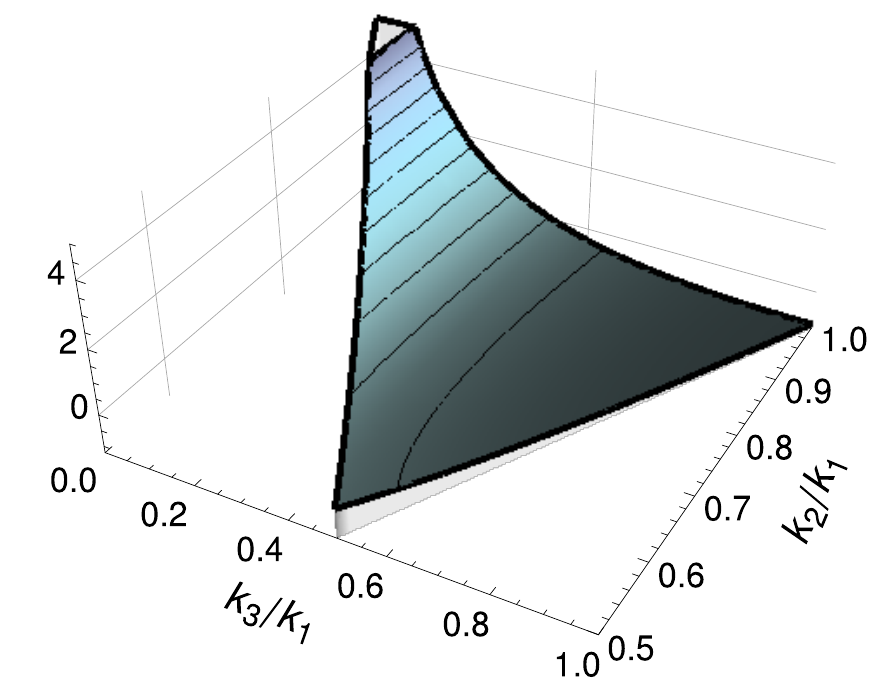}
 }
 \subfigure[~$\Lop{\FRAC{K_4}{K}}$]{
  \label{fig:log_kay_shape_3}
  \includegraphics[width=0.24\textwidth]{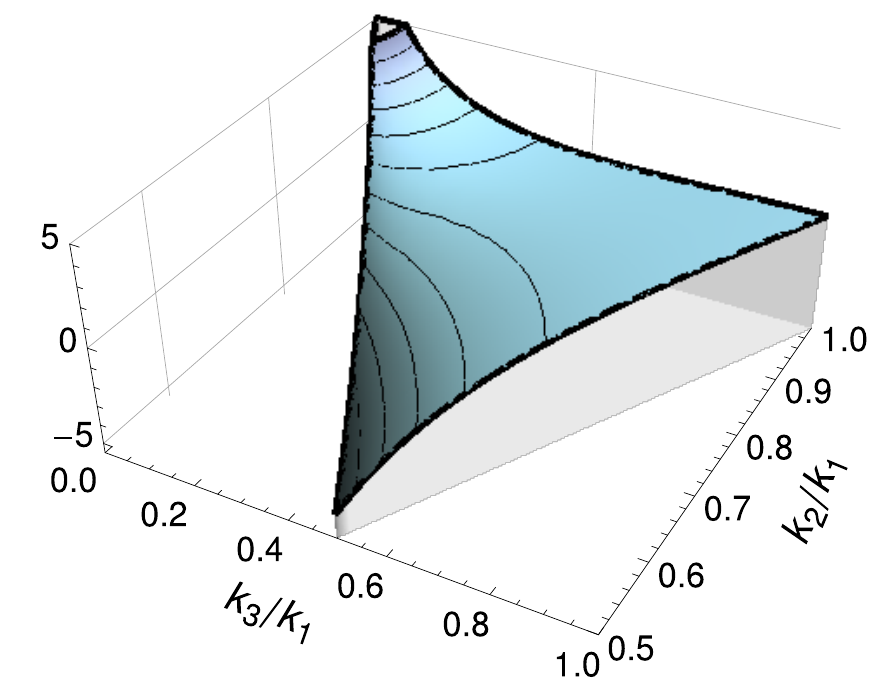}
 }
 \subfigure[~$\Lop{\FRAC{K_{14}}{K^2}}$]{
  \label{fig:log_kay_shape_4}
  \includegraphics[width=0.24\textwidth]{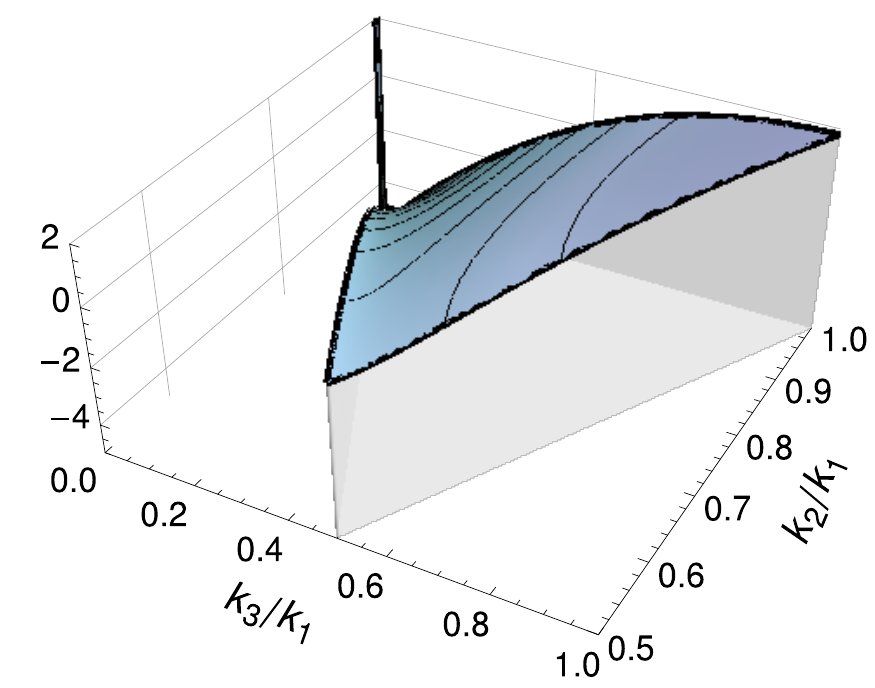}
 }
 \subfigure[~$\Lop{\FRAC{K_6}{K^2}}$]{
  \label{fig:log_kay_shape_5}
  \includegraphics[width=0.24\textwidth]{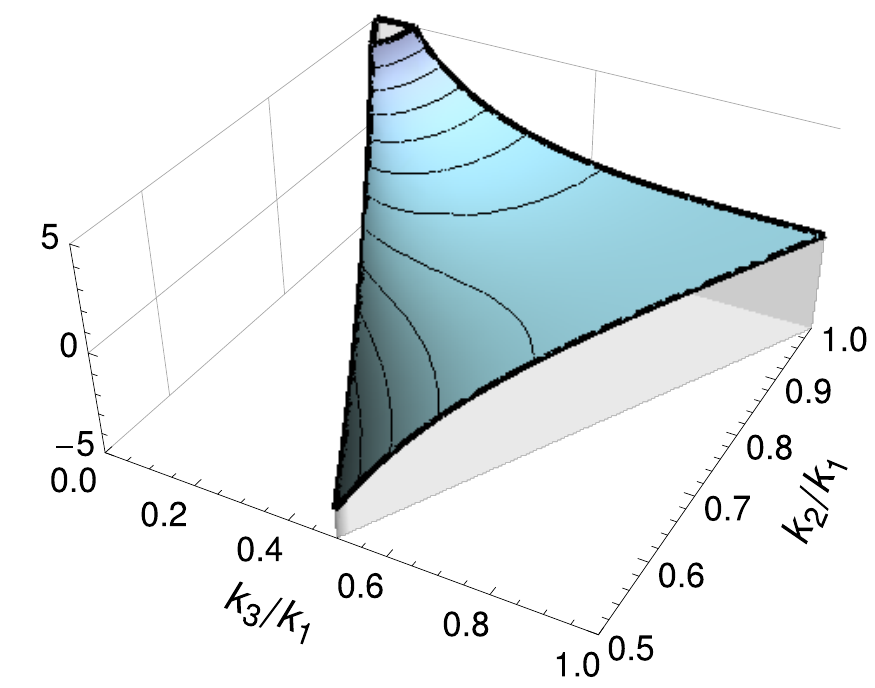}
 }
 \subfigure[~$\Lop{\FRAC{K_{123}}{K^3}}$]{
  \label{fig:log_kay_shape_6}
  \includegraphics[width=0.24\textwidth]{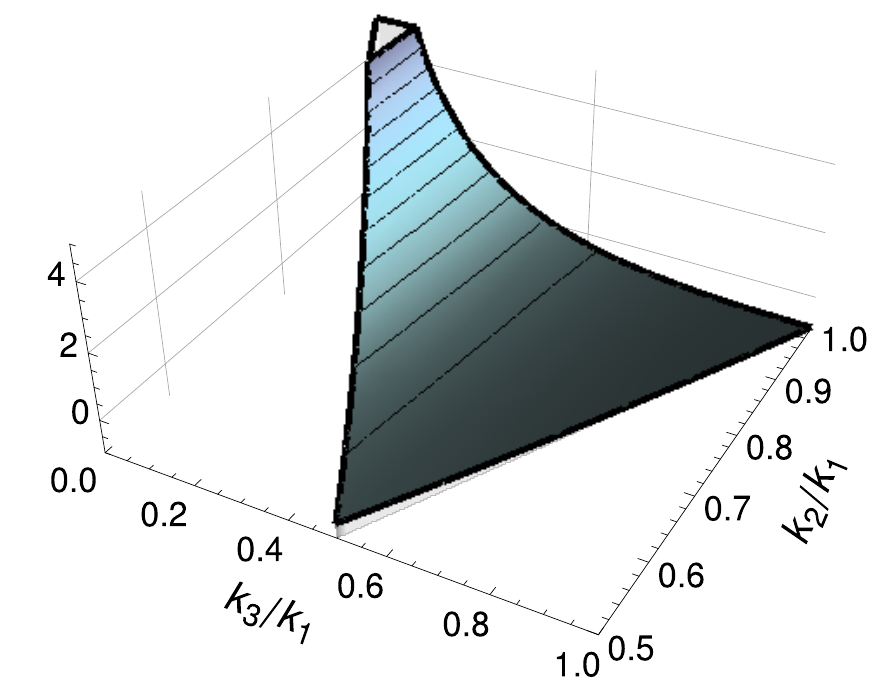}
 }
 \caption{Plots of the basic normalized and divergent-free contributions $\Lop{\Ashape{i}^\Idx{lc}}$, where $i = 1,\ldots,6$, which are the result of the $\Lop{\cdot}$ operator, defined beneath \eqRef{eq:defLogKKays}, on the terms in \eqRef{eq:defBasisKay} except $\Ashaperm{local} = K_3$. These shapes are chosen to minimize the correlation between the six basic shapes. (Note the different scales of the vertical axis.)}
 \label{fig:log_kay_shapes}
\end{figure}

In the limit of exact scale invariance, the Noller and Magueijo bispectrum reduces to the one by Khoury and Piazza \cite{Khoury:2008wj}. In this limit, the $(K_{111} / 2 K^3)^{n_s - 1}$ prefactor vanishes for all shape functions in \eqRef{eq:NollerMagueijoShape}. Hence, the full bispectrum  $\Ashaperm{NM}$ is a linear superposition of the base $K$ terms with coefficients depending on the parameter set $\{ \epsilon, c_s, f_X \}$.

Since we are discussing the observationally motivated case of a small deviation from scale invariance, one can expand the full bispectrum around $\boldalpha_1 = n_s - 1 = 0$.\footnote{Note that the expansion coefficients $\alpha_n$, see \eqRef{eq:orthoExpand}, (and all derived variations such as $\tilde \alpha_n$ and $\alpha^\idx{i}_n$) should not be confused with the model parameters $\boldalpha_1$ and $\boldalpha_2$ in \eqRef{eq:alphas}. Aiming at full consistency with the literature, we cannot avoid this conflicting notation. We distinguish the model parameters $\boldalpha_1$ and $\boldalpha_2$ by a  bold face in order to guard against misunderstandings.} The leading order contribution is the scale-invariant case discussed in \cite{Khoury:2008wj}, whose basic terms are given in \eqRef{eq:defSystematicKaysNM}. To subleading order, \ie\ $\Ord{\boldalpha_1}$, a linear combination of the basic $K$ terms and an additional set of base shapes that are the product of the ones in \eqRef{eq:defSystematicKaysNM} and $\log( K_{111} / K^3)$ arise. Such terms are also present in the bispectrum by Chen \etAl\ \cite{Chen:2006nt}.

\begin{figure}[t]
 \centering
 \includegraphics[width=0.9\textwidth]{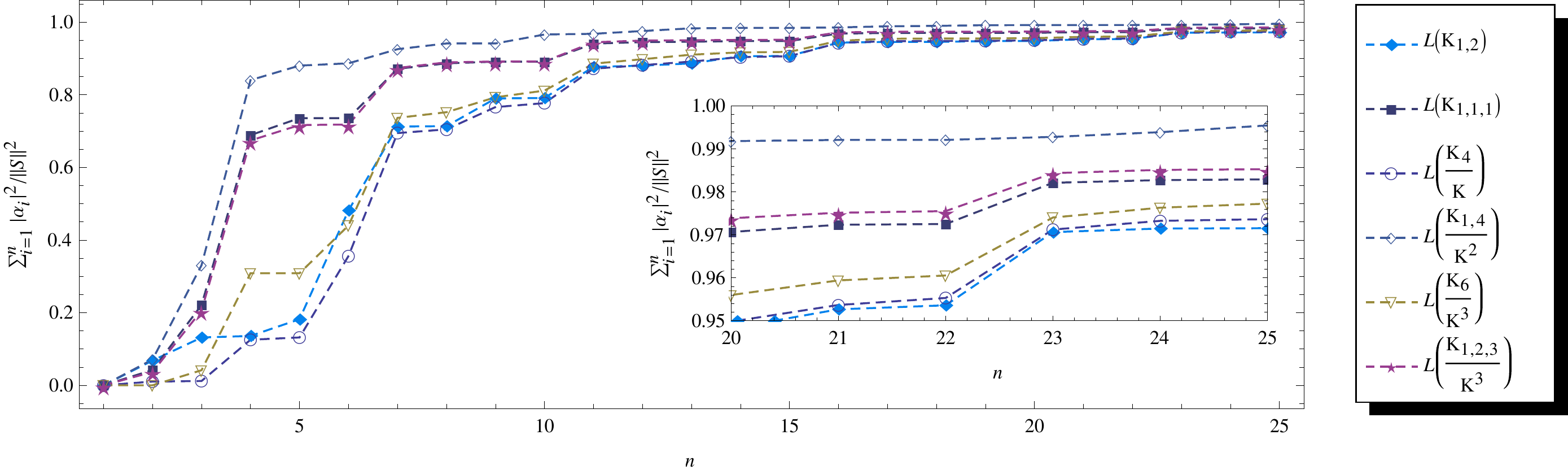}
 \caption{Convergence of the expansion series, \ie\ $\sum_{n=0}^{n_\mathrm{max}} \alpha^\idx{i}_n \conj{\alpha^\idx{i}_n} / \iprod{ \Sshape[\tilde \sshape]{i}^\Idx{log} }{ \Sshape[\tilde \sshape]{i}^\Idx{log} }$, of the terms in \eqRef{eq:defLogKKays}. The inset shows that $\Ord{20}$ shapes are sufficient to achieve better than $95\%$ convergence.}
 \label{fig:log_kays_conv}
\end{figure}

Because the multiplication with the logarithm should not drastically affect the overall shape, we expect that these new terms are strongly correlated with their ``mother terms''. In order to quantify this conjecture, we employ the set of least correlated, independent basic terms in \eqRef{eq:defBasisKay} and calculate the correlation matrix for these terms and the corresponding additional contributions,
\begin{align}
 \Lop{K_{12}}, \quad \Lop{K_{111}}, \quad \Lop{\frac{K_4}{K}}, \quad \Lop{\frac{K_{14}}{K^2}}, \quad \Lop{\frac{K_6}{K^3}}, \quad \text{and} \quad \Lop{\frac{K_{123}}{K^3}};
 \label{eq:defLogKKays}
\end{align}
here the operator $\Lop{\cdot}$ removes the local contribution from a given shape, multiplies the remainder with  $\log(K_{111} / K^3)$ and  normalizes the product using the $\normEtc{\cdot}$ operator. Note that $\FNL{local}$ is non-divergent for all terms in \eqRef{eq:defLogKKays}. The convergence of the expansion series of the shape functions
\begin{align}
 \Sshape[\tilde \sshape]{i}^\Idx{log} = \Lop{\Ashaperm{i}^\idx{lc}} / K_{111}, 
\end{align}
 is again rapid,  \figRef{fig:log_kays_conv}, and we achieve better than $95\%$ convergence using $20$ mode functions for all terms in \eqRef{eq:defLogKKays}.

The full bispectrum up to first order in $\boldalpha_1$ contains only the terms in \Eqs{\eqref{eq:defBasisKay}} and \eqref{eq:defLogKKays}, \ie
\begin{equation}
 \begin{aligned}
  \Ashaperm{NM}(\boldalpha_1, \boldalpha_2, c_s, f_X) = \; &f_\mathrm{local} \; \Ashaperm{local} + f_\mathrm{const} \; \Ashaperm{const} \\
  &+ \textstyle \sum_i f_i \; \normEtc{\Ashape{i}^\Idx{lc}} + \sum_i g_i \; \Lop{\Ashape{i}^\Idx{lc}} + \Ord{\boldalpha_1^2},
 \end{aligned}
 \label{eq:NM_shape_ordalpha1}
\end{equation}
where the $f_{\ldots}$ and $g_i$ are of order $\Ord{\boldalpha_1}$.  The $\Sshaperm{div} = K_3 \log(K_{111}/K^3)$ shape, whose $\FNL{local}$ is not defined because $\log(K_{111}/K^3) \limarrow{k_3 \to 0} \infty$, is absent.\footnote{Naively, there is a formal $K_3 \log(K_{111}/K^3)$ term in the $\Ord{\boldalpha_1}$ bispectrum which is, however, canceled by the $\Lop{\cdot}$ operator.} Hence, $\FNL{local}$ of the whole expression up to first order in $\boldalpha_1$ is proportional to $f_\mathrm{local}$, which can be derived from the consistency relation in \eqRef{eq:CreminelliZaldarriaga} by Creminelli and Zaldarriaga \cite{Creminelli:2004yq} (\ie\ the local contribution is absent in the case of exact scale invariance). In the following, it is important to keep in mind that even though the almost scale-invariant case leads to a local contribution, it is suppressed by the smallness of $n_s - 1$.

\begin{figure}[t]
 \centering
 \includegraphics[width=0.4\textwidth]{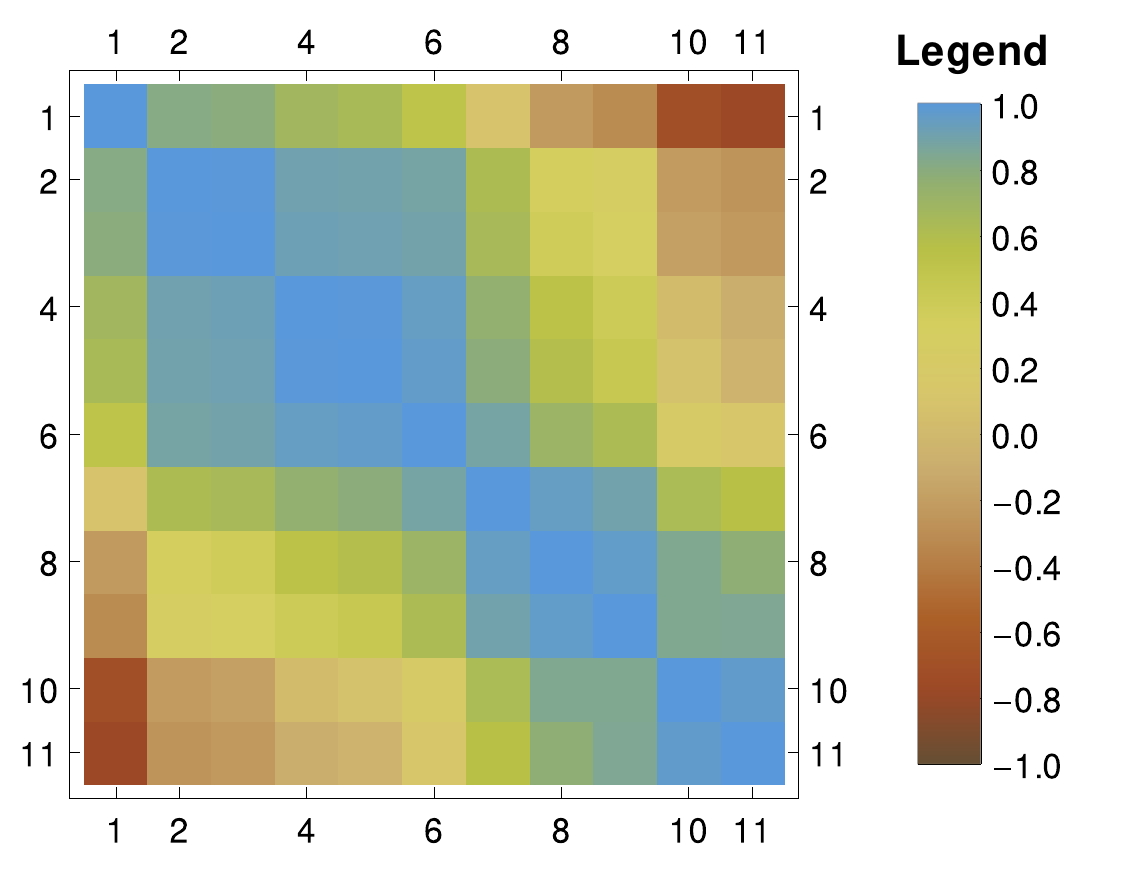}
 \caption{Correlation matrix $\{ C_{ij}^\Idx{log} \}_{i=1,\ldots,10, \, j=1,\ldots,10}$ of the terms in \eqRef{eq:defLogKKays}, ordered as in \eqRef{eq:logKKaysOrder}.}
 \label{fig:log_kays_corr}
\end{figure}

The correlation matrix in \figRef{fig:log_kays_corr} is best visualized if we order shapes as
\begin{equation}
 \begin{array}{l}
  1 \leftrightarrow -\normEtc{K_{12}}; \quad 2 \leftrightarrow -\normEtc{K_{14}/K^2}, \; 3 \leftrightarrow -\Lop{K_{14}/K^2}; \\
  4 \leftrightarrow \Lop{K_{111}}, \; 5 \leftrightarrow \Lop{K_{123}/K^3}, \; 6 \leftrightarrow \normEtc{K_{123}/K^2}; \\
  7 \leftrightarrow \Lop{K_6/K^3}, \; 8 \leftrightarrow \Lop{K_4/K}, \; 9 \leftrightarrow \normEtc{K_6/K^3}; \\
  10 \leftrightarrow \Lop{K_{12}}, \; 11 \leftrightarrow \normEtc{K_4/K},
 \end{array}
 \label{eq:logKKaysOrder}
\end{equation}
where correlated groups are indicated by spacing and punctuation. 
The structure of the correlation matrix proves, as conjectured, the similarity of the additional shapes to their mother terms (or another basic term).

For example, $\normEtc{K_{14}/K^2}$ and $\Lop{K_{14}/K^2}$ have a $99.9\%$ correlation, but there are three exceptions from the general rule: the shapes $\Lop{K_{111}}$, $\Lop{K_{12}}$ and $\Lop{K_4/K}$.

$\Lop{K_{111}}$ is, by definition, orthogonal to $K_{111}$. This shape is $99.8\%$ correlated with $\Lop{K_{123}/K^3}$, which itself is only $96.1\%$ correlated with its mother term $\normEtc{K_{123}/K^3}$. This relationship is apparent in the plots (compare \figRef{fig:log_kay_shapes}(b, f) and \figRef{fig:kay_shape_6}).

$\Lop{K_4/K}$ shares a higher correlation ($96.2\%$) with $\normEtc{K_6/K^3}$ than with $\Lop{K_6/K^3}$ ($89.8\%$). This is related to the behavior of the shapes in the squeezed limit (compare \figRef{fig:log_kay_shapes}(c, e) and \figRef{fig:kay_shape_3}).

The mother term of the former, $\normEtc{K_4/K}$, is $96.9\%$ correlated with $\Lop{K_{12}}$, because the divergence of $S$ in the squeezed limit is suppressed. The latter is only $77.4\%$ correlated with its mother term, $\normEtc{K_{12}}$.

\subsubsection{Convergence of the Noller and Magueijo example models}

Because observations show a nearly scale-invariant power spectrum \eqRef{eq:WMAP7ns}, we restrict our analysis to models that satisfy this constraint, taking $n_s = 0.96$ as a reference value. Consequently, the $(K_{111}/K^3)^{n_s - 1}$ factor in \eqRef{eq:NollerMagueijoShape} leads to a divergent contribution to the bispectrum, which cannot be subtracted easily 
and we have to fall back to the clipping method introduced in \sectRef{sec:divShapes}.\footnote{Substraction was possible in the last section,  since the bispectrum had been expanded to first order in $\boldalpha_1$ only; here, we would like to perform the modal expansion of the full bispectrum, that is, to go beyond first order in $\boldalpha_1$.}

\begin{figure}[t]
 \centering
 \subfigure[~General DBI expansion]{
  \label{fig:NM_coeffs_gendbi}
   \includegraphics[width=0.45\textwidth]{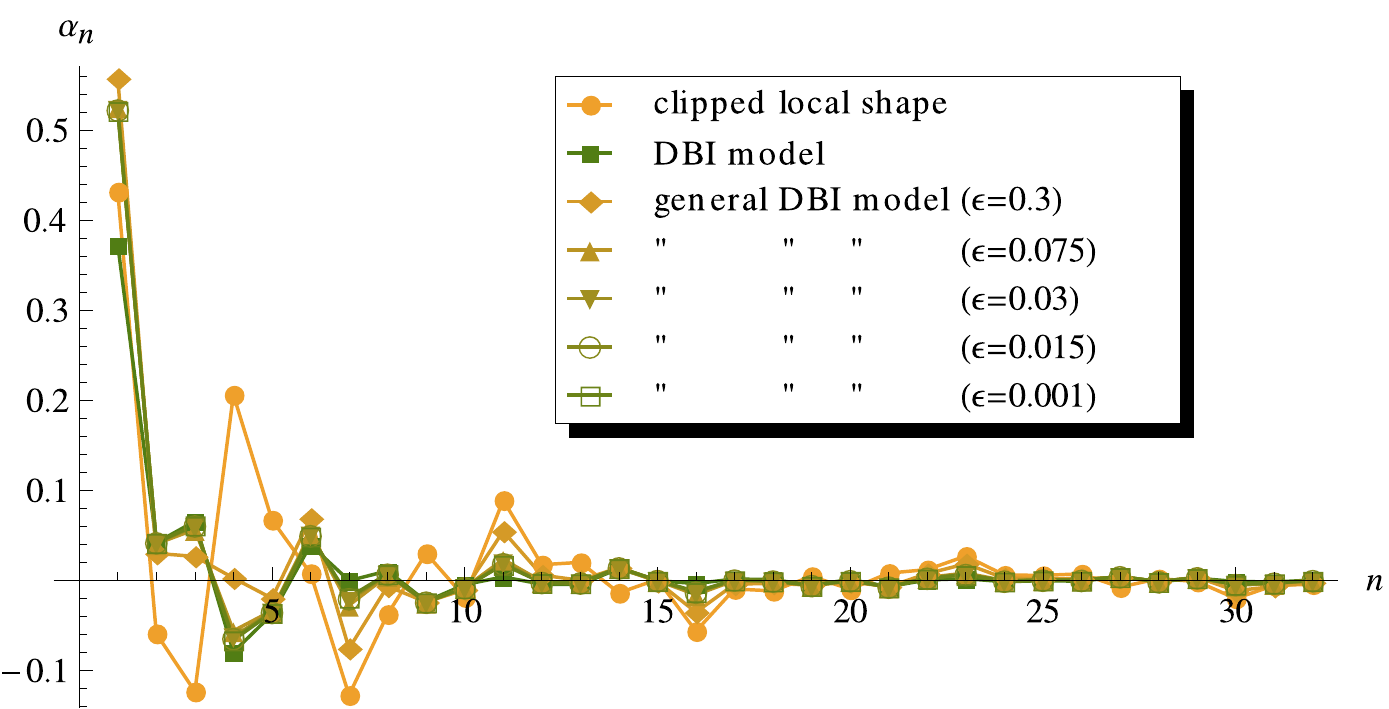}
 }
 \subfigure[~General DBI convergence]{
  \label{fig:NM_conv_gendbi}
   \includegraphics[width=0.45\textwidth]{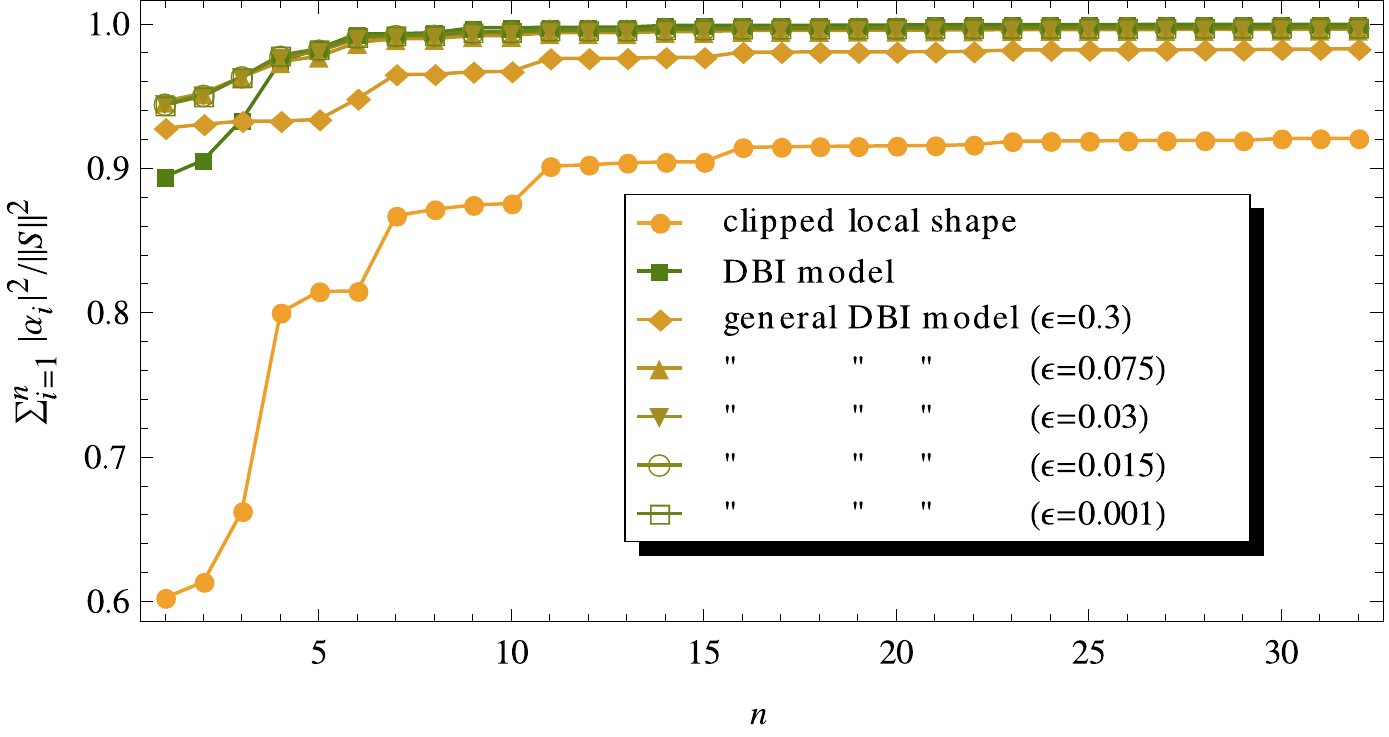}
 }
 \subfigure[~Negative $f_X$ expansion]{
  \label{fig:NM_coeffs_negfx}
   \includegraphics[width=0.45\textwidth]{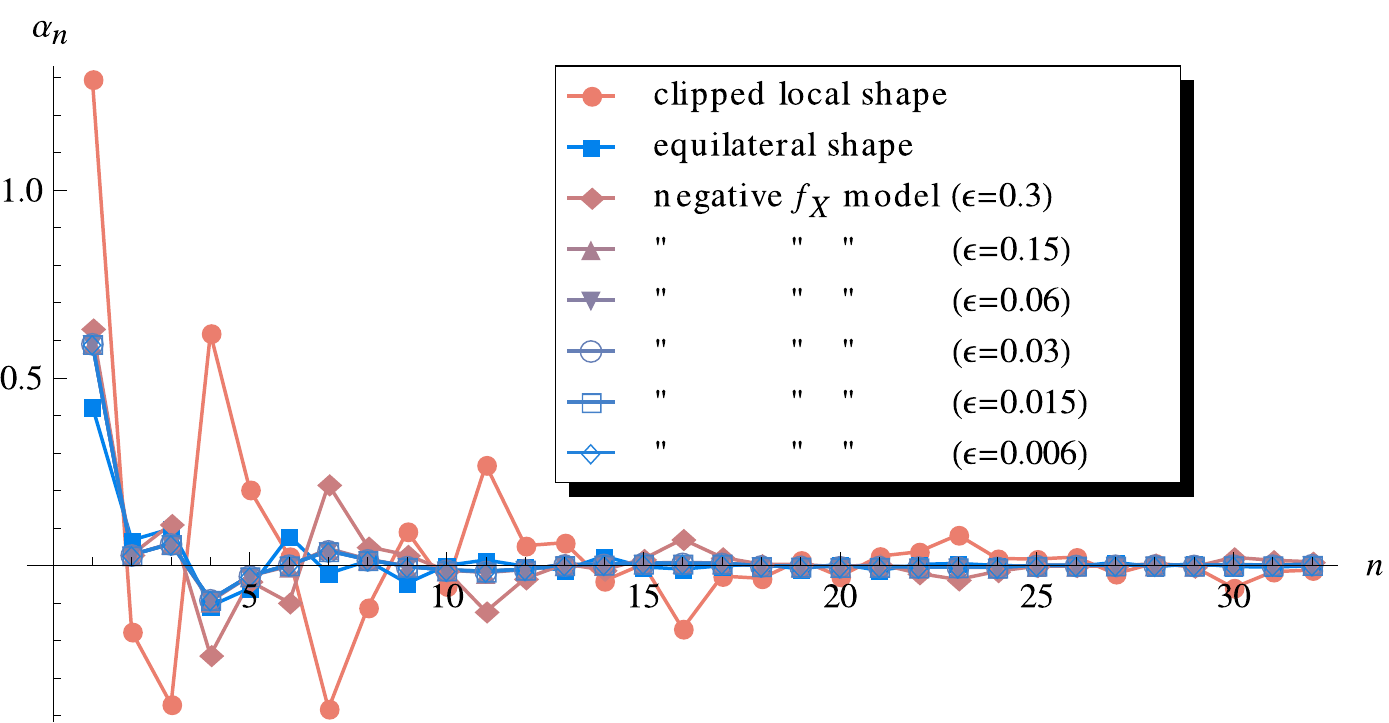}
 }
 \subfigure[~Negative $f_X$ convergence]{
  \label{fig:NM_conv_negfx}
   \includegraphics[width=0.45\textwidth]{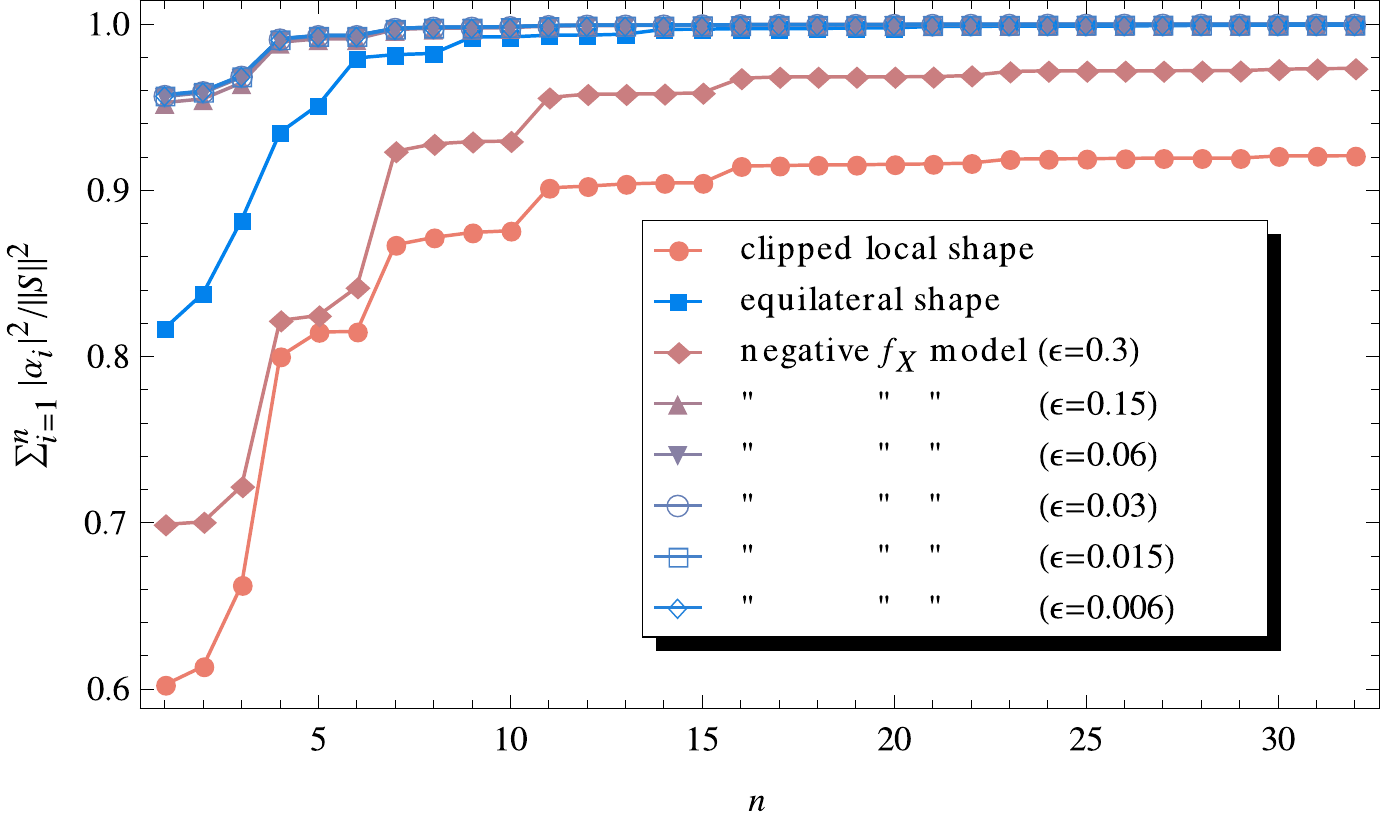}
 }
 \caption{Coefficients and convergence of the expansion series, \ie\ $\sum_{n=0}^{n_\mathrm{max}} \alpha^\idx{i}_n \conj{\alpha^\idx{i}_n} / \iprod{\tilde{\Sshape{i}}}{\tilde{\Sshape{i}}}$, of the bispectra for the models defined in \sectRef{sec:slowRollViolatingModeExpansion} for selected model parameters.}
 \label{fig:NM_expansion}
\end{figure}

We calculated the expansion series of the two proposed example models, the general DBI model and the so-called negative $f_X$ model, for the set of parameters given in \tabRef{tab:model_parameters}. The expansion coefficients of the normalized shape functions,
\begin{align}
 \Sshaperm{NM}(k_1,k_2,k_3) = \frac{1}{N} \frac{\Ashaperm{NM}(k_1, k_2, k_3)}{(k_1 k_2 k_3)^2}, \quad \text{where} \quad \Sshaperm{NM}(k,k,k) = 1,
\end{align}
in the case of the general DBI model are given in \figRef{fig:NM_coeffs_gendbi} for selected values of the slow roll parameter $\epsilon$; they are compared to the usual DBI model and the clipped local model (divided by a factor of 3 in order to have compatible amplitudes).

In the same manner, we give the expansion coefficients of the negative $f_X$ model in \figRef{fig:NM_coeffs_negfx}, compared to the equilateral and the clipped local model (the latter is again divided by a factor of 3).

In line with the dependence of the equilateral non-linearity parameter $\FNL{equi}$ on $\epsilon$ (see \figRef{fig:NM_fNLequi}), we find that the relative equilateral contribution decreases with increasing $\epsilon$. The local contribution is almost constant because the spectral index is unchanged. We did not include error bars in the plots of the expansion coefficients in \figRef{fig:NM_expansion}(a, c), which are given by the accuracy of \texttt{Mathematica}'s numerical integration utility. In order to have physically significant error bars for the mode coefficients relevant for CMB data comparison, one could derive the uncertainty from simulated maps \cite{Fergusson:2009nv}.

The relative increase of the local contribution explains why the correlation of the expansion series with the full bispectrum, shown in \figRef{fig:NM_expansion}(b, d), decreases for larger $\epsilon$. For the largest deviation from slow roll, we still achieve better than $92\%$ convergence using $32$ modes for both example models; we achieve almost $93\%$ using the full $n_\mathrm{max} = 53$ modes (not shown).

\begin{figure}[t!]
 \centering
 \subfigure[~base term contribution]{
  \label{fig:NM_expect_gendbi}
  \includegraphics[width=0.7\textwidth]{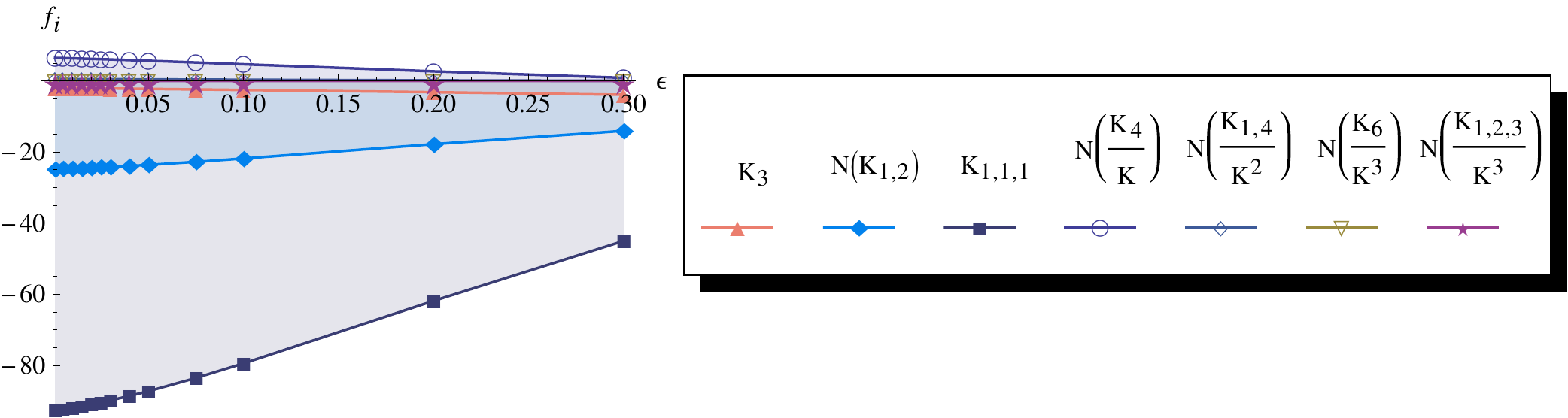}
 }
 \subfigure[~$\log(\cdot)$ term contribution]{
  \label{fig:NM_expectlogk_gendbi}
  \includegraphics[width=0.7\textwidth]{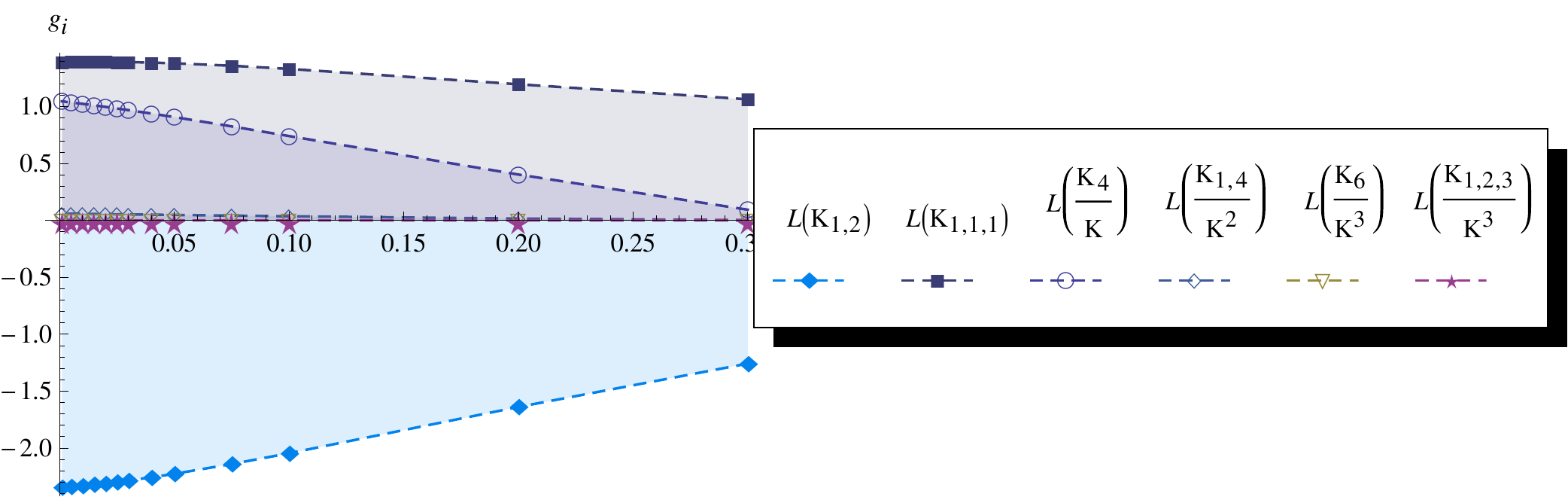}
 }
 \subfigure[~reconstructed contributions]{
  \label{fig:NM_reconst_gendbi}
  \begin{minipage}[c]{0.47\textwidth}
   \includegraphics[width=1\textwidth]{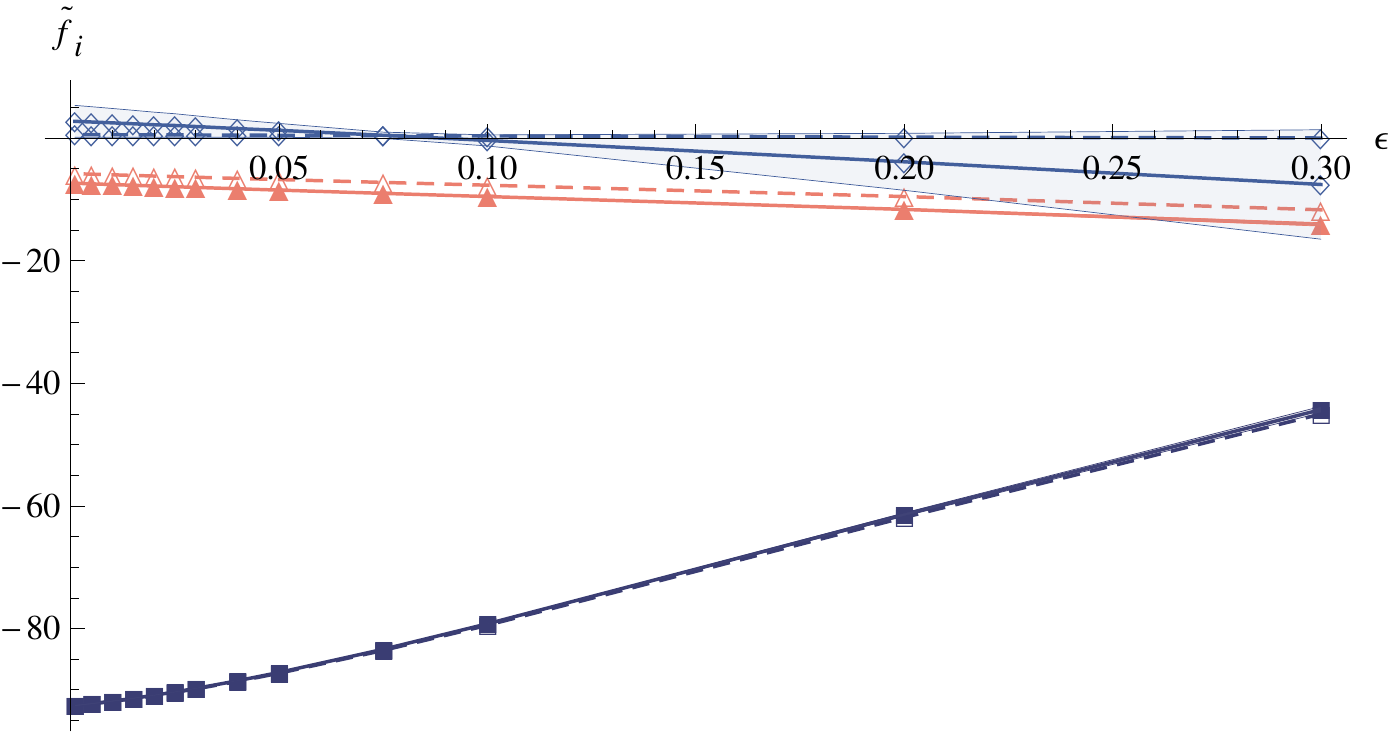}\\[.25cm]
   \includegraphics[width=1\textwidth]{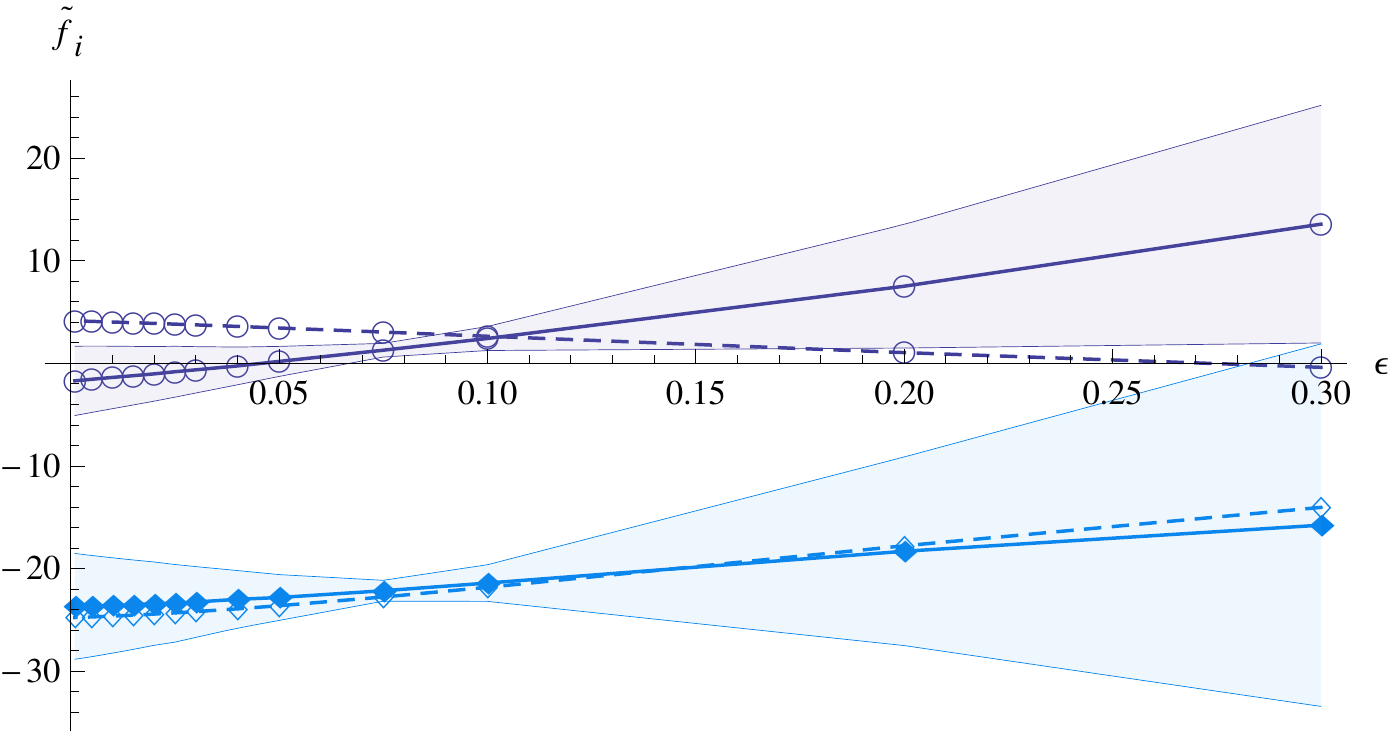}
  \end{minipage}
  \begin{minipage}[c]{0.47\textwidth}
   \vspace{1cm}
   \includegraphics[width=1\textwidth]{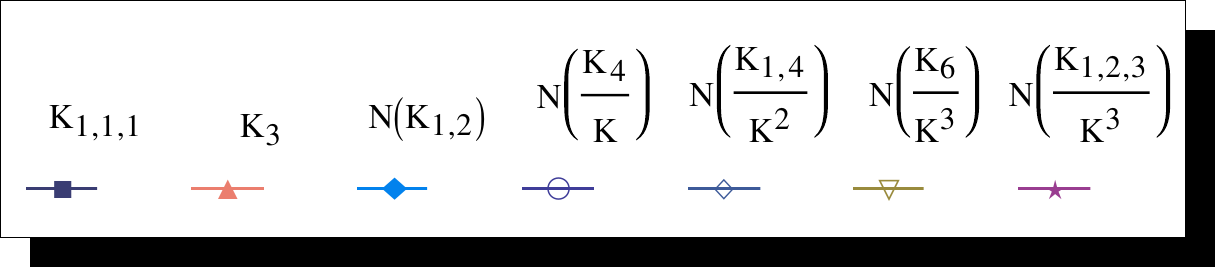}\\[1cm]
   \includegraphics[width=1\textwidth]{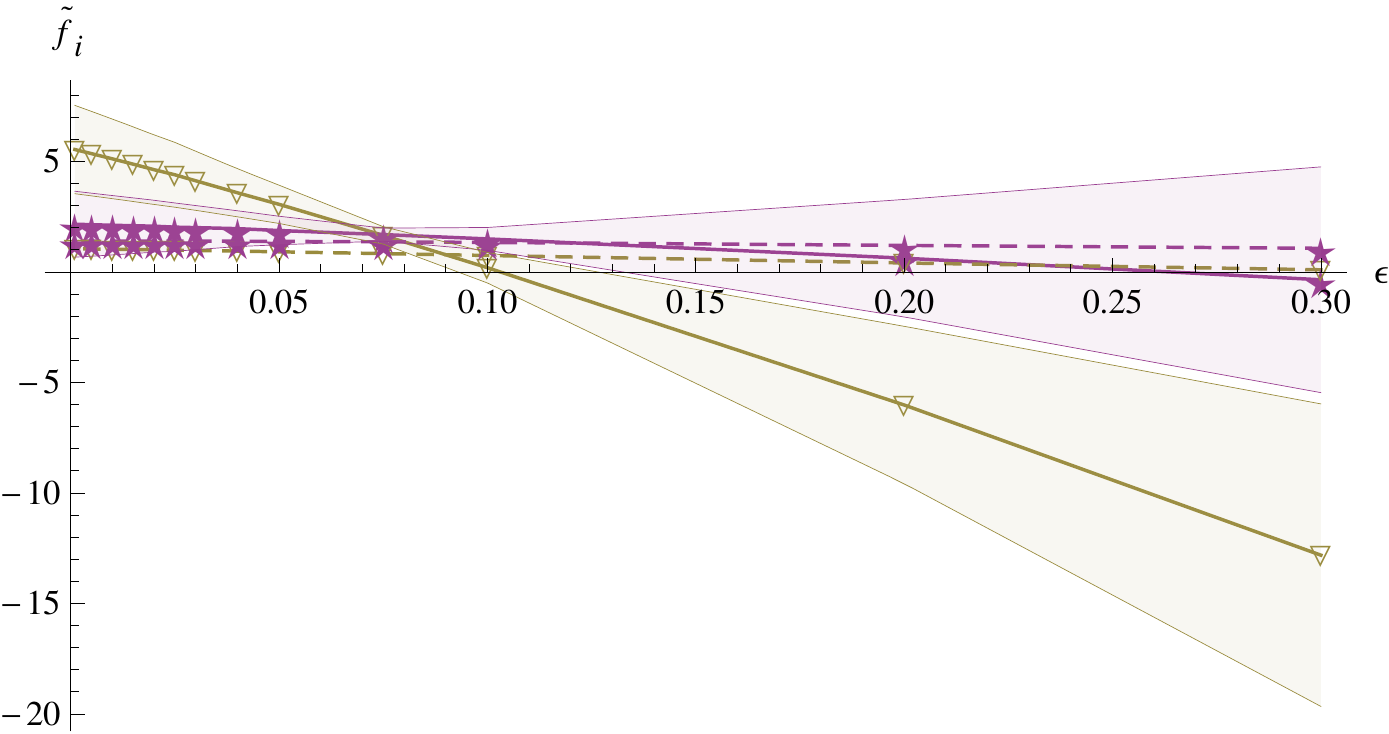}
  \end{minipage}
 }
 \caption{\textbf{(a, b)} Contributions of the terms in \eqRef{eq:NM_shape_ordalpha1} to the bispectrum amplitudes for the general DBI model defined in \sectRef{sec:slowRollViolatingModeExpansion} for a certain set of example parameters. \textbf{(c)} Results of the linear regression on the expansion coefficients $\alpha_n$ as in \eqRef{eq:linRegress}.}
 \label{fig:NM_model_decomp}
\end{figure}

\subsubsection[Consistency of the relative basic contributions]{Consistency of the relative contributions of the basic terms to the full bispectrum}

In order to compare the expansion coefficients with our expectations, we expand the full bispectrum as given in \eqRef{eq:NollerMagueijoShape} around $\boldalpha_1 = n_s - 1$. We decompose the bispectrum amplitudes of order $\Ord{\boldalpha_1}$ into the \emph{least correlated} basic terms and the derived $\Lop{\cdot}$-type shapes as in \eqRef{eq:NM_shape_ordalpha1} in order to improve the reconstruction that we employ to verify our results.

In \figRef{fig:NM_model_decomp} we plot the contributions of the basic terms to the zeroth and first order bispectra for the discussed set of model parameters.

\begin{figure}[t!]
 \centering
 \includegraphics[width=0.9\textwidth]{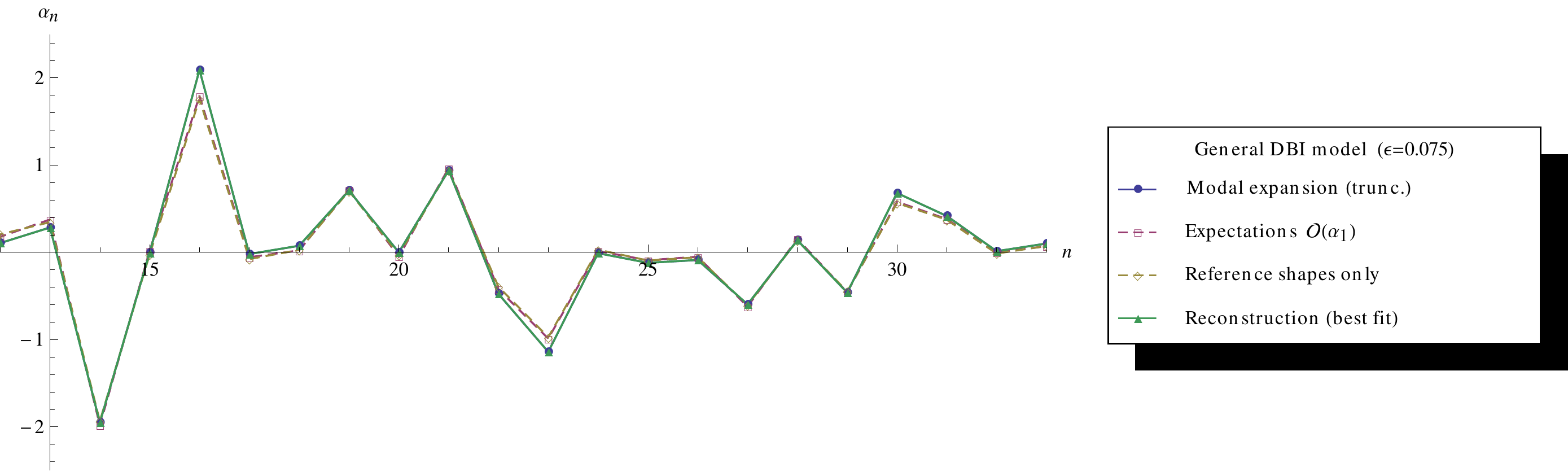}
 \caption{Calculated, predicted (up to $\Ord{\boldalpha_1}$) and recovered (from a linear regression using reference shapes) expansion coefficients of the general DBI model with $\epsilon = 0.075$. We plot the coefficients of the bispectrum amplitude (not normalized) in order to compare with predictions in the range $n = 13, \ldots, 33$.}
 \label{fig:NM_reconstr_alphas_negfx}
\end{figure}

During the discussion of the $\Lop{\cdot}$-type shapes in \sectRef{sec:approx_si}, we learned that we cannot distinguish certain shapes due to their high correlation. We thus take one representative from each group and estimate the contribution of representatives by a linear regression on the expansion coefficients. The result of that procedure can be seen in \figRef{fig:NM_reconst_gendbi}. The dashed line is the expected combined contribution for each group and the full lines are the results of the regression on the $\tilde f_i$ parameters,
\begin{align}
 \argmin_{\tilde f_i} \left( \alpha_n - \sum_i \tilde f_i \; \alpha^\idx{i}_n \right)^2 \; \forall n \le n_\mathrm{max} = 52,
\end{align}
where the $\alpha_n^\idx{i}$ are the expansion coefficients of the shapes representing a group of highly correlated shapes (compare \figRef{fig:log_kays_corr}); representatives of our ``binned'' shape classes are: the constant mode, the clipped local template, $\normEtc{K_{12}}$, $\normEtc{K_{4}}$, $\normEtc{K_{14}}$, $\normEtc{K_6/K^3}$, and $\normEtc{K_{222}/K^3}$.  The numerical inversion of the correlation matrix would be unpredictable if we had chosen to reconstruct the individual contributions of the highly correlated shapes. The result of this regression yields\footnote{We employ no weights for the coefficients $\alpha_n$, since we assume them to have comparable errors.}
{
 \allowdisplaybreaks
 \begin{align}
  \tilde f_i &= \sum_{j=1}^{n_\mathrm{shapes}} \left(C^{-1}\right)_{ij} d_j, & & & \text{where}
  \label{eq:linRegress} \\
  C_{ij} &= \sum_{n=1}^{n_\mathrm{max}} \alpha^\idx{i}_n \alpha^\idx{j}_n & \; \forall i,j &= 1,\ldots,n_\mathrm{shapes} & \text{and} \\
  d_j &= \sum_{n=1}^{n_\mathrm{max}} \alpha^\idx{j}_n \alpha_n & \; \forall j &= 1,\ldots,n_\mathrm{shapes}.
 \end{align}
}

\begin{figure}[t!]
 \centering
 \subfigure[~base term contribution]{
  \label{fig:NM_expect_negfx}
  \includegraphics[width=0.7\textwidth]{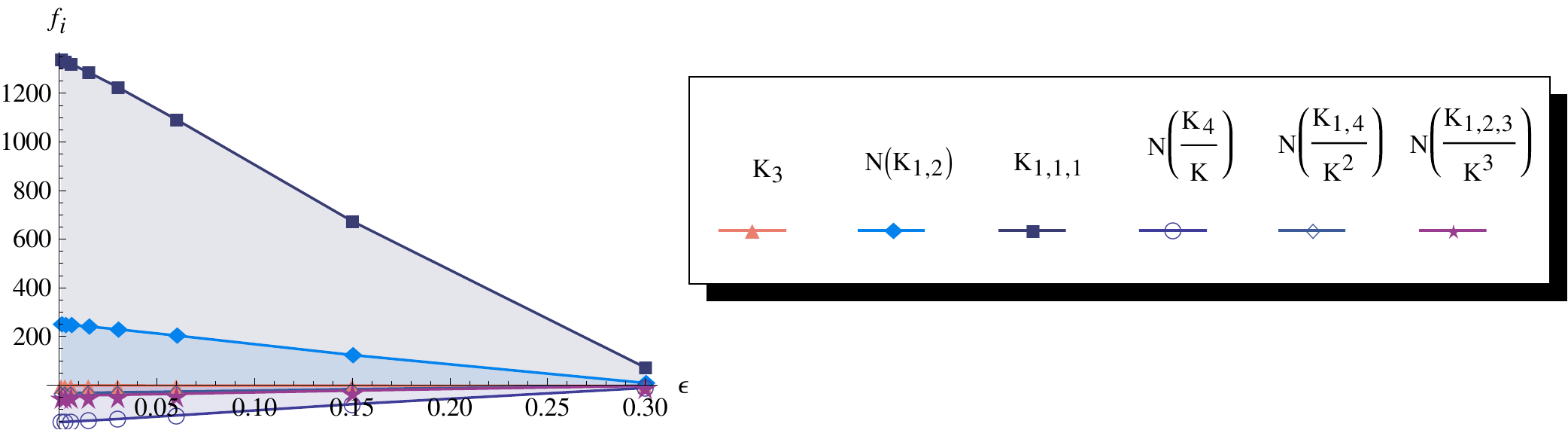}
 }
 \subfigure[~$\log(\cdot)$ term contribution]{
  \label{fig:NM_expectlogk_negfx}
  \includegraphics[width=0.7\textwidth]{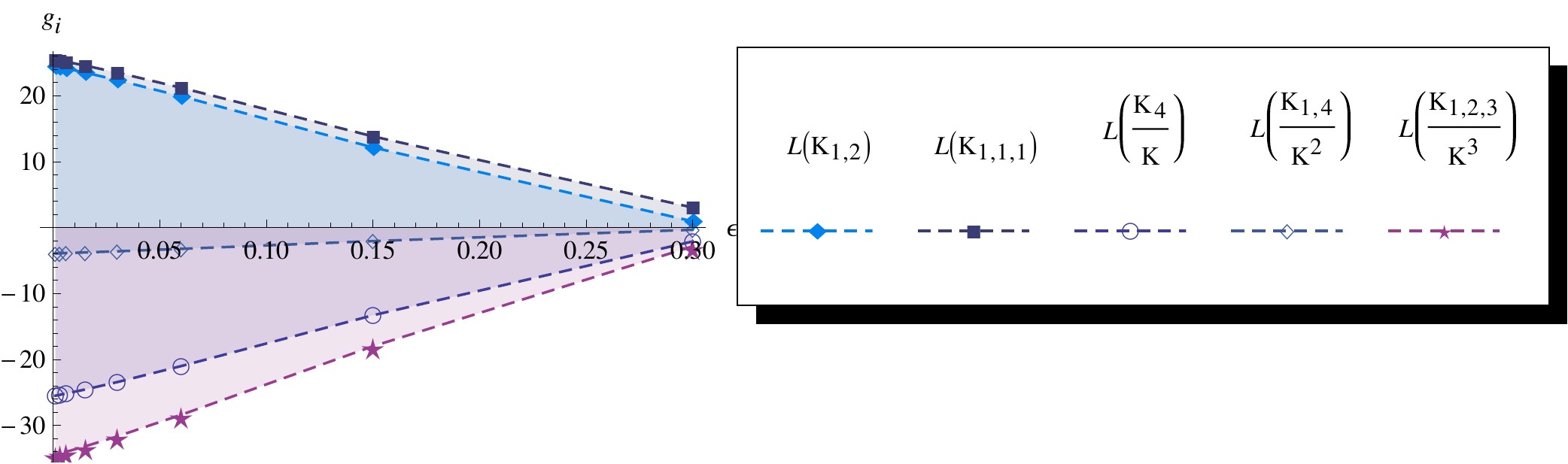}
 }
 \subfigure[~reconstructed contributions]{
  \label{fig:NM_reconst_negfx}
  \begin{minipage}[c]{0.47\textwidth}
   \includegraphics[width=1\textwidth]{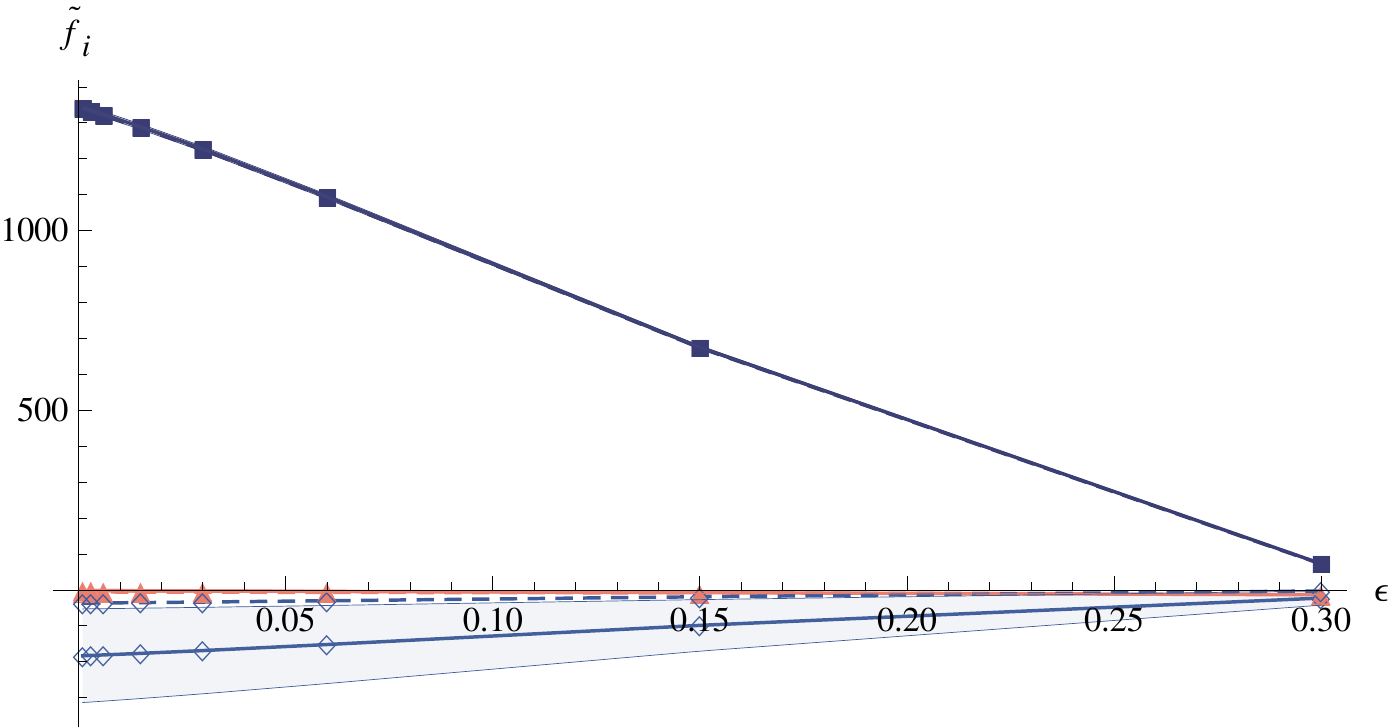}\\[.25cm]
   \includegraphics[width=1\textwidth]{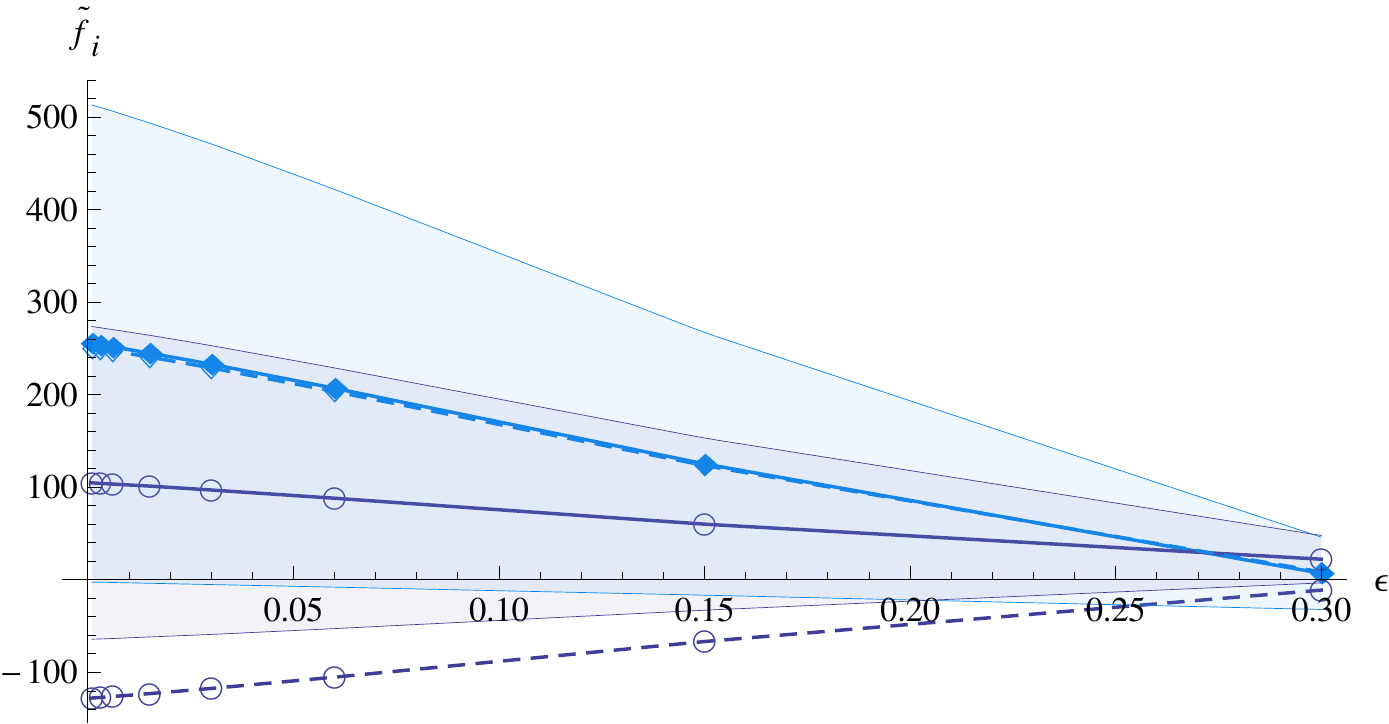}
  \end{minipage}
  \begin{minipage}[c]{0.47\textwidth}
   \vspace{1cm}
   \includegraphics[width=1\textwidth]{ImagesJCAP/NollerMagueijo_ModelDecompLeastLegend_GenDBI.pdf}\\[1cm]
   \includegraphics[width=1\textwidth]{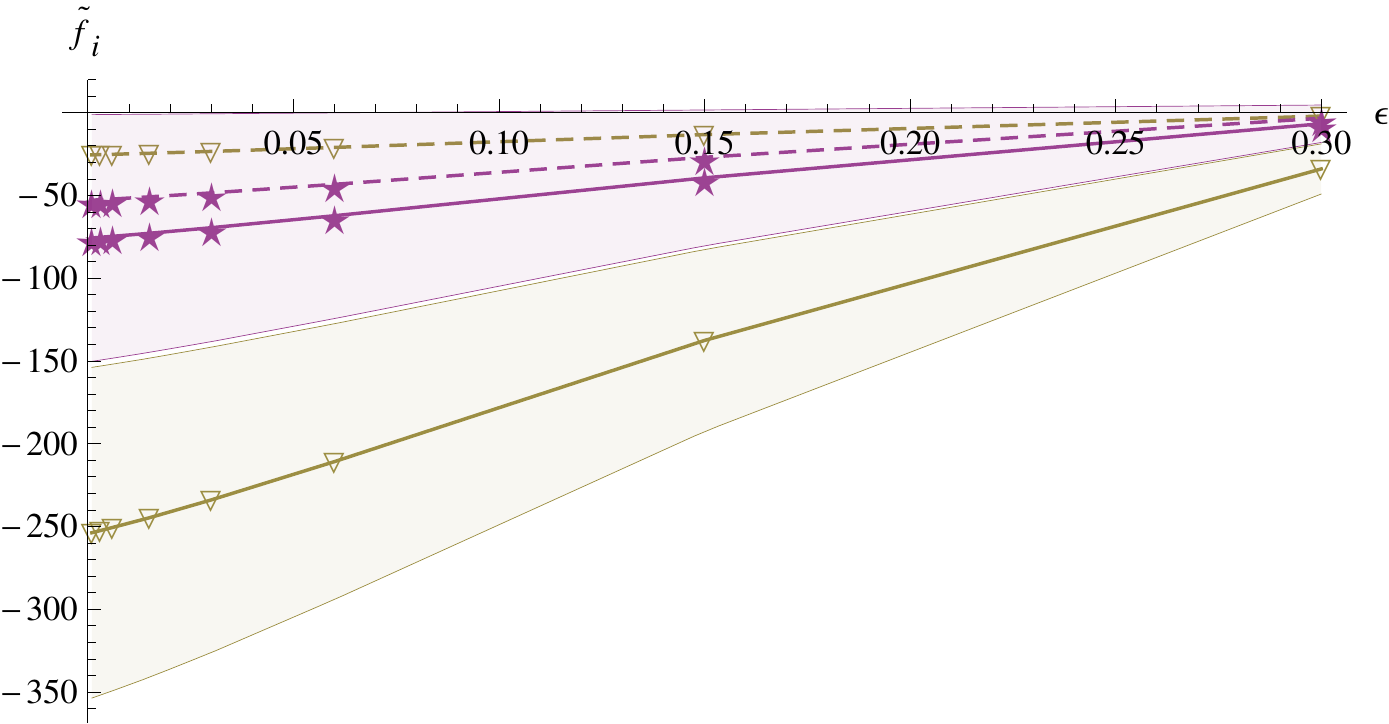}
  \end{minipage}
 }
 \caption{\textbf{(a,b)} Contributions of the terms in \eqRef{eq:NM_shape_ordalpha1} to the bispectrum amplitudes for the negative $f_X$ model defined in \sectRef{sec:slowRollViolatingModeExpansion} for a certain set of example parameters. \textbf{(c)} Results of the linear regression on the expansion coefficients $\alpha_n$ as in \eqRef{eq:linRegress}.}
 \label{fig:NM_model_decomp_negfx}
\end{figure}

The reconstructed coefficients deviate slightly from the predictions as shown in \figRef{fig:NM_reconst_gendbi}, but the order of the contributions has been fully recovered. The two largest contributions, the constant mode and the $\normEtc{K_{12}}$ shape (corresponding to the equilateral and orthogonal template) are well reconstructed. The $95\%$ confidence intervals (assuming Gaussian distributed $\alpha_n$) of the linear regression results are indicated by the shaded regions.\footnote{We use \texttt{Mathematica}'s \texttt{LinearFitModel}. The confidence intervals of the parameters are computed using student-t statistics, where the parameter's standard errors are taken from the diagonal terms of the covariance matrix.} Corresponding to the large correlations among the base shapes, the confidence intervals are of the same order as the reconstructed values for all shapes except for the (orthogonal) constant mode, the (clipped) local template, and the $N[K_{12}]$ shape, probably caused by a systematic effect due to the employed clipping mechanism.

Why are the $\normEtc{K_{14}/K^2}$ and $\normEtc{K_6/K^3}$ contributions not well recovered? There are systematic deviations for both contributions that can be explained as follows: first of all, our predictions are first order in $\boldalpha_1$ only, which can lead to small deviations. 
Second, we neglect the differences between the ``binned'' and referenced shapes. We find that this effect is insignificant for the general DBI model, as the expected mode coefficients and the sum over binned amplitudes as well as reference mode coefficients are almost the same, see \figRef{fig:NM_reconstr_alphas_negfx} for an exemplary reconstruction for $\epsilon = 0.075$. However, the differences between coefficients of the $\Ord{\boldalpha_1}$ prediction and the original expansion coefficients can explain the systematic deviation of the reconstructed contributions.

\begin{figure}[t!]
 \centering
 \includegraphics[width=0.9\textwidth]{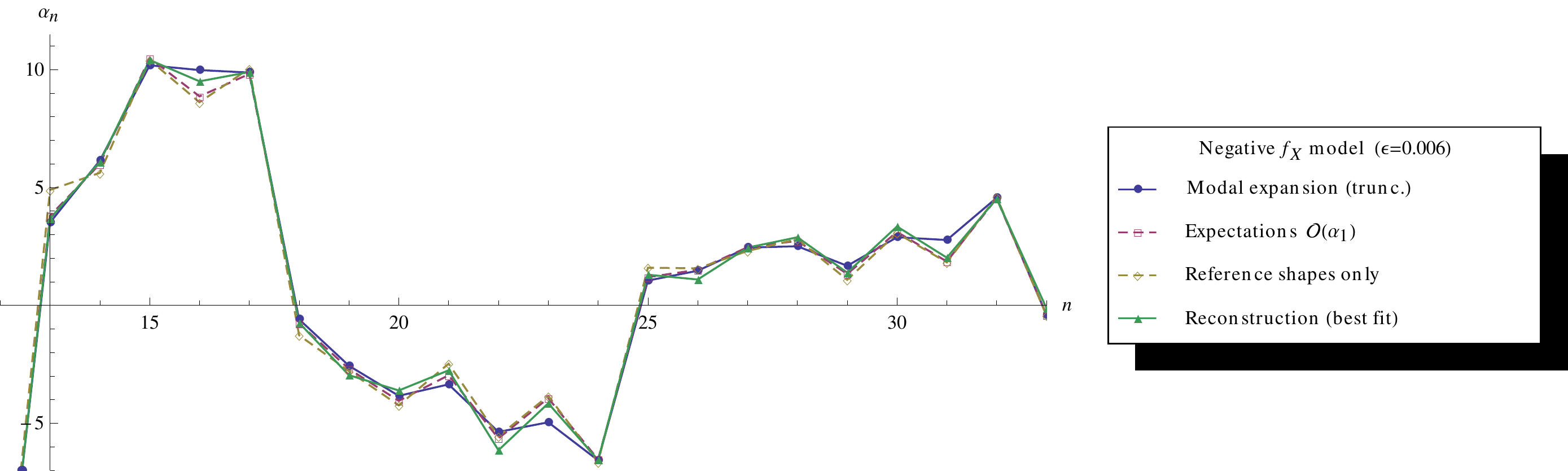}
 \caption{Calculated, predicted (up to $\Ord{\boldalpha_1}$) and recovered (from a linear regression using reference shapes) expansion coefficients of the negative $f_X$ model with $\epsilon = 0.006$. We plot the coefficients of the bispectrum amplitude (not normalized) in order to compare with predictions in the range $n = 13, \ldots, 33$.}
 \label{fig:NM_reconstr_alphas_gendbi}
\end{figure}

We likewise reconstructed the bispectrum amplitudes for the negative $f_X$ model. Again, the order of the contributions is fully recovered and the two largest contributions are well recovered quantitatively (see \figRef{fig:NM_model_decomp_negfx}). The local contribution is, as expected, relatively small compared to the equilateral one.

We can explain the large confidence intervals for the base shapes with the same reasoning as for the general DBI mode, that is by large correlations. Systematic deviations are present for the $\normEtc{K_{14}/K^2}$ and $\normEtc{K_6/K^3}$ contributions, caused again by the same systematic effects explained above. Looking at the calculated, expected and reconstructed coefficients for the exemplary case $\epsilon = 0.006$ (see \figRef{fig:NM_reconstr_alphas_gendbi}), we see that binning does not alter results significantly, but the expansion coefficients of the full bispectrum amplitude deviate from the $\Ord{\boldalpha_1}$ expectation, leading again to a systematic bias in the reconstruction.

We conclude that  the reconstruction of dominant contributions in expanded bispectra, the standard templates in the case at hand, is not only qualitatively, but also quantitatively possible. The reconstruction of the remaining subleading basic shapes is more challenging. The main drawback we encountered is the large correlation between basic shapes, which weakens constraints; furthermore, we found systematics in the not exactly scale-invariant cases, which should be incorporated into the confidence intervals.

\subsection{Summary of our results}
\label{sec:summary}

We first identified the basic contributions to the bispectrum amplitude $\ashape$ of scale-invariant fast-roll models, which are of the form $K_3$, $K_{pq} / K^{p+q-3}$ and $K_{rst} / K^{r+s+t-3}$. These shapes are degenerate (linearly dependent) so that their relative amplitude in an observed bispectrum cannot be separated. We found that naively chosen linearly independent basic terms are prone to high correlations: in \figRef{fig:base_NM_kays_corr_ordered}, it is clear that three terms in \eqRef{eq:defSystematicKaysNM} are of the equilateral type and can hardly be distinguished from the equilateral template. In this case,  we are essentially left with the three common templates (local, equilateral and orthogonal) that measure the divergent contribution removed from our $\normEtc{K_{\ldots}}$ shapes, the equilateral-type shapes and the constant mode. Thus, the contribution of $\Ashape{6} = K_{222} / K^3$ is the only one left to be analyzed after measuring the three common $\fNL$s, because $\Ashape{5}$ does not contribute to the bispectrum amplitude in \eqRef{eq:NollerMagueijoShape}.

Fortunately, this predicament can be alleviated  considerably: to optimize observability and separability of the shape constituents, we identified the least correlated basic shapes (factor analysis) that spans the same space of allowed shapes, see \eqRef{eq:kayInterdep}. This step avoids block structures in the correlation matrix (see \figRef{fig:base_kays_corr}), leading to an invertible matrix. As a consequence, we can estimate the amplitudes of the local template, the constant mode and all five remaining independent shape constituents in a set of expansion coefficients by a linear regression. Furthermore, as these least correlated shapes are also well described by a truncated expansion series (rapid convergence), one can reconstruct expansion coefficients of general bispectra in scale-invariant single-field models fast, which is crucial  for efficient MCMC simulations

Shifting our focus to almost scale-invariant models, we took into account first order corrections in $n_s - 1$, leading to subleading contributions in the bispectrum  of the $\Lop{K_{\ldots}}$ form. To first order, we show that these subleading contributions fall into subclasses composed of the least correlated basic shape constituents, as these terms are strongly correlated with either their ``mother terms'' or another basic shape.

The convergence of the expansion for our example models from \tabRef{tab:model_parameters} is challenged by $\FNL{local}(\Ashaperm{NM}) = \infty$ (unless only terms up to first order in $\Ord{\boldalpha_1}$ are taken into account), which required the use of a clipping technique. The correlation of the expansion series and the full bispectrum plotted in figures~\ref{fig:NM_conv_gendbi} and \ref{fig:NM_conv_negfx} shows that the more relevant the local contribution is, the lower the degree of convergence becomes.

Nevertheless, we showed that reasonably good convergence can still be achieved for two exemplary model families (not exactly scale-invariant, not fine tuned for convergence in any way): they can be described by $\Ord{50}$ expansion coefficients with a loss of information of at most $10\%$ -- even the largest deviation from slow roll (having the highest relative local contribution) is $90\%$ correlated with the full bispectrum while truncating the expansion series at $n_\mathrm{max} = 53$.  We find that contributions of the dominant constant mode, the $\normEtc{K_{12}}$ shape, and (to a lesser degree) the local template can be recovered quantitatively, while a reconstruction is less successful for the subdominant basic shapes (even if the least correlated basic shapes are used).  

The dominant contributions are exactly the ones that are well described by the projected non-linearity parameters $\FNL{equi}$, $\FNL{ortho}$, and $\FNL{local}$; hence, for this particular class of single-field models it is sufficient to focus on these three well known non-linearity parameters. 
The correlation between subclasses of the remaining shapes leads to large uncertainties on the best-fit parameter. In addition, we find systematic effects complicating reconstruction.

To summarize, we provided a complete dissection of possible bispectra in nearly scale-invariant, single-field models of inflation. An important tool are modal techniques, which were used systematically. Further, we would like to stress the importance of using the least correlated basic shape constituents as an intermediate step, both for subsequent efficient MCMC simulations as well as for reconstruction. The toolchain employed in this section is completely general and can be applied to more complicated bispectra (primordial, CMB or late time).

\subsection{Relationship of our analysis to recent work}
\label{sec:recent_work}

While completing this paper, a related article \cite{Ribeiro:2011ax} appeared. Ribeiro and Seery discuss the phenomenology of non-Gaussianities of general scalar field models, whose curved background extensions maintain second-order field equations and stress tensors. The corresponding Lagrangian was first written down by Horndeski \cite{Horndeski:1974} and recently resurfaced within the so-called \emph{Galileon inflationary model}, where a Galilean shift symmetry is required. This model gained popularity over the last years \cite{Burrage:2010cu} (see \cite{Creminelli:2010qf} for related work). The possible NG bispectrum shapes for general Horndeski models are identified in \cite{Gao:2011qe, DeFelice:2011uc, RenauxPetel:2011sb}.

Ribeiro and Seery \cite{Ribeiro:2011ax} relate a newly found orthogonal shape to previously described orthogonal models --- \emph{orthogonal to the equilateral and the so-called orthogonal template} --- by means of a mode decomposition; hence no qualitatively new non-Gaussianity shapes appear in these models. The cases discussed in \cite{Ribeiro:2011ax} are related but not identical to the ones investigated here. Furthermore, our main focus has been the systematic dissection of bispectra. Ribeiro and Seery emphasize a particular new shape that appears in their models and investigate similarities to related shapes.

\subsection{Outlook to application in N-body simulations}
\label{sec:outlook}

Non-Gaussianities are present in large scale structure surveys, but it is challenging to disentangle late-time non-linearities from primordial signals. Nevertheless, observational bounds based on the scale dependent bias \cite{Dalal:2007cu,Matarrese:2008nc,Slosar:2008hx,Xia:2011hj} are becoming competitive with CMB results; to improve upon these bounds, N-body simulations are needed to shed light onto the dynamical evolution of non-Gaussianities. Current simulations use primarily non-Gaussianities of the local type to set the initial conditions for the gravitational potential. However, if non-Gaussianities are observed, it becomes of prime interest to discriminate between different primordial shapes and thus be able to invoke more general initial conditions.

Just as for CMB physics, factorizable templates (local, equilateral and orthogonal) lead to reduced computational effort. However, for more general shapes brute force methods appeared to be the only option \cite{Wagner:2010me, Wagner:2011wx} for a while. 
Given the tools employed in this paper, it is not surprising that one can remove this hurdle by applying modal techniques, as shown recently in \cite{Regan:2011zq}.\footnote{The kernel used in \cite{Regan:2011zq}, which affects the modal decomposition, is only one possibility; further details are given in \cite{Fergusson:2010ia}.} 

We are currently implementing \cite{Steffen}  modal techniques for the bispectrum  as well as the trispectrum  and aim to run large simulations in the foreseeable future with the goal to follow the evolution of shapes as non-linearities grow, ultimately leading to the use of  N-body simulations in conjunction with LSS surveys to constrain primordial non-Gaussianities and thus inflationary models.

\section{Oscillatory and other features in the bispectrum, alternative mode functions}
\label{sec:oscFeat_altModes}

In \SectRef{sec:scaleDepFeat}, a specific scale-dependent model has been introduced, the sharp feature model. Due to slow convergence \wrt\ the polynomial mode functions, see \figRef{fig:poly_alphaFeat_nonconvergence}, a set of alternative mode functions (Fourier modes) was proposed by Meerburg \cite{Meerburg:2010ks}.

In this section, we review Meerburg's choice of Fourier modes and explain why a selected subset of the mode functions increases the correlation between the truncated expansion series and the full amplitude. Furthermore, we apply the alternative modal expansion to a recently discovered shape that stems from  particle production during inflation.

\subsection{Fourier mode functions}
\label{sec:MeerburgModes}

After reviewing the construction of the alternative mode functions, we discuss the improved convergence for oscillatory bispectrum features.

\subsubsection{Construction of the orthonormal basis}

\begin{figure*}[t]
 \centering
 \subfigure[~real part]{
  \label{fig:fourier_qn_re}
  \includegraphics[width=0.5\textwidth]{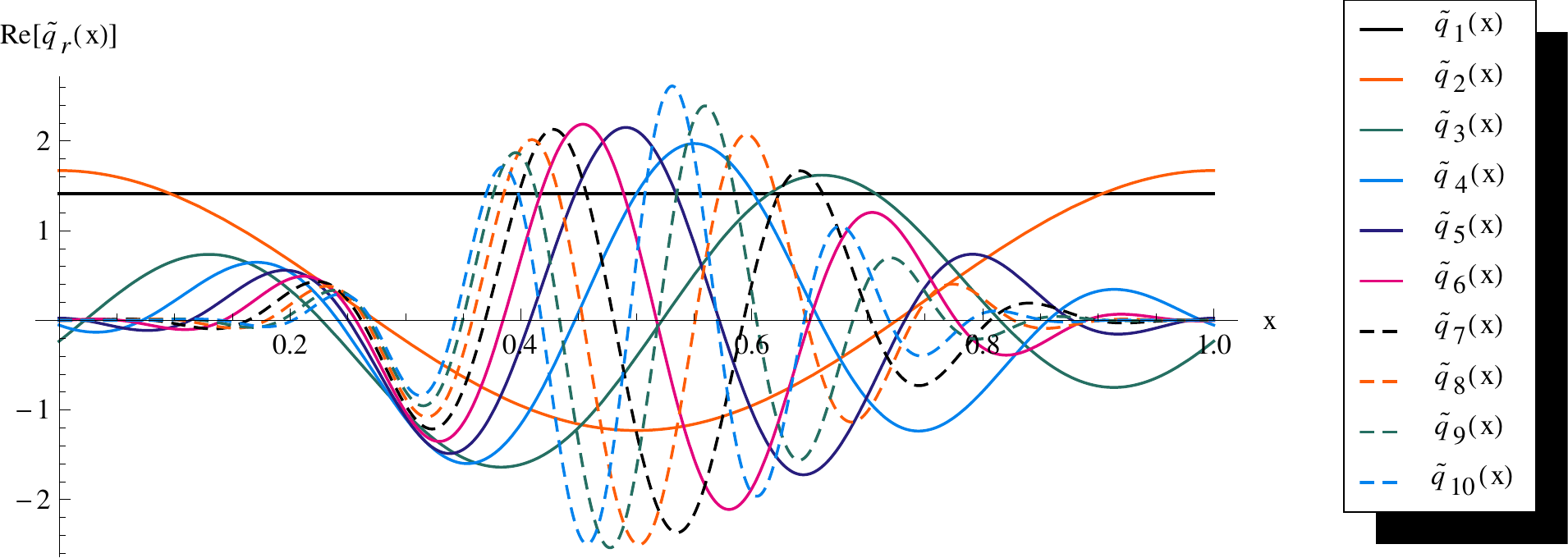}
 }
 \subfigure[~imaginary part]{
  \label{fig:fourier_qn_im}
  \includegraphics[width=0.4\textwidth]{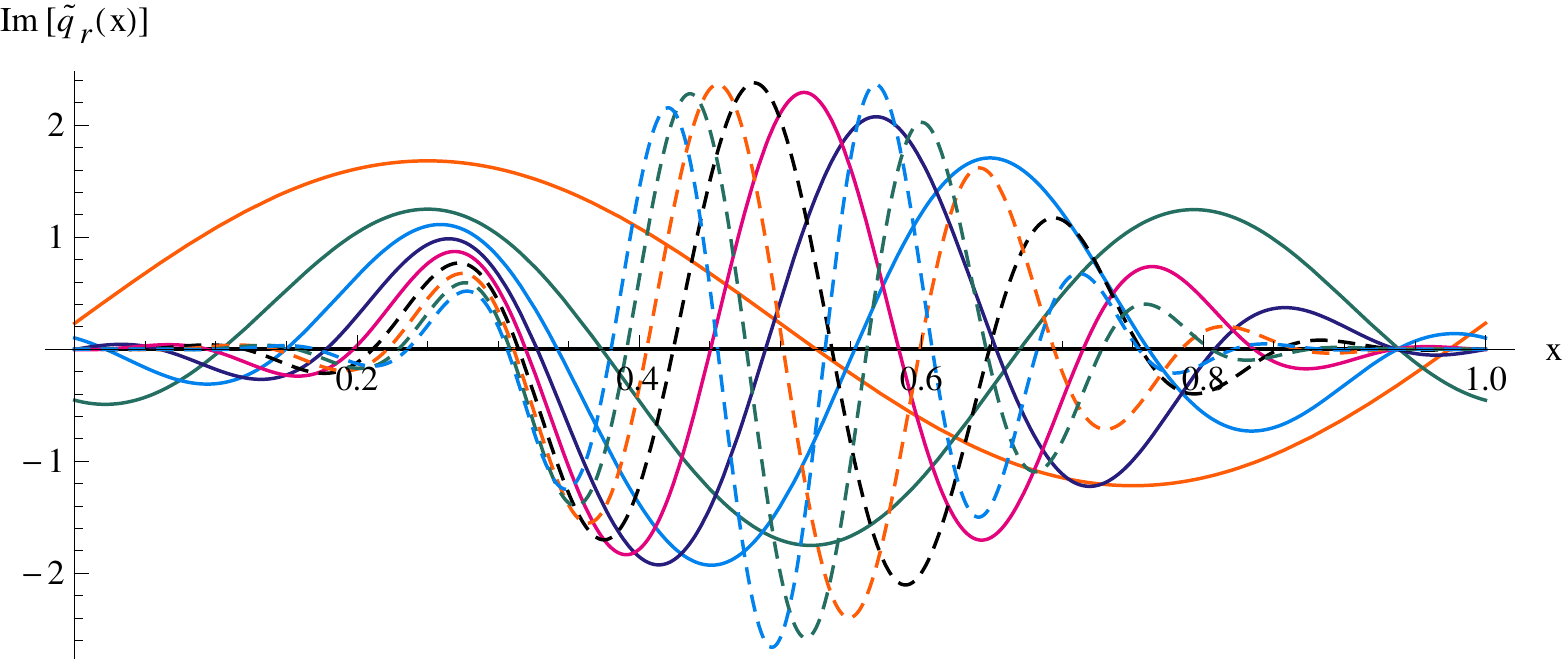}
 }
 \caption{Plot of the first ten orthonormal one-dimensional Fourier modes, $\tilde q_r$ (see \cite{Meerburg:2010ks}, \fig{6}) using a Fourier standard basis instead of monomials.}
 \label{fig:fourier_qn}
\end{figure*}

Using Fourier modes as in \cite{Meerburg:2010ks}, the construction of the orthonormal basis in \sectRef{sec:orthoBasis} needs some minor modifications. First, we replace the monomial standard basis in \eqRef{eq:1dMonomialBasis} by a Fourier standard basis,
\begin{align}
 \tilde e_r \defeq \EE{2 \pi \ii r x}, \quad r \in \N_0,
 \label{eq:1dFourierBasis}
\end{align}
in the definition of the orthonormal one-dimensional functions (orthonormal \wrt\ the weight function $\tilde w$). Since the  mode functions become complex, the second argument  in all inner products $\iprod{f}{g}$ is complex-conjugated.

Denoting all functions based on these Fourier modes by a tilde, we plot the tetrahedal Fourier modes $\tilde q_r(x)$ for $r = 1, \ldots, 10$ in \figRef{fig:fourier_qn}. The next steps are in complete analogy to the polynomial case in \sectRef{sec:orthoBasis}. We apply the symmetrization procedure to compute the $\tilde \modeQ_{prs}(x,y,z)$ and Gram-Schmidt orthonormalization in order to obtain the orthonormal Fourier mode functions $\tilde \orthoR_n(x,y,z)$ on the tetrapyd domain.

\begin{figure*}[t]
 \centering
 \includegraphics[width=0.9\textwidth]{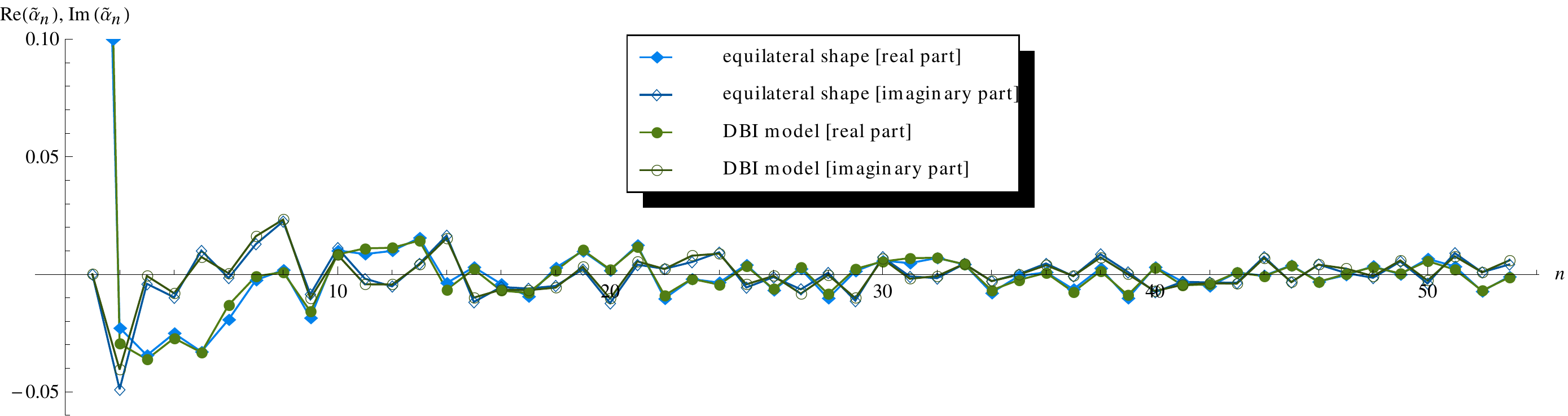}
 \caption{Plot of the expansion coefficients for the equilateral shape and the DBI model \wrt\ the Fourier mode functions $\tilde\orthoR_n$. Note that the constant mode contributions lie outside the plot range at $\tilde \alpha_1 = 0.424$ and $\tilde \alpha_1 = 0.373$, respectively.}
 \label{fig:fourier_alphaEquiVsDBI}
\end{figure*}

\subsubsection{Convergence properties of common shapes}

Because we expand a real shape function $S$ in complex modes, the modal coefficients
\begin{align}
  \tilde \alpha_n \equiv \tilde \alpha^{\tilde \orthoR}_n \defeq \iprod{\tilde \orthoR_n}{\sshape}
\end{align}
fulfill $\tilde \alpha_n^\ast = \tilde \alpha_{-n}$.\footnote{This equality is the reason, why we still can take $r \ge 1$ and $n \ge 1$.} Parseval's theorem (see \eqRef{eq:parsevalTheorem} for the polynomial case) needs a slight modification, \ie
\begin{align}
 \iprod{S}{S} = \tilde \alpha_1^2 + 2 \sum_{n=2}^\infty \re \left[ \tilde \alpha_n \sconj{\tilde \alpha_n} \right].
\end{align}

\begin{figure*}[t]
 \centering
 \subfigure[~equilateral shape and DBI model]{
  \label{fig:fourier_equiDBI_convergence}
  \includegraphics[width=0.7\textwidth]{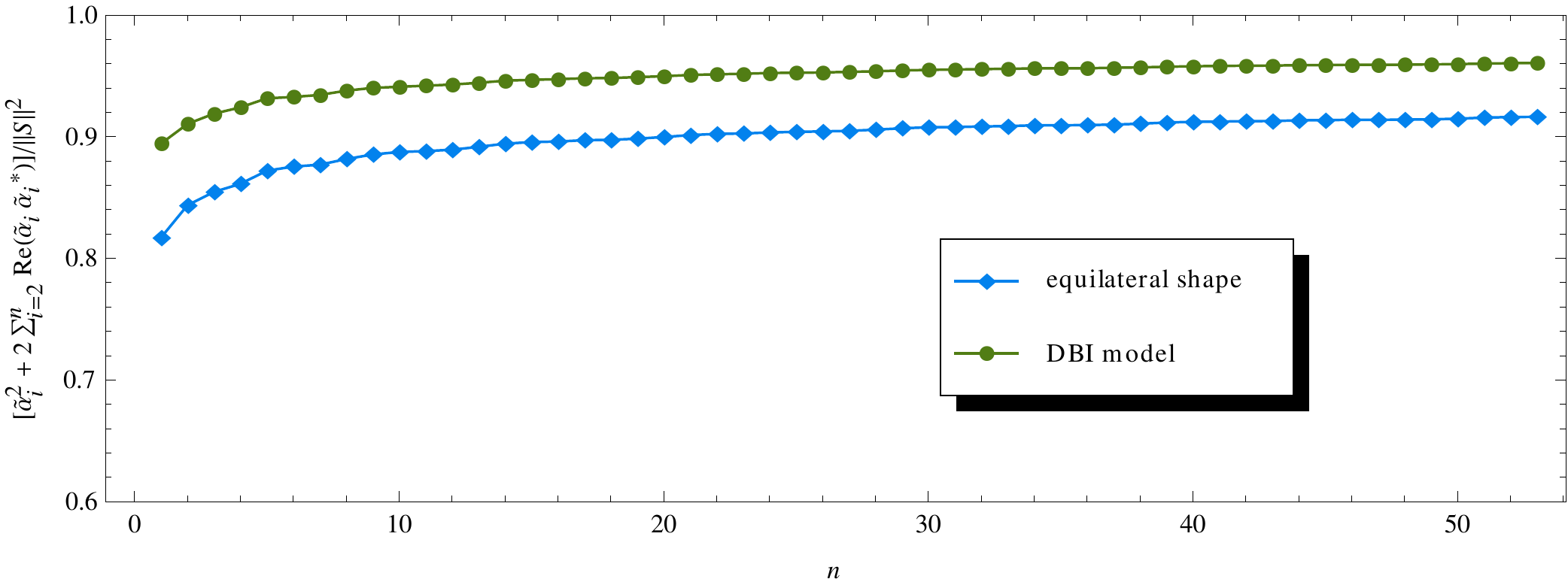}
 }
 \subfigure[~feature shape]{
  \label{fig:fourier_feat_convergence}
  \includegraphics[width=0.7\textwidth]{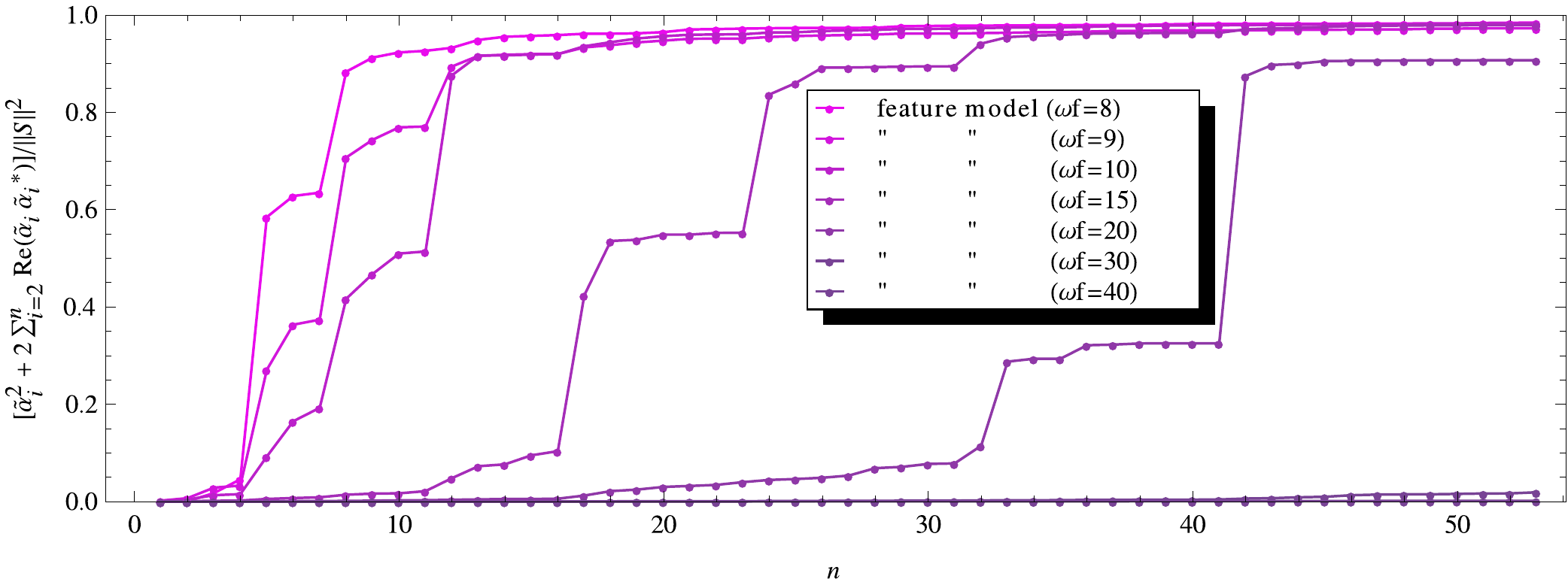}
 }
 \caption{Plot of the sum of squared coefficients \textbf{(a)} for the equilateral shape, see \eqRef{eq:equi_shape}, and the DBI model, see \eqRef{eq:dbi_shape}, and \textbf{(b)} for the feature model as given in \eqRef{eq:Sfeat} for $k_\ast = k_\mathrm{max} / \omega_f$, where $\omega_f \in \{8, 9, 10, 15, 20, 30\}$, and $\delta = 0$. Both expansions are \wrt\ to the Fourier mode functions.} 
 \label{fig:fourier_alpha_convergence}
\end{figure*}

Due to the oscillatory nature of the Fourier modes, we cannot expect a good convergence of our highly correlated test shapes, the equilateral / DBI shapes in \eqs\ \eqref{eq:equi_shape} and \eqref{eq:dbi_shape}. In fact, apart from the constant mode coefficient $\tilde \alpha_1$ (accounting for $81.7\%$ and $89.4\%$, respectively), the first 53 Fourier modes do not contribute much, as shown in \figRef{fig:fourier_alphaEquiVsDBI}. We find $91.6\%$ and $96.1\%$ correlation, respectively, between the truncated expansions ($n_\mathrm{max} = 53$ modes) and the respective full shape.

Hence, the degree of convergence is worse compared to the polynomial expansion for these shapes, as shown in \figRef{fig:fourier_equiDBI_convergence}. In terms of the convergence indicator $\varepsilon$ introduced in \sectRef{sec:convProp}, we find $\varepsilon^\Idx{equi} \approx 0.084$ and $\varepsilon^\Idx{DBI} \approx 0.040$ after $n_\mathrm{max} = 53$ modes, clearly less satisfactory than the polynomial expansion.

However, the replacement of polynomials by Fourier modes pays off in the case of localized and/or oscillatory contributions to the bispectrum, such as in feature models \cite{Chen:2006xjb,Chen:2008wn} and resonant non-Gaussianities \cite{Chen:2008wn, Chen:2010bka,Flauger:2009ab, Flauger:2010ja}. We agree with Meerburg \cite{Meerburg:2010ks} regarding the improved convergence properties, \eg, for the sharp feature models defined in \eqRef{eq:Sfeat}. We plot the convergence of the Fourier mode expansion series for the cases $k_\ast = k_\mathrm{max} / \omega_f$, where $\omega_f \in \{10, 20, 30, 40\}$, in \figRef{fig:fourier_feat_convergence}. In comparison to the polynomial expansion in \figRef{fig:poly_feature}, we see that particular modes improve convergence drastically, while the majority do not. As Meerburg pointed out, this is due to the fact that the sharp feature model is analytically expandable in separable trigonometric functions by using trigonometric identities (\eq{22} in \cite{Meerburg:2010ks}). For lower frequencies, the convergence is improved compared to the polynomial case shown in \figRef{fig:poly_alphaFeat_nonconvergence}. For higher frequencies both expansions are futile, since the oscillations are too rapid to be recovered by the first mode functions. 

Meerburg investigated two other classes of models. The first group encompasses the resonant-type non-Gaussianities that arise due to a periodic feature in the inflaton potential \cite{Chen:2008wn, Chen:2010bka,Flauger:2009ab,Flauger:2010ja} (present, \eg, in monodromy inflation \cite{McAllister:2008hb,Flauger:2009ab}). He took
\begin{align}
 \Sshaperm{res} = \sin \left( \omega_r \ln \frac{k_1 + k_2 + k_3}{k_\mathrm{max}} \right)
 \label{eq:resonant_shape}
\end{align}
as an example shape and found good convergence for the lower part of the allowed frequency range ($20 \le \omega_r \le 10^3$). The number of modes needed is decreased by a factor of $5$. However, he did not achieve more than $15\%$ convergence using $80$ Fourier modes for $\omega_r = 80$, with worse results for higher frequencies. As a consequence, non-Gaussian signals of the resonant type might not be detected within an analysis that is constrained to $\mathcal{O}(100)$ modes, even if Fourier modes are used.

The second group are non-Gaussian shapes that arise in theories with initial state modifications \cite{Chen:2006nt,Holman:2007na,Meerburg:2009ys,Meerburg:2009fi} where small deviations from the Bunch Davies vacuum can lead to large NG signals. The reconstruction of such models is challenging with either basis, because a large number of features coupled with rapid oscillations appear. Furthermore, features in different regions of the tetrapyd have different effective frequencies, introducing a scale dependence (see \cite{Meerburg:2010ks}, \fig{1}).

\subsubsection{Pre-ordering mode functions \label{sec:pre-ordering}}

For resonant bispectra of the type in \eqRef{eq:resonant_shape} (not analytically separable via trigonometric identities) only a shape-specific fraction of mode numbers (see \fig{7} in \cite{Meerburg:2010ks}) is relevant. For this particular shape, the relevant modes are those that oscillate along the $K = k_1 + k_2 + k_3$ direction, consistent with the analytic expression, see \eqRef{eq:resonant_shape}.

Thus, if a shape is restrained to some specific form, one may simply pre-order the mode functions according to their relative contribution to the overall convergence so that considerable less modes are needed for the same degree of convergence.

Maybe more importantly is the application of such a technique to the actual data, such as the CMB bispectrum, which contains known acoustic oscillations. Applying  a factor analysis to the PLANCK data in order to identify the important modes in a polynomial expansion is currently being considered.

\subsection{Particle-production scenarios}
\label{sec:partProd}

We would like to extend the analysis of oscillating, scale-dependent shapes to signals that arise due to particle-production scenarios during inflation. Because of interactions with additional fields, non-inflaton particles can be produced at special points on the inflaton trajectory. The backscattering of these particles leaves a trail in the bispectrum of curvature perturbations in the form of localized features (see \cite{Barnaby:2010sq} for a review). Such events are common in trapped inflation \cite{Kofman:2004yc,Sanchez:2006eq,Green:2009ds,Battefeld:2010sw}, monodromy inflation \cite{Silverstein:2008sg}, or modulated perturbations from extra species point encounters \cite{Langlois:2009jp,Battefeld:2011yj,Cook:2011hg}. 

In order to examine those scale-dependent bispectra, we focus on a shape derived by Barnaby \etAl\ \cite{Barnaby:2009mc,Barnaby:2009dd} for the case of a simple prototype interaction
\begin{equation}
 \Lagr_\mathrm{int} = - \frac{g^2}{2} (\phi - \phi_0)^2 \chi^2
\end{equation}
between the inflaton, $\phi$, and an iso-inflaton, $\chi$ (see \cite{Barnaby:2011vw} for a more realistic model, containing interactions of a pseudo-scalar inflaton with gauge fields). The shapes of the resulting non-Gaussianities are discussed in \cite{Barnaby:2010ke} (see \cite{Barnaby:2011pe} for a recent moment analysis). The imprints in the primordial bispectrum share qualitative features with the ansatz
\begin{align}
 \bspect{\phi}(k_1, k_2, k_3) \propto \prod_{i=1}^3 \left[ \EE{- \Frac{\pi k_i^2}{3 k_\ast^2}} \frac{ 1 - \cos\left( \sqrt{ k_i^2 + m^2 } t \right) }{ k_i^2 + m^2 } \right],
 \label{eq:partProdShape}
\end{align}
where $k_\ast$ defines the characteristic scale at which the feature occurs.\footnote{The full, rather cumbersome bispectrum is also derived in \cite{Barnaby:2010ke}, \eq{81}.} The interaction takes place near the point $\phi = \phi_0$, so that we can take
\begin{align}
 \phi(t) \simeq \phi_0 + v t.
\end{align}
Following \cite{Barnaby:2010ke}, the characteristic scale is defined as $k_\ast = \sqrt{g \abs{v}}$, so that $k_\ast \sim 30 H$ in the case $g^2 \sim 0.1$. Assuming a standard chaotic inflation model, $V(\phi)= m^2 \phi^2 / 2$ with $m \simeq 10^{-6} \sqrt{8 \pi} M_p$ and $\phi = 3.2 \sqrt{8 \pi} M_p$, Barnaby used lattice simulations to extract the probability density function (PDF) for $\delta \phi$ from the simulated data. In addition, he was able to calculate the bispectrum analytically. \EqRef{eq:partProdShape} is a simplified expression that resembles the qualitative behavior, but should not be used for comparison with data. Backscattering also affects the power spectrum, leading to a ``bump feature'' at a wavenumber $k_\mathrm{bump}$ specific to the location of the interaction on the inflaton trajectory. Depending on the relation of the wavenumber to the observable window, the shape of the contribution to the bispectrum differs (\fig{5} in \cite{Barnaby:2010ke}). The feature in the shape function is localized in any case, and oscillatory for $k_\mathrm{bump} \ll k_\mathrm{max}$ (lower right plot of \fig{5} in \cite{Barnaby:2010ke}), corresponding to $k_\mathrm{max} \sim \EE{2} k_\mathrm{bump}$).

Taking $\phi_0 = 3.2\sqrt{8 \pi} M_p$ and $m = 10^{-6}\sqrt{8 \pi} M_p$, while using $g^2 = 0.01$ and $k_\ast = 30 H$, we analyzed the shape in \eqRef{eq:partProdShape} at the time $t = \frac 3 H$ and at the location $k_\mathrm{max} = \EE{2} k_\mathrm{bump}$ with our mode expansion tool-chain and tested it for convergence, both with tetrahedral polynomials and oscillating Fourier modes.

\begin{figure*}[t]
 \centering
 \subfigure[~shape]{
  \label{fig:shape_partProd}
  \includegraphics[width=0.27\textwidth]{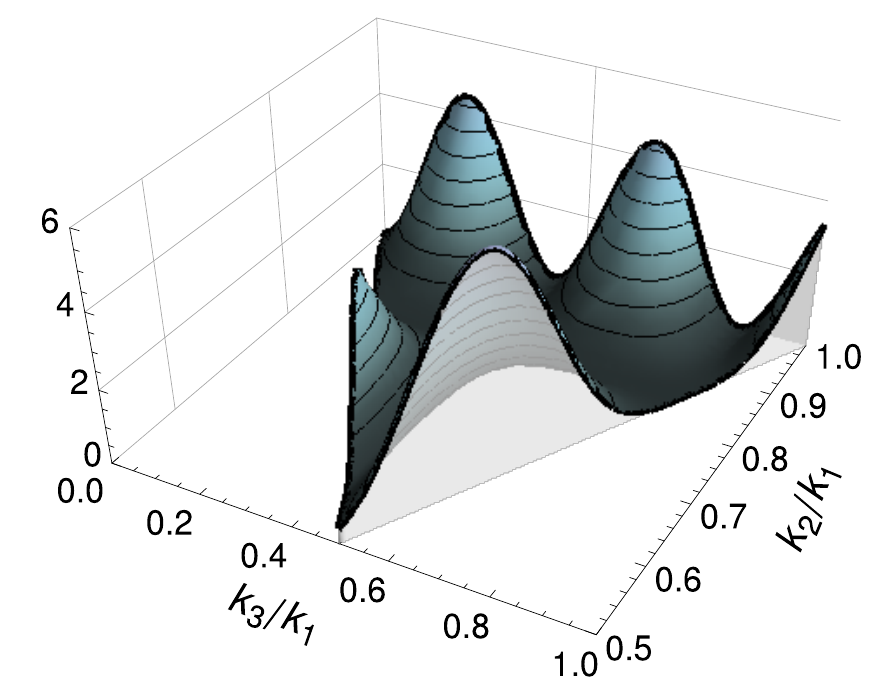}
 }
 \subfigure[~convergence]{
  \label{fig:poly_fourier_alphaBarnaby_convergence}
  \includegraphics[width=0.68\textwidth]{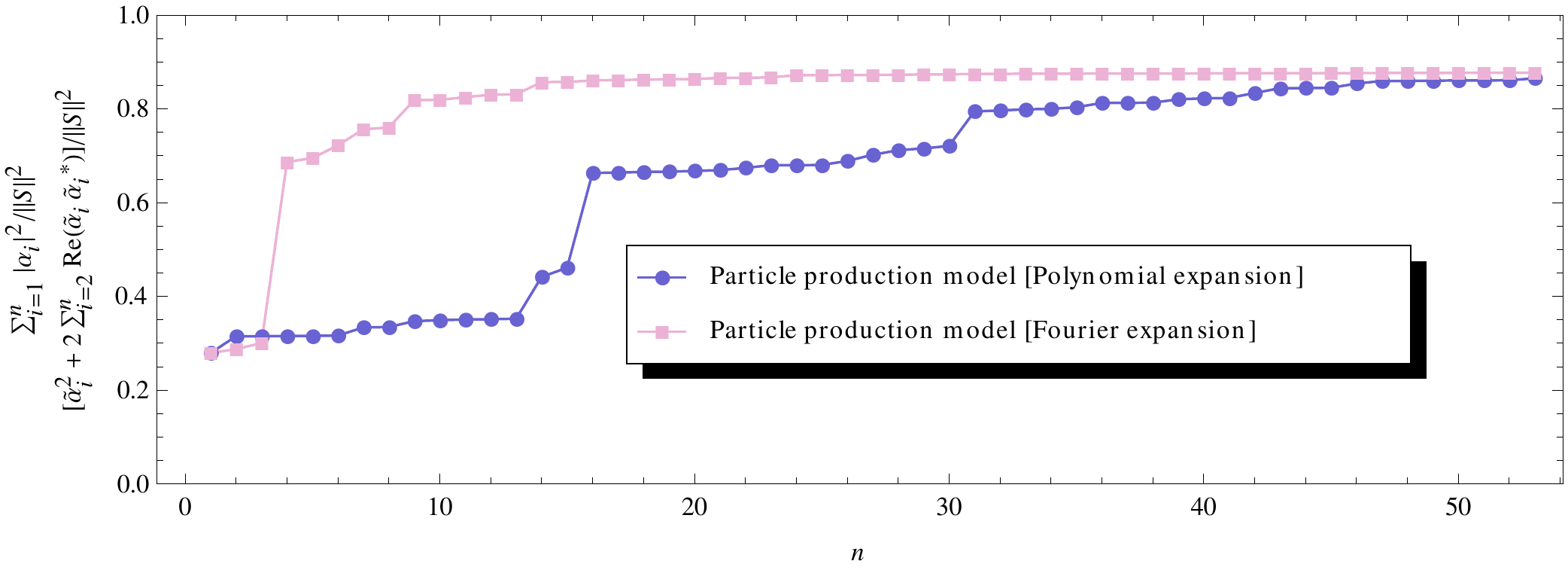}
 }
 \caption{Plot of the sum of squared coefficients for the particle production model as proposed in \cite{Barnaby:2010ke}.}
 \label{fig:poly_alphaBarnaby}
\end{figure*}

It is evident from \figRef{fig:poly_alphaBarnaby} that both expansions achieve a higher than $85\%$ convergence using $n_\mathrm{max} = 53$ modes. This could be expected, because the ansatz in \eqRef{eq:partProdShape} is already separable and of oscillatory form. This explains the sharp rise in \figRef{fig:poly_fourier_alphaBarnaby_convergence} at the fourth Fourier mode; the exponential damping along the $K = k_1 + k_2 + k_3$ direction prevents a faster convergence of the expansion. The polynomial expansion converges slower but reaches almost the same level using $n_\mathrm{max} = 53$.

We would like to emphasize  that the Fourier expansion is able to tell apart the sharp feature model from the particle production scenario by the indices of the relevant modes: these are the modes along the $K$ direction, like the fifth mode, for the feature models, whereas our specific choice of a particle production scenario peaks at the fourth mode (which is $(0,0,2)$, see appendix \ref{sec:slicingOrdering}). As argued in \sectRef{sec:pre-ordering}, a good expansion should be re-ordered, assigning lower indices to modes that are most likely relevant for a particular shape family. These relevant mode coefficients should then be determined from the data and compared to the predictions.

We plan to perform a more detailed investigation of the full bispectrum resulting from both singular or repeated particle production events during inflation in a future publication. In that regard, the relatively good convergence of the Fourier expansions is encouraging.

\section{Conclusion}
\label{sec:conclusion}

Confronted with increasingly involved models of the early universe and equally complex non-Gaussianities, \ie\ the bispectrum shape, it is a daunting task to tell models apart based on limited observations of the cosmic microwave background radiation or large scale structure.

We provided an in-depth investigation of a complex, non-separable primordial shape family (in $k$-space) originating in general single-field models of inflation (Lorentz invariant Lagrangian, at most first order in field derivatives \cite{Noller:2011hd}, reviewed in this paper). We disentangled  the general bispectrum shape into a minimal set of simple basic shape constituents (independent of model parameters), chosen to minimize correlation among each other by means of a factor analysis. These shapes show rapid convergence in a polynomial modal expansion on the tetrapyd domain (expansion techniques of \cite{Fergusson:2009nv} are reviewed in this paper). Once expanded, the corresponding modal coefficients of a general shape can be computed efficiently via a simple sum, a great advantage for subsequent MCMC simulations. Furthermore, the basic shape constituents offer physical insights into the general shape, since they can (in principle) be discriminated in observations. Roughly, the least correlated basic shapes interpolate between the orthogonal mode functions (optimally constrained by the data, but opaque w.r.t.\ physical processes) and the full primordial shape function  (imprint of a cosmological model but not directly accessible ).

We investigated two concrete, slow-roll violating model families (general DBI and $f_X$ models \cite{Noller:2011hd}) resulting in large non-Gaussianities. We provided modal expansions of these shapes and identified the dominant basic shape contributions. We find new shapes that are, however, closely correlated to known templates or suppressed for the models under consideration. Hence, in this concrete class a restriction to common separable templates (local, orthogonal, equilateral) provides a good approximation. The techniques demonstrated in this case study are readily applied to other, more complex models (in preparation) and can also be used for late-time shapes in multipole space, as needed for comparison with the CMB data.

In addition, we briefly summarize current work on N-body simulations, for which we employ the same modal techniques. 

However, not all primordial shapes show the needed rapid convergence in a polynomial modal expansions. As a concrete, hitherto untreatable case we discuss the bispectrum originating from particle production during inflation (more precisely caused by the associated backscattering of produced particles from the inflaton condensate), as present in trapped inflation or monodromy inflation. This shape contains oscillatory as well as localized features. These can be recovered if Fourier modes are chosen instead of polynomials, which we show explicitly. Further, model discrimination is simpler, since key frequencies are readily identified in a Fourier expansion. Hence, it might be prudent to perform future data analysis with more than one basis set that complement each other, such as polynomials and Fourier modes. In addition, proper ordering of modes w.r.t.\ their relative importance to the data could improve convergence and thus efficiency.

\acknowledgments
We would like to thank N.~Barnaby, D.~Battefeld, J.~Fergusson, S.~Klemer, M.~Liguori, E.~Lim, J.~Niemeyer, J.~Noller, D.~Seery, P.~Shellard and R.~Ribeiro, for many useful discussions.

T.~B.~would like to thank the institute for AstroParticule et Cosmologie (APC) and Portsmouth University for hospitality.

\appendix

\section{Slicing ordering of the mode functions}
\label{sec:slicingOrdering}

We use the so-called \emph{slicing ordering} of the mode functions $\modeQ_n = \modeQ_{prs}$; up to $n_\mathrm{max = 53}$ it is (\cf\ \cite{Fergusson:2009nv}, \eq{58})
\begin{align}
  1 &\to \{0, 0, 0\}, &  2 &\to \{0, 0, 1\}, &  3 &\to \{0, 1, 1\}, &  4 &\to \{0, 0, 2\}, &  5 &\to \{1, 1, 1\}, \notag \\
  6 &\to \{0, 1, 2\}, &  7 &\to \{0, 0, 3\}, &  8 &\to \{1, 1, 2\}, &  9 &\to \{0, 2, 2\}, & 10 &\to \{0, 1, 3\}, \notag \\
 11 &\to \{0, 0, 4\}, & 12 &\to \{1, 2, 2\}, & 13 &\to \{1, 1, 3\}, & 14 &\to \{0, 2, 3\}, & 15 &\to \{0, 1, 4\}, \notag \\
 16 &\to \{0, 0, 5\}, & 17 &\to \{2, 2, 2\}, & 18 &\to \{1, 2, 3\}, & 19 &\to \{0, 3, 3\}, & 20 &\to \{1, 1, 4\}, \notag \\
 21 &\to \{0, 2, 4\}, & 22 &\to \{0, 1, 5\}, & 23 &\to \{0, 0, 6\}, & 24 &\to \{2, 2, 3\}, & 25 &\to \{1, 3, 3\}, \notag \\
 26 &\to \{1, 2, 4\}, & 27 &\to \{0, 3, 4\}, & 28 &\to \{1, 1, 5\}, & 29 &\to \{0, 2, 5\}, & 30 &\to \{0, 1, 6\}, \notag \\
 31 &\to \{0, 0, 7\}, & 32 &\to \{2, 3, 3\}, & 33 &\to \{2, 2, 4\}, & 34 &\to \{1, 3, 4\}, & 35 &\to \{0, 4, 4\}, \notag \\
 36 &\to \{1, 2, 5\}, & 37 &\to \{0, 3, 5\}, & 38 &\to \{1, 1, 6\}, & 39 &\to \{0, 2, 6\}, & 40 &\to \{0, 1, 7\}, \notag \\
 41 &\to \{0, 0, 8\}, & 42 &\to \{3, 3, 3\}, & 43 &\to \{2, 3, 4\}, & 44 &\to \{1, 4, 4\}, & 45 &\to \{2, 2, 5\}, \notag \\
 46 &\to \{1, 3, 5\}, & 47 &\to \{0, 4, 5\}, & 48 &\to \{1, 2, 6\}, & 49 &\to \{0, 3, 6\}, & 50 &\to \{1, 1, 7\}, \notag \\
 51 &\to \{0, 2, 7\}, & 52 &\to \{0, 1, 8\}, & 53 &\to \{0, 0, 9\}. 
 \label{eq:slicingOrdering}
\end{align}
Notice that we, in contrast to \cite{Fergusson:2009nv}, start at index $n=1$.

\section{Shorthand notation for the bispectrum shapes}
\label{sec:shapeShorthand}

In order to keep the different combinations of wavenumbers that contribute to the shape functions simple, we follow \cite{Fergusson:2008ra} and list the simplest terms that are consistent with the symmetries as follows:
\begin{align}
 K_p &\defeq \sum_i k_i^p, & K &\equiv K_1 = k_1 + k_2 + k_3, \\
 K_{pq} &\defeq \frac{1}{\Delta_{pq}} \sum_{i \not= j} k_i^p k_j^q \\
 K_{pqr} &\defeq \frac{1}{\Delta_{pqr}} \sum_{i \not= j \not= l} k_i^p k_j^q k_l^r,
\end{align}
where $\Delta_{pq} = 1 + \delta_{pq}$ and $\Delta_{pqr} = \Delta_{pq} (\Delta_{qr} + \delta_{pr})$. $\delta_{pq}$ is the Kronecker symbol.

\subsection{Conversion of \altPdfText{$K$}{K}-terms}
\label{sec:KtermInterdep}

Not all of the allowed $K$-terms in \eqRef{eq:allowedKays} are independent of each other. Here we present the conversion for all terms up to degree 6
\begin{align}
 \left( \begin{array}{c}
  \FRAC{K_{13}}{K} \\
  \FRAC{K_4}{K} \\
  \FRAC{K_{112}}{K} \\
  \FRAC{K_{14}}{K^2} \\
  \FRAC{K_5}{K^2} \\
  \FRAC{K_{113}}{K^2} \\
  \FRAC{K_{122}}{K^2} \\
  \FRAC{K_{15}}{K^3} \\
  \FRAC{K_{33}}{K^3} \\
  \FRAC{K_{114}}{K^3} \\
  \FRAC{K_{123}}{K^3} \\
  \FRAC{K_{24}}{K^3}
 \end{array} \right) = \left( \begin{array}{ccccccc}
  0 & 1 & -2 & -2 & 0 & 0 & 0 \\
  1 & -1 & 2 & 2 & 0 & 0 & 0 \\
  0 & 0 & 1 & 0 & 0 & 0 & 0 \\
  0 & 1 & -4 & 2 & -5 & 0 & 0 \\
  1 & -2 & 6 & 0 & 5 & 0 & 0 \\
  0 & 0 & 1 & -2 & 2 & 0 & 0 \\
  0 & 0 & 0 & 1 & -1 & 0 & 0 \\
  1 & -2 & 6 & 0 & 5 & -1 & 0 \\
  \FRAC{1}{2} & -\FRAC{3}{2} & 6 & -\FRAC{9}{2} & 9 & -\FRAC{1}{2} & \FRAC{9}{2} \\
  0 & 0 & 1 & -3 & 3 & 0 & 3 \\
  0 & 0 & 0 & 1 & -1 & 0 & -3 \\
  -1 & 3 & -12 & 8 & -16 & 1 & -6
 \end{array} \right) \cdot \left( \begin{array}{c}
  K_3 \\
  K_{1,2} \\
  K_{1,1,1} \\
  \FRAC{K_{22}}{K} \\
  \FRAC{K_{23}}{K^2} \\
  \FRAC{K_6}{K^3} \\
  \FRAC{K_{222}}{K^3}
 \end{array} \right).
 \label{eq:kayInterdep}
\end{align}
expressed in terms of ad hoc chosen basic shapes (they happen to be highly correlated and thus in retrospective not an ideal choice).

\section{The clipping mechanism}
\label{sec:clipping_mechanism}

\begin{figure*}[bthp]
 \centering
 \subfigure[~convergence]{
  \label{fig:truncMech_local_conv}
  \includegraphics[width=0.7\textwidth]{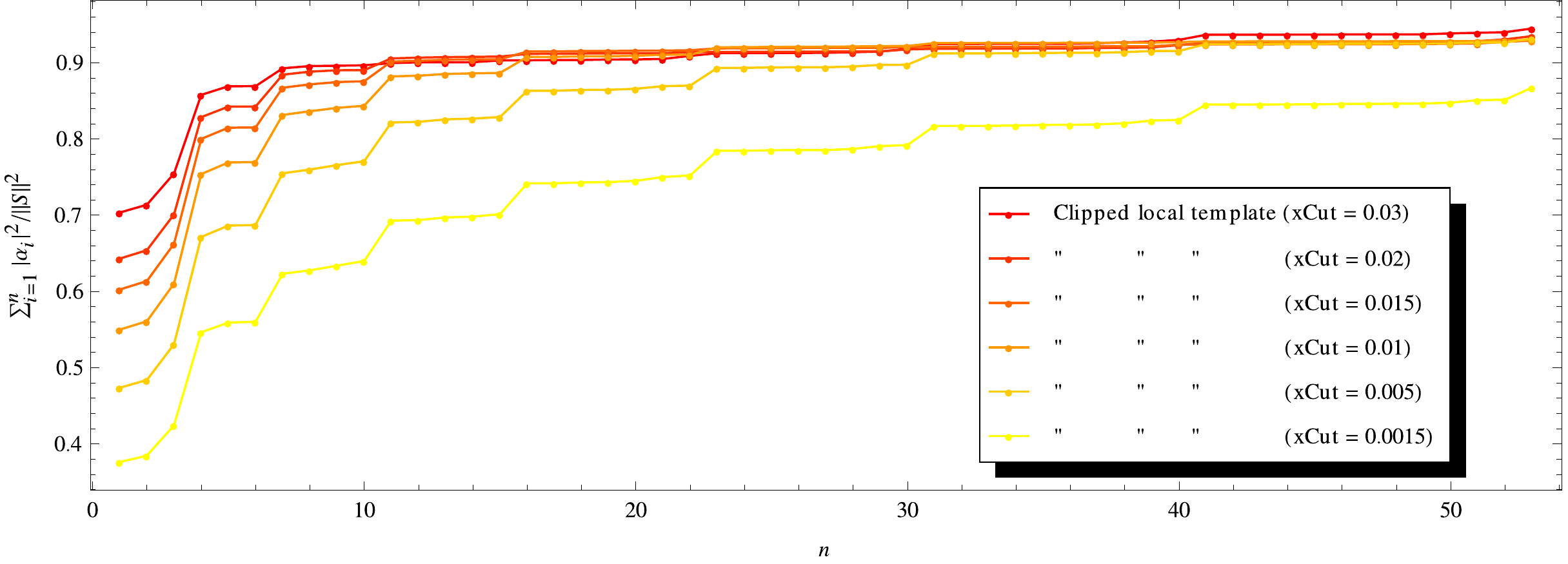}
 }
 \subfigure[~correlation]{
  \label{fig:truncMech_local_corr}
  \includegraphics[width=0.7\textwidth]{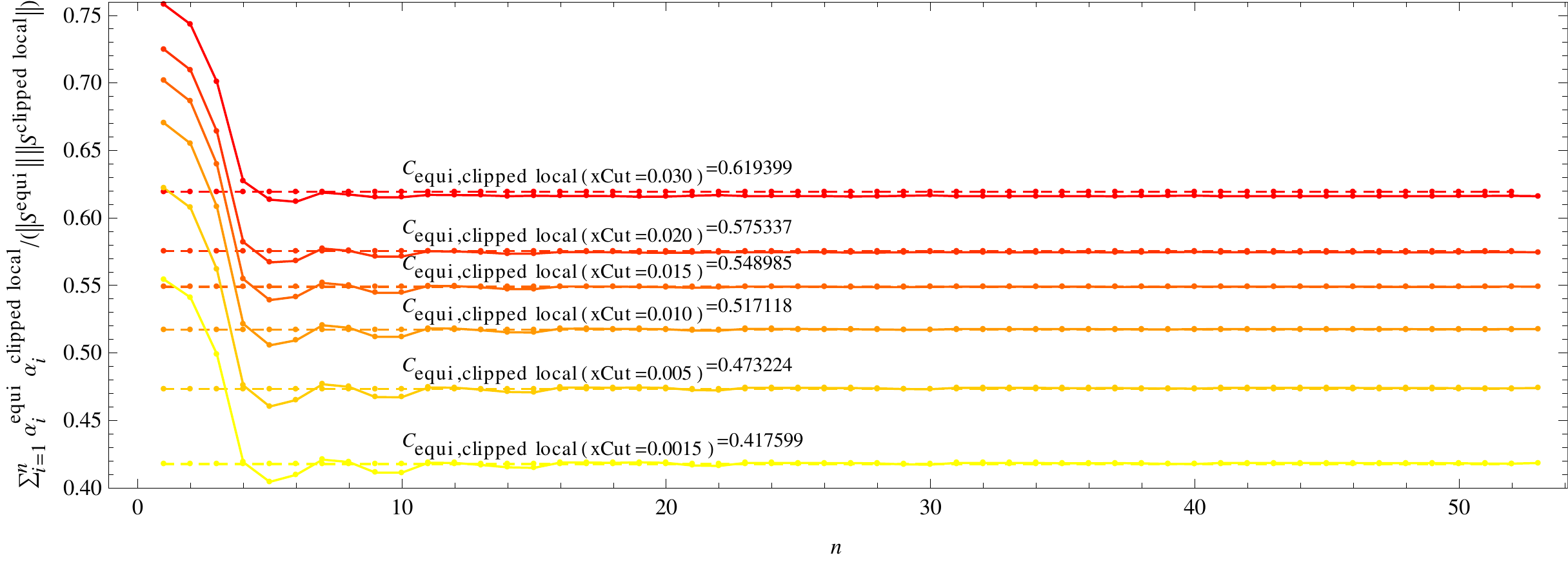}
 }
 \caption{\textbf{(a)} Plot of the sum of squared coefficients for the clipped local shape using different limits. \textbf{(b)} Plot of the correlation between the clipped shapes and the equilateral shape.}
 \label{fig:truncMech_local}
\end{figure*}

We checked the reliability of the clipping mechanism\footnote{We chose the name ``clipping mechanism'' for the truncation of a shape to avoid 
 misinterpretation with the truncation of an expansion series to some finite $n$.} proposed by Fergusson and Shellard \cite{Fergusson:2008ra} by applying different cuts, \ie
\begin{align}
 \frac{k_1}{k_2 + k_3} < x_\mathrm{cut} \quad \text{where} \quad x_\mathrm{cut} \in \{ 0.030, 0.020, 0.010, 0.005, 0.015 \},
\end{align}
to the local shape. The convergence of the expansion series in \figRef{fig:truncMech_local_conv} indicates that the smaller the cut region the worse the convergence. However, the cut area should not be too large either in order to retain the physically relevant region in \eqRef{eq:kminFergShellard}, and to maintain a distinguishable local shape. The correlation with the equilateral shape, see \figRef{fig:truncMech_local_corr}, is a good indicator in that regard. Note that the partial correlation sums feature good convergence properties because the cut region does not contribute much to the correlation.

The proposed cut $x_\mathrm{cut} = 0.015$, which Fergusson and Shellard selected based on an investigation of the late-time behaviour, is satisfactory.

\section{Late-time shapes}
\label{sec:late-time}

We focused on the early-time bispectrum in this paper, given as
\begin{align}
 \bspect{\comovR}(k_1, k_2, k_3) = \frac{N}{(k_1 k_2 k_3)^2} \sshape(k_1, k_2, k_3).
\end{align}
However, for completeness, we would like to mention the relation to the CMB bispectrum (or late-time bispectrum)
\begin{align}
 B_{l_1, l_2, l_3} = \sum_{m_i} \wignertj{l_1}{l_2}{l_3}{m_1}{m_2}{m_3} \average{ a_{l_1 m_1} a_{l_2 m_2} a_{l_3 m_3} },
\end{align}
 which is a convolution of the early-time bispectrum with radiation transfer functions \cite{Komatsu:2001rj}
\begin{equation}
 \begin{aligned}
  b_{l_1 l_2 l_3} = N \powerfrac{2}{\pi}{3} \Intd{x} x^2 \int &\dabl k_1 \dabl k_2 \dabl k_3 \, \sshape(k_1, k_2, k_3) \\
  & \times g_{T l_1}(k_1) g_{T l_2}(k_2) g_{T l_3}(k_3) j_{l_1}(k_1 x) j_{l_2}(k_2 x) j_{l_3}(k_3 x),
 \end{aligned}
 \label{eq:redLateBispecFromS}
\end{equation}
where $g_{T l}(k)$ is the temperature radiation transfer function for the $l$th multipole moment, $j_l(x)$ the corresponding Bessel function and we introduced the \emph{reduced bispectrum} via the relation
\begin{align}
 B_{l_1, l_2, l_3} \eqdef \mathcal{G}^{l_1 l_2 l_3}_{m_1 m_2 m_3} b_{l_1 l_2 l_3}.
\end{align}
Here, $\mathcal{G}^{l_1 l_2 l_3}_{m_1 m_2 m_3}$ is a geometrical factor that depends only on the multipole moments and relates the isotropic reduced bispectrum to the full bispectrum. The convolution in \eqRef{eq:redLateBispecFromS} leads to ``acoustic peaks'' in the late-time bispectrum, as discussed by Fergusson \etAl\ in \cite{Fergusson:2010dm}.

\subsection{\altPdfText{$l$}{l}-domain expansion}
\label{sec:lSpaceExp}

Fergusson \etAl\ \cite{Fergusson:2009nv} identify the tetrahedal polynomials in the multipole domain,
\begin{equation}
 \begin{aligned}
  l_1, l_2, l_3 &\le l_\mathrm{lmax}, \quad l_1, l_2, l_3 \in \N, \\
  l_1 &\le l_2 + l_3 \text{ for } l_1 \ge l_2, l_3, \quad + \text{ cyclic perms.}, \\
  l_1 + l_2 + l_3 &= 2 n, \quad n \in \N,
 \end{aligned}
 \label{eq:multipoleDomain}
\end{equation}
where $l_\mathrm{max} = 2000$ is the maximal wavenumber observable by PLANCK. Motivated by the late-time estimator for $\fNL$, an appropriate weight functions for this domain is
\begin{align}
 \bar w_{l_1 l_2 l_3} = \frac{1}{4 \pi} \prod_i (2 l_i + 1) \wignertj{l_1}{l_2}{l_3}{0}{0}{0}^2,
 \label{eq:lspaceWeight}
\end{align}
which can be approximated by
\begin{equation}
 \begin{aligned}
  \bar w(l_1, l_2, l_3) = \frac{1}{2 \pi^2} \frac{\prod_i (2 l_i + 1) (2 l + \frac 1 3)}{\prod_i (2 l - 2 l_i + \frac 1 3)} \sqrt{ \frac{\prod_i (2 l - 2 l_i + \frac 1 6)}{2 l + \frac 7 6} } \frac{\EE{2 l [\log(2 l) - \log(2 l + 1)] + 1}}{2 l + 1}, \\
  \quad l_1 + l_2 + l_3 = 2l, \; l \in \N.
  \label{eq:lspaceWeightApprox}
 \end{aligned}
\end{equation}
Note that there are some minor typographical errors in the version given by Shellard \cite{Fergusson:2009nv}, \eq{46}. Compare the differences in \figRef{fig:lspace_weight}.

\begin{figure*}[bthp]
 \centering
 \includegraphics[width=0.4\textwidth]{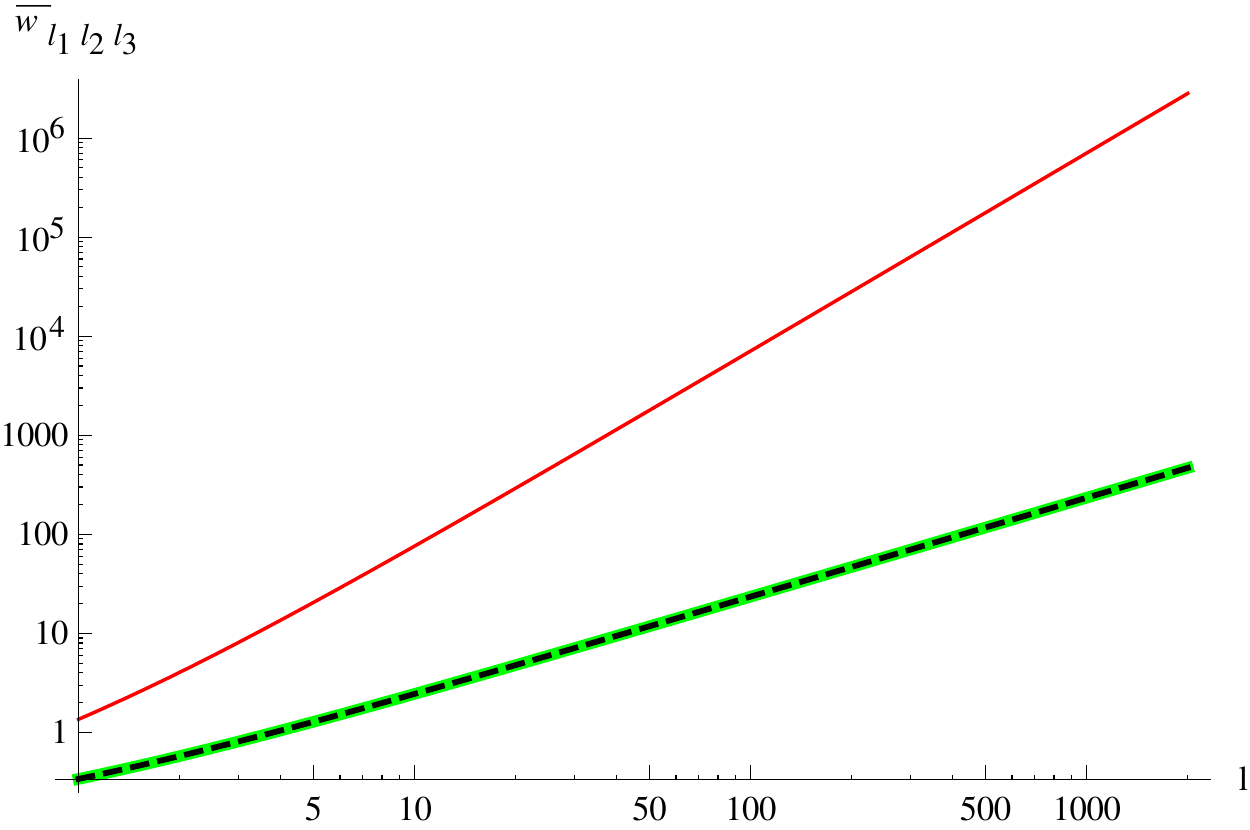}
 \caption{Plot of the $l$-space weight function for the $l_1 = l_2 = l_3 = l$ case: 1), the original version, see \eqRef{eq:lspaceWeight} (black, dashed), 2), our approximation in \eqRef{eq:lspaceWeightApprox} (green, thick), 3), the misprinted approximation in \cite{Fergusson:2009nv}, \eq{46} (red), which, as we suppose, should be corrected to read as in \eqRef{eq:lspaceWeightApprox}.}
 \label{fig:lspace_weight}
\end{figure*}

Using the approximation, they observed that the 1-d polynomials $\bar q_r(x)$ are almost the same as the $K$-domain tetrahedral polynomials in \figRef{fig:poly_qn} (see \cite{Fergusson:2009nv}, \fig{6}, for a comparison). The next step is to use these polynomials to create the separable functions $\bar\modeQ_n = \bar q_{\{p} \bar q_r \bar q_{s\}}$.

\bibliography{bib_FastRollModeExpansion}

\end{document}